\documentclass{article}

\usepackage[utf8]{inputenc}
\usepackage[T1]{fontenc}
\PassOptionsToPackage{round,authoryear}{natbib}
\usepackage[preprint]{neurips_2026}
\usepackage{amsmath,amssymb,amsfonts,amsthm,mathtools}
\usepackage{bm}
\usepackage{booktabs}
\usepackage{enumitem}
\usepackage{xurl}
\usepackage{hyperref}
\usepackage{xcolor}
\usepackage[patch=none]{microtype}
\usepackage{graphicx}
\usepackage{placeins}
\usepackage{needspace}
\hypersetup{
    pdftitle={Decision-Relevant Forecast Calibration for Predictive Control},
    pdfauthor={Victor M. Preciado},
    colorlinks=true,
    citecolor=blue,
    linkcolor=blue,
    urlcolor=blue
}

\newcommand{\R}{\mathbb{R}}

\newcommand{\E}{\mathbb{E}}
\newcommand{\Prob}{\mathbb{P}}
\newcommand{\calA}{\mathcal{A}}

\newcommand{\calD}{\mathcal{D}}

\newcommand{\calG}{\mathcal{G}}
\newcommand{\calH}{\mathcal{H}}

\newcommand{\calP}{\mathcal{P}}
\newcommand{\calQ}{\mathcal{Q}}
\newcommand{\calR}{\mathcal{R}}
\newcommand{\calS}{\mathcal{S}}

\newcommand{\calZ}{\mathcal{Z}}
\newcommand{\argmax}{\operatorname*{arg\,max}}
\newcommand{\argmin}{\operatorname*{arg\,min}}

\newcommand{\val}{\operatorname{val}}

\newcommand{\diag}{\operatorname{diag}}
\newcommand{\supp}{\operatorname{supp}}
\newcommand{\clip}{\operatorname{clip}}
\newcommand{\Var}{\operatorname{Var}}
\newcommand{\Span}{\operatorname{span}}
\newcommand{\Proj}{\operatorname{Proj}}
\newcommand{\indic}{\mathbf{1}}

\theoremstyle{plain}
\newtheorem{theorem}{Theorem}[section]
\newtheorem{proposition}[theorem]{Proposition}
\newtheorem{corollary}[theorem]{Corollary}
\newtheorem{lemma}[theorem]{Lemma}

\theoremstyle{definition}
\newtheorem{definition}[theorem]{Definition}
\newtheorem{assumption}[theorem]{Assumption}
\newtheorem{example}[theorem]{Example}
\newtheorem{remark}[theorem]{Remark}

\AddToHook{env/theorem/before}{\Needspace{5\baselineskip}}
\AddToHook{env/proposition/before}{\Needspace{5\baselineskip}}
\AddToHook{env/corollary/before}{\Needspace{5\baselineskip}}
\AddToHook{env/lemma/before}{\Needspace{5\baselineskip}}
\AddToHook{env/assumption/before}{\Needspace{5\baselineskip}}
\AddToHook{env/definition/before}{\Needspace{5\baselineskip}}
\AddToHook{env/example/before}{\Needspace{5\baselineskip}}
\AddToHook{env/remark/before}{\Needspace{5\baselineskip}}

\title{Decision-Relevant Forecast Calibration\\
for Predictive Control}

\author{Victor M. Preciado\\
\small Department of Electrical and Systems Engineering\\
\small Department of Computer and Information Science\\
University of Pennsylvania}
\date{November 15, 2026}

\begin{document}
\maketitle

\begin{abstract}
Calibration evidence for one predictive controller need not remain valid when the frequencies of operating contexts change. Starting from the established decision-calibration identity for selected-plan value, we develop an audit and certification procedure for a declared family of context-density shifts with a frozen predictor/controller and unchanged conditional outcome law. Coverage requires the calibrated span to contain density-weighted selected-plan sensitivities. A residual Gram matrix determines the minimum augmentation rank and an optimal budget of tests for uncovered $L_2$ geometry. This geometric optimum need not minimize bias over probability-feasible conditional means. For finite contexts, an asymmetric absolute-loss dual evaluates that feasible bias, and directional linear programs with a geometric covering factor enclose its continuous worst-shift value. Intersecting simultaneous moment intervals with the conditional-mean domain also gives a probability-feasible certificate and a linear program for selecting an operating shift after validation.

A designed 240-context study requires three additional directions for its three-feature shift family; two spectral directions remove at least $80.8\%$ of the probability-feasible worst-case bias. The designed transfer-validation study uses 100,000 episodes per law. A separate misspecified logistic forecaster is trained on physical-covariate outcomes. Matched rank-ten sensitivity and generic contextual corrections reduce expected cost by $2.590\%$ and $3.239\%$, with the generic method winning all 200 calibration refits. On 100,000 new episodes drawn after freezing the validation protocol, both corrections certify absolute evaluation bias below the prespecified illustrative tolerance $0.04$ throughout the declared velocity-shift interval, while the base has no certified shift at that tolerance. The synthetic law was available when this design was chosen; the primary outcome is highly predictable under that law. The probability-feasible certificate gives no numerical improvement over the moment rectangle in this sample. The contribution is reusable evaluation and audit capability, with explicit geometric and statistical limits, rather than recalibration superiority. The evidence uses synthetic finite contexts, known context frequencies and committed plans; evaluation bias, oracle regret, constraint risk and feedback deployment remain distinct.
\end{abstract}


\section{Introduction}
\label{sec:introduction}

A predictive controller chooses a plan using forecasts of future demand, disturbances, or other agents. Forecast validation usually measures prediction error or calibration across forecast values. These criteria need not determine the error in the utility assigned to the selected plan. The relevant question is which residual moments suffice for that decision problem.

We study a forecast interface of the form $U_h(a,p)=c_h(a)+r_h(a)^\top p$. Here $h$ is the observed history, $a$ is a feasible finite-horizon plan, and $p$ summarizes the future uncertainty needed to evaluate that plan. The nominal controller selects $\pi_{\rm nom}(h,v)$ by optimizing against the learned forecast $v=f(h)$. Its utility-sensitivity field is $g_{\rm nom}(h,v)=r_h(\pi_{\rm nom}(h,v))$. The expected evaluation error is therefore the moment $\E[g_{\rm nom}(H,V)^\top(Y-V)]$, rather than a generic prediction loss.

\subsection{A scalar example}
\label{sec:main-idea}
Consider a bounded disturbance $W_t$, an input in a compact interval, the dynamics $x_{t+1}=x_t+u+W_t$, and the realized utility $-(x_{t+1}-x_{\rm ref})^2-\lambda_u u^2$, with $\lambda_u>0$. Set $z_h(u)=x_t+u-x_{\rm ref}$ and $Y=(W_t,W_t^2)^\top$. The conditional expected utility is
\begin{equation}
\E[\mathcal U^{\rm real}(u,W_t)\mid H=h]
=-z_h(u)^2-\lambda_u u^2-2z_h(u)\E[W_t\mid h]-\E[W_t^2\mid h].
\label{eq:intro_expected_utility_moments}
\end{equation}
The two moments are a utility-sufficient summary: replacing them by a forecast $p=(p_1,p_2)^\top$ gives
\begin{equation}
U_h(u,p)=-z_h(u)^2-\lambda_u u^2-2z_h(u)p_1-p_2,
\qquad r_h(u)=(-2z_h(u),-1)^\top.
\label{eq:intro_forecast_utility_affine}
\end{equation}
The selected input is $u^{\rm nom}=\pi_{\rm nom}(h,v)$. Its forecast-dependent evaluation error is exactly $r_h(u^{\rm nom})^\top(Y-v)$. Residuals may cancel within a forecast bin while this weighted residual does not, because the weight changes with the state and selected input. Conversely, a residual component orthogonal to this weight does not change the expected utility of this fixed plan. The resulting condition is
\begin{equation}
\E_\mu[g_{\rm nom}(H,V)^\top(Y-V)]=0.
\label{eq:intro_decision_calibration}
\end{equation}
We call it sensitivity-weighted decision calibration. Its precise scope is specified in Section~\ref{sec:setup}.

\subsection{Contributions and use}
\label{sec:contributions}
For a single frozen controller under one operating law, the relevant diagnostic is one scalar residual moment. Its decision-calibration meaning is established in the literature. The sensitivity-span characterization applies Hilbert-space orthogonality to identify which test classes determine that moment and to construct an admissible exposed residual when they do not. The plug-in minimax consequence uses the partial-calibration invariance argument; the duality development supplies explicit attainment and policy-recovery qualifications. These results organize the interface and its limits, including the distinction between evaluation bias and oracle-plan regret.

The main extension concerns reuse of calibration evidence across a declared family of operating laws. A change in the frequency of high-velocity initial contexts, for example, changes how residuals are weighted even when the forecaster, controller, and conditional disturbance law are unchanged. The additional fields are products $w g_{\rm nom}$, where $w$ is a context-density ratio. Section~\ref{sec:operating_transfer} characterizes their coverage, computes the largest uncovered shift through an outcome-free Gram matrix, and determines how many additional independent moments are necessary. With a budget of $b$ new tests, singular directions minimize the remaining worst-shift $L_2$ exposure. This is a control-specific audit and test-design calculation using established change-of-measure and singular-value tools, rather than a new general transfer principle.

The constructive MPC basis now has a specified function: its product closure covers an operating-law class beyond the single source moment. Polynomial degrees add, so a general quadratic sensitivity requires cubic tests for affine density ratios; the original quadratic library suffices only under a verified stronger affine-sensitivity structure. The proposed spectral test budget is geometric; it need not minimize either probability-feasible worst-case bias or statistical confidence width. Section~\ref{sec:feasible_bias} supplies the missing feasible-bias audit: an asymmetric absolute-loss dual for binary conditional means, a directional enclosure of the continuous shift maximum, and a mean-domain confidence set that can be queried by linear programming. These constructions use established convex duality and covering geometry; their purpose is to distinguish and compute the decision-relevant guarantees rather than equate the three design criteria. A separate signed-moment confidence set yields a computable certified set of operating shifts from one independent episode sample, without treating repeated shift queries as independent validation experiments.

The numerical evidence separates constructed geometry, fitted-forecast performance, and independent validation. The fixed-law exact-interface study illustrates the organizing geometry, including its small benefit, adverse or inconclusive outcomes, and conservative certificate. The transfer study demonstrates a source-calibrated field that fails under reweighting, a structured-calibrated law that remains invariant, and the predicted test-budget curve. A separately trained forecaster study uses physical-covariate errors and a matched-capacity generic recalibrator. The generic method wins, which limits algorithmic claims while showing that the decision-value error can arise without sensitivity planting. The probability-feasible audit quantifies how much each added test reduces the admissible bias in utility units, with an outer-approximation error bound. A prospectively fixed validation experiment checks operating choices for the learned pipelines and reports nonempty certificates at an illustrative tolerance, together with the lack of improvement from the mean-domain intersection. Retrospective repeated-sampling diagnostics quantify how often those certificates are nonempty and distinguish a predictable known-law illustration from evidence of practical sample efficiency. None of these committed-plan studies establishes a closed-loop control guarantee.

\paragraph{Notation.}
We fix the decision time when writing $(H,V,Y)$. Uppercase letters are random variables and lowercase letters are realizations. Vector dimensions are determined by the forecast interface, not by the state dimension.
\begin{table}[t]
\centering\small
\caption{Principal notation. Role-specific subscripts distinguish the interior margin, ridge penalty, and confidence level.}
\label{tab:notation}
\begin{tabular}{p{0.25\linewidth}p{0.66\linewidth}}
\toprule
Symbol & Meaning\\
\midrule
$H,h,\calH$ & Random history, realized history, and history space.\\
$N;\ L_{\rm ep};\ M$ & Prediction horizon; episode length; number of independent calibration episodes.\\
$V,v=f(h)$ & Random forecast and its realized value.\\
$Y;\ \Delta=Y-V$ & Future utility statistic and forecast residual.\\
$q(h,v)$ & Candidate conditional mean of $Y$; $\mu$ fixes the law of $(H,V)$.\\
$g_{\rm nom};\ \Psi$ & Selected-plan utility sensitivity; certified vector tests.\\
$\xi(h,v);\ \delta_i$ & Feasible conditional-mean perturbation; tolerance of calibration moment $i$.\\
$\rho_{\rm int};\ \kappa_b;\ \zeta$ & Relative-interior margin; bin ridge penalty; confidence failure probability.\\
$e_j;\ q^{\rm true};\ q^{\rm adv}$ & Coordinate unit vector; true conditional mean; adversarial conditional mean.\\
\bottomrule
\end{tabular}
\end{table}

\section{Related Work}
\label{sec:related_work}

\paragraph{Calibration for decision classes.}
Multicalibration strengthens calibration by testing predictions over specified subpopulations or functions \citep{hebertjohnson2018multicalibration,gopalan2022lowdegree}. Decision calibration tailors those tests to downstream choices \citep{zhao2021decisioncalibration}. Its accurate-loss-estimation identity already gives the selected-utility moment for one context-independent loss. Omniprediction asks for one predictor that can be post-processed for a family of losses \citep{gopalan2022omnipredictors}; outcome-indistinguishability methods identify statistical tests that imply such loss guarantees \citep{gopalan2023loss}. In particular, binary decision outcome indistinguishability weights residuals by the loss difference between outcomes. These are close conceptual predecessors. We express those directions through a predictive controller's selected plan, and study how their coverage changes across a specified operating-law family. The initial Hilbert-space span argument supplies organizing geometry, not a new principle of decision calibration or omniprediction. For nonlinear outcome losses, \citet{tang2026dimension} analyze the cost of feature lifting and dimension-free smooth-response alternatives; our finite utility-statistic interface does not evade those complexity questions.

\paragraph{Partial calibration and contextual minimax decisions.}
\citet{kiyani2026partial} derive minimax decisions under partial calibration and show when the plug-in best response is minimax-optimal. The same invariance argument accommodates observed context $h$ and utility $U_h(a,p)$; history dependence by itself is not a separate minimax theorem. We use it as a consequence of a control-specific test choice. Our duality statement makes measurable selection, primal and multiplier attainment, relaxation of nonconvex plans, and saddle-compatible recovery explicit. The examples in Section~\ref{sec:robust} explain why those policy-recovery qualifications are necessary. The operating-shift results ask a different design question: which additional tests cover the products of a fixed controller's sensitivity with a specified density-ratio family, and how should a limited test budget be allocated?

\paragraph{Multicalibration and changes of operating law.}
Universal adaptability uses multicalibration to transfer inferences from a source population to target populations \citep{kim2022adaptability}. Density-ratio test classes and broader forms of invariance are developed by \citet{wu2024bridging}; \citet{ye2026boosting} analyze transfer through cumulative audit classes and their products with density-ratio approximations. We do not claim the change-of-measure identity or product-test sufficiency as new. The contribution here is a predictive-control audit and test-design calculation: a residual Gram matrix gives the worst uncovered operating shift, the minimum number of additional independent moments, and the optimal geometric exposure after a prescribed number of additions. Explicit-MPC structure supplies finite polynomial libraries for those products. A simultaneous moment confidence set then certifies a continuum of specified operating choices from one validation sample. The conditional outcome law must remain unchanged; this is not a guarantee for arbitrary policy-induced distribution changes. When the shift metric is the Gram matrix of the centered density features, its ellipsoid is a $\chi^2$-divergence ball restricted to their span. Variance-sensitive distributionally robust optimization provides related context \citep{duchi2019variance}; the present audit allows a general metric and concerns projected sensitivity products. We do not identify every divergence-constrained ambiguity set with an exact variance formula. The probability-feasible audit in Section~\ref{sec:feasible_bias} uses standard support-function and linear-program duality \citep{rockafellar1970convex} to evaluate a different residual geometry from the spectral relaxation, with a certified directional approximation of its continuous shift maximum.

\paragraph{Partial identification and moment-based robust optimization.}
Bounding a linear target functional subject to known moments of an unknown function is an established partial-identification problem. In a causal setting, \citet{mogstad2018instrumental} construct such bounds for treatment parameters and compute finite-dimensional versions by linear programming. The probability-feasible audit has the same mathematical structure, with a conditional mean $q$ and controller-dependent target fields in place of marginal treatment responses. It also fits moment-based distributionally robust optimization, whose tractable robust counterparts are developed by \citet{wiesemann2014distributionally}. On finite binary contexts, the identification is explicit: assign joint masses $p_{j1}=\mu_jq_j$ and $p_{j0}=\mu_j(1-q_j)$. Calibration moments become linear constraints on these masses, and the signed evaluation bias is a linear objective. We use this standard machinery to compute controller/shift audits and operating choices from simultaneous validation evidence. Neither sharp LP bounds from moment information nor their robust-counterpart construction is claimed as a new general principle.

\paragraph{Statistical and sequential scope.}
\citet{noarov2023scope} relate the possibility of multicalibrating statistical properties to property elicitation. \citet{noarov2025sequential} construct online forecasts with low conditional bias and downstream regret guarantees in an adversarial sequential environment. \citet{marx2025arbitrary} provide calibrated probabilistic forecasts for arbitrary sequences and evaluate downstream energy decisions. Our finite-sample results concern certification of a fixed forecast/controller interface under specified sampling and transfer laws. Independent episodes supply a direct population interpretation. Variance-sensitive radii use the established empirical Bernstein inequality of \citet{maurer2009empirical}; neither that concentration result nor the singular-value approximation used for geometric test design \citep{eckart1936approximation} is claimed as new. A sequential conditional-moment diagnostic requires its own link to the target operating law. We do not replace online prediction or regret guarantees.

\paragraph{Forecast value and elementary decisions.}
The distinction between statistical forecast quality and decision value has a long history. \citet{murphy1993good} distinguishes consistency, quality, and value, while cost--loss analysis evaluates probabilistic forecasts through users' economic decisions \citep{richardson2000skill}. For point forecasts, \citet{ehm2016quantiles} represent consistent quantile and expectile scores as mixtures of elementary scores and use Murphy diagrams to compare forecast rankings. The analogy is that a structured decision family identifies the forecast directions worth testing. Our span condition instead concerns vector residual moments induced by a selected predictive-control plan, under an affine utility representation.

\paragraph{Decision-focused learning and control-relevant models.}
Decision-focused learning trains predictors through their downstream objectives \citep{donti2017task,wilder2019melding,elmachtoub2022smart}; \citet{mandi2024decision} survey its methods and empirical tradeoffs. Identification for control, value-aware model learning, and loss-calibrated inference likewise measure model error through its use \citep{gevers2005identification,hjalmarsson2005experiment,bombois2006least,farahmand2017value,lacostejulien2011approximate}. Value equivalence preserves model predictions only for specified policies and value functions, with linear-span consequences \citep{grimm2020value}. Utility-weighted forecast calibration also appears in constrained portfolio decisions \citep{wright2026utility}. We study a post-hoc certificate for a given predictor and controller, including its reuse over a specified context-frequency family. It does not imply that a sensitivity-based recalibration algorithm outperforms generic context regression. The matched-capacity learned-forecast comparison in Section~\ref{sec:learned_forecast_experiment} tests that distinction. Learning-based and differentiable MPC provide complementary model-learning and sensitivity-computation tools \citep{hewing2020learning,amos2018differentiable}.

\paragraph{Predictive control and uncertainty interfaces.}
Robust and stochastic MPC encode uncertainty through disturbance sets or stochastic models \citep{bemporad1999robust,mayne2000constrained,mesbah2016stochastic,rawlings2017mpc}; distributionally robust methods use probability ambiguity sets \citep{calafiore2006distributionally,esfahani2018wasserstein}. Here the ambiguity set is induced by certified calibration moments. Conformal prediction supplies prediction sets and risk guarantees \citep{vovk2005algorithmic,shafer2008tutorial,angelopoulos2024conformalrisk}, including safe-planning applications \citep{lindemann2023safecomplete} and a decision-theoretic interpretation for risk-averse choices \citep{kiyani2025rac}. Decision-calibrated prediction sets in power-system operations target downstream constraint reliability \citep{stratigakos2026decision}. Our certificates concern expected plan evaluation and do not order constraint-violation risk. Incentive compatibility for strategic agents is a separate requirement from forecast reliability \citep{bergemann2026mechanism}.

\paragraph{Organization.}
Section~\ref{sec:setup} defines the forecast interface and operating law. Sections~\ref{sec:calibration}--\ref{sec:decision_calibration} develop the calibration tests and sensitivity-span characterization. Sections~\ref{sec:robust}--\ref{sec:approx} give robust and approximate consequences, including control-native test constructions. Section~\ref{sec:empirical} treats finite-sample certification. Section~\ref{sec:operating_transfer} develops reusable operating-shift certificates, optimal $L_2$-geometric test augmentation, and a separate probability-feasible audit. Section~\ref{sec:experiments} retains the original exact-interface illustration; Sections~\ref{sec:operating_transfer_experiment}--\ref{sec:fresh_shift_validation} evaluate transfer, probability-feasible bias, independently learned forecast errors, and prospectively fixed validation of operating choices. Section~\ref{sec:discussion} discusses the scientific scope.

\section{Predictive Control with Learned Forecasts}
\label{sec:setup}

This section introduces the formal predictive-control setting used throughout the paper. We describe the closed-loop system, the information histories available to the controller, and the learned forecasts that enter the predictive-control problem. We then specify the future uncertainty statistic that the forecast is meant to approximate, define finite-horizon control plans and their forecast-parametrized utilities, and introduce the nominal plug-in predictive controller. Finally, we define the closed-loop distribution with respect to which calibration and reliability are assessed, thereby setting up the partial-calibration framework developed in the following sections.

\subsection{Closed-loop system and histories}

We consider a discrete-time controlled system
\begin{equation}
    \bm{x}_{t+1}=F(\bm{x}_t,\bm{u}_t,\bm{w}_t),
    \label{eq:system}
\end{equation}
where $\bm{x}_t\in\mathcal{X}$ is the state, $\bm{u}_t\in\mathcal{U}$ is the control input applied at time $t$, and $\bm{w}_t\in\mathcal{W}$ is a realized value of an exogenous disturbance or environment variable. 
At time $t$, before selecting the next input $\bm{u}_t$, the controller has access to the realized closed-loop history
\[
    h_t=(\bm{x}_0,\bm{u}_0,\bm{x}_1,\bm{u}_1,\ldots,\bm{x}_t)
    \in\mathcal{H}_t.
\]
Given this realized history, a learned forecaster produces a vector-valued prediction of control-relevant statistics of future uncertainty:
\[
    v_t=f(h_t)\in\mathcal{P}\subseteq\mathbb{R}^d.
\]
We call $\mathcal{P}$ the \emph{forecast-statistic space}, and assume that it is compact and convex. The forecast $v_t$ need not be a full probability distribution. It may consist of mode probabilities, conditional moments, basis-function coefficients, predicted residual statistics, risk summaries, or any other finite-dimensional statistic used by the predictive controller. When the closed-loop history is viewed as a random object $H_t$, the forecaster induces the random forecast
\[
    V_t=f(H_t),
\]
whose distribution is determined by the closed-loop distribution of $H_t$.

\subsection{Future uncertainty and utility-sufficient summaries}

We next specify the object that the learned forecaster is meant to predict. Let $\bm{W}_{t:t+N-1}$ denote the future disturbance sequence over a prediction horizon of length $N$, and let $\bm{Z}_t$ collect any additional future exogenous quantities that may affect the predictive-control problem, such as obstacle trajectories, human intentions, demand signals, environment modes, or model residuals. We introduce a measurable lifting map
\[
    \varphi:
    \mathcal{W}^N \times \mathcal{Z}
    \to
    \mathcal{Y}\subseteq\mathcal{P},
\]
which extracts the $d$-dimensional statistic of future uncertainty used by the predictive controller. The corresponding random lifted statistic is
\[
    Y_t
    =
    \varphi(\bm{W}_{t:t+N-1},\bm{Z}_t)
    \in\mathcal{Y}.
\]
Thus, $Y_t$ is the random future quantity that will be revealed only after the future disturbances and exogenous variables are realized. Its realized value is denoted by $y_t$. By contrast, the forecast produced at time $t$ is denoted by $v_t=f(h_t)$; it is available before the future uncertainty is realized and is intended to approximate the ideal conditional statistic $\bm{p}_t^\star=\mathbb{E}[Y_t\mid H_t=h_t]$. In this sense, $y_t$ is what eventually occurs, while $v_t$ is the learned prediction used by the controller when choosing its current plan.

We choose the benchmark control-relevant forecast to be the conditional mean of the lifted statistic. This choice is natural because the conditional expectation is the minimizer, in the mean-squared-error sense, among all $H_t$-measurable predictors of $Y_t$. Thus, for a realized history $h_t$, we define
\[
    \bm{p}_t^\star
    =
    \mathbb{E}[Y_t\mid H_t=h_t].
\]
The learned forecast $v_t=f(h_t)$ is intended to approximate this benchmark statistic, but in general it need not be equal to $\bm{p}_t^\star$. Other forecast targets, such as conditional quantiles, risk measures, or distributional summaries, could also be considered if the predictive controller is formulated around those objects. The central question in our work is not whether the forecaster is globally correct as a model of future uncertainty, but whether its errors are calibrated in the decision-sensitive directions that matter for predictive control.

\subsection{Finite-horizon plans and affine utility}

A finite-horizon control plan is a sequence of future inputs chosen using the information available at time $t$. We denote such a plan by
\[
    \bm{a}_t
    =
    (\bm{u}_{t|t},\bm{u}_{t+1|t},\ldots,\bm{u}_{t+N-1|t})
    \in\mathcal{A}(h_t),
\]
where $\bm{u}_{k|t}$ denotes the input planned for future time $k$ using the history $h_t$ available at time $t$. The set $\mathcal{A}(h_t)$ contains the plans that are feasible from the current history, taking into account the system constraints, information available to the controller, and the prediction horizon of length $N$. We write $\bm{a}$ for a generic feasible plan.

The predictive controller assigns a scalar utility to each feasible plan. This utility depends on three objects: the realized history $h$, the candidate plan $\bm{a}$, and the forecasted statistic $\bm{p}$ that summarizes the relevant future uncertainty. We write this forecast-parametrized utility as
\[
    U_{h}(\bm{a},\bm{p}).
\]
Throughout the main theory, we assume that, after lifting the future uncertainty if necessary, this utility is affine in the forecasted statistic:
\begin{equation}
    U_{h}(\bm{a},\bm{p})
    =
    c_{h}(\bm{a})
    +
    \bm{r}_{h}(\bm{a})^\top \bm{p}.
    \label{eq:affine_utility}
\end{equation}
Here $c_{h}(\bm{a})$ is the part of the utility that does not depend on the forecast, while
\[
    \bm{r}_{h}(\bm{a})\in\mathbb{R}^d
\]
is the vector of coefficients multiplying the forecasted statistic. Equivalently, $\bm{r}_{h}(\bm{a})$ is the sensitivity of the predicted utility of plan $\bm{a}$ to perturbations in $\bm{p}$: a forecast error $\bm{\delta}$ changes the predicted utility of the fixed plan $\bm{a}$ by
\[
    U_{h}(\bm{a},\bm{p}+\bm{\delta})
    -
    U_{h}(\bm{a},\bm{p})
    =
    \bm{r}_{h}(\bm{a})^\top \bm{\delta}.
\]
Thus, the vector $\bm{r}_{h}(\bm{a})$ identifies the directions in the forecasted statistic that matter for the utility of the plan under consideration.

\begin{remark}[An exact affine interface]
\label{rem:affine_interface}
The representation in~\eqref{eq:affine_utility} is exact when finite-mode probabilities or a finite collection of moments reproduce the conditional expected utility of every feasible fixed plan. Its validity must be checked for the chosen uncertainty model and horizon. Affinity of a convenient planning surrogate by itself does not establish that it evaluates the physical model's expected utility.
\end{remark}

The affine representation in \eqref{eq:affine_utility} defines how a candidate plan is evaluated once a forecasted statistic is supplied. Thus, given the realized history $h_t$ and the learned forecast $v_t$, the nominal predictive controller selects a feasible plan by solving the  following forecast-based optimization problem:
\begin{equation}
    \pi_{\rm nom}(h_t,v_t)
    \in
    \arg\max_{\bm{a}\in\mathcal{A}(h_t)}
    U_{h_t}(\bm{a},v_t).
    \label{eq:nominal_predictive_controller}
\end{equation}
This controller evaluates each feasible plan using $v_t$ as the control-relevant summary of future uncertainty. After the sequence of inputs in a plan $\bm{a}$ is selected, the predictive controller does not commit to the entire sequence. Instead, it applies only the first input in the selected plan to the system. The state is then observed again, the history is updated with the new state and applied input, a new forecast is produced, and a new finite-horizon optimization problem is solved. This receding-horizon implementation is the standard operating principle of model predictive control \citep{mayne2000constrained,rawlings2017mpc}: later entries of the plan are provisional, guiding the current decision by anticipating future consequences, but are recomputed as new information becomes available. The measurable-plan conditions in Assumption~\ref{assumption:regularity} justify a measurable nominal optimizer. The certificate evaluates that selected plan even when a receding-horizon implementation later recomputes it.

\subsection{Closed-loop distribution}

We now specify the probability distribution with respect to which calibration and reliability are assessed. Fix a time index $t$, or more generally an operating distribution obtained by aggregating closed-loop data over several time indices or operating conditions. Under the operating regime of interest, the closed-loop evolution induces the random history
\[
    H_t
    =
    (\bm{X}_0,\bm{U}_0,\bm{X}_1,\bm{U}_1,\ldots,\bm{X}_t),
\]
and the learned forecaster maps this random history to the random forecast
\[
    V_t=f(H_t).
\]
We let $\mu$ denote the joint law of the information available to the predictive controller and the forecast produced from that information:
\[
    \mu
    :=
    \mathcal{L}(H_t,V_t).
\]
Thus, expectations and calibration statements with respect to $\mu$ are taken over the histories and forecasts that arise under the closed-loop operating regime.

The operating regime that induces $\mu$ may correspond to a baseline controller, a previously deployed controller, or the predictive controller being certified. Unless otherwise stated, all calibration and optimality statements in this paper are relative to this closed-loop distribution $\mu$.

The true conditional mean of the future statistic, given the history and the forecast, is
\[
    q^{\rm true}(h,v)
    =
    \mathbb{E}\!\left[
    Y_t
    \,\middle|\,
    H_t=h,
    V_t=v
    \right],
\]
defined for $(h,v)$ on the support of $\mu$. Since
$V_t=f(H_t)$, conditioning on both variables is equivalent,
on this support, to conditioning on the history alone. Nevertheless, writing
$q^{\rm true}$ as a function of both $h$ and $v$ is useful
because the calibration constraints below will be expressed as tests of histories
and forecasts.

Full closed-loop calibration would require
\[
    q^{\rm true}(h,v)
    =
    v
    \qquad
    \mu\text{-a.s.}
\]
This condition is often too strong for learned forecasters used inside predictive controllers. We therefore work with partial calibration conditions that test selected moments of the forecast residual
\[
    \bm{\Delta}_t
    =
    Y_t-V_t .
\]
Below, the time index is suppressed when an operating law is fixed: $H,V,Y$ denote the history, forecast, and outcome, and $h,v,y$ their realizations.

Throughout the source ambiguity analysis, $\mu$ fixes the law of $(H,V)$ while the conditional mean map varies. Section~\ref{sec:operating_transfer} explicitly introduces target context laws through specified density ratios, with the pipeline and conditional outcome law held fixed. Expectations containing $Y$ also average its specified conditional law. The certificate controls expected selected-plan utility under that law. Replanned values and oracle-plan regret are separate quantities (Section~\ref{sec:replanning_regret}); changing the controller can also change the operating law. A trajectory-safety or stability guarantee requires further control assumptions. The numerical study evaluates the finite-horizon interface directly.

\paragraph{Measurability and integrability.}
The following conditions make the expectations and measurable optimizers used below well defined. They also identify the additional regularity needed for the adversarial duality theorem.
\begin{assumption}[Measurable compact plans and an integrable envelope]
\label{assumption:regularity}
The history space is standard Borel, and the operating law $\mu$ on $\calZ=\calH\times\calP$ is a probability measure, completed when necessary. The set $\calP\subset\R^d$ is nonempty, compact, and convex. All feasible plans lie in a common compact metric space $A_0$. The correspondence $h\mapsto\calA(h)\subseteq A_0$ has a measurable graph and nonempty compact values. The functions $c_h(a)$ and $r_h(a)$ in~\eqref{eq:affine_utility} are jointly measurable in $(h,a)$ and continuous in $a$ for each $h$. There is a nonnegative $b\in L_1(\mu)$ such that
\[
 |c_h(a)|+\|r_h(a)\|_2\le b(h,v)
 \quad\text{for all }a\in\calA(h),\quad\mu\text{-a.e. }(h,v).
\]
The finite calibration tests $\psi_i$ are bounded and measurable. Any randomness used to produce the forecast is included in the information history before the outcome is revealed.
\end{assumption}
The measurable maximum theorem gives measurable maximizers of $U_h(a,p)$ over $\calA(h)$ and measurable maximized values; see Theorem 18.19 of \citet{aliprantis2006infinite}. Its application uses the measurable feasible correspondence, compact values, and the stated measurable/continuous utility, rather than compactness alone. The envelope implies integrability uniformly over feasible plans and $p\in\calP$. The stronger boundedness and interior conditions used in the sensitivity-span theorem are stated separately. Randomized plans, when needed for minimax analysis, are probability measures supported on $\calA(h)$, and their utility is the expectation of~\eqref{eq:affine_utility}.

A feasible deterministic policy is a measurable map $\pi(h,v)\in\calA(h)$. A relaxed policy is a measurable probability kernel supported on that same feasible set.

\section{Closed-Loop Partial Calibration}
\label{sec:calibration}

\subsection{Sensitivity tests}

Let
\begin{equation}
    \Psi=\{\psi_1,\ldots,\psi_k\}
\end{equation}
be a finite class of measurable test functions
\begin{equation}
    \psi_i:\calH\times\calP\to\R^d.
\end{equation}
Each test defines a scalar residual moment
\begin{equation}
    \psi_i(H,V)^\top(Y-V).
\end{equation}
The tests may encode forecast bins, state regions, controller modes, active constraints, safety regimes, or local cost sensitivities.

\begin{definition}[$\Psi$-calibration under the operating distribution]
A forecaster $f$ is \emph{$\Psi$-calibrated relative to $\mu$} if
\begin{equation}
    \E_\mu\left[\psi_i(H,V)^\top(Y-V)\right]=0,
    \qquad i=1,\ldots,k.
    \label{eq:psi_calibration_true}
\end{equation}
Equivalently, writing $q^{\rm true}(h,v)=\E[Y\mid H=h,V=v]$, the condition is
\begin{equation}
    \E_\mu\left[\psi_i(H,V)^\top(q^{\rm true}(H,V)-V)\right]=0,
    \qquad i=1,\ldots,k.
\end{equation}
\end{definition}

The use of vector-valued tests is important. A scalar calibration test $g(H,V)$ enforces a vector moment $\E[g(H,V)(Y-V)]=0$ only if we include all coordinate directions. In control, we often want to enforce projected moments such as
\begin{equation}
    \E\left[r_H(a)^\top(Y-V)\right]=0,
\end{equation}
which are naturally represented by vector tests $\psi(H,V)=r_H(a)$.

\subsection{Calibration-induced ambiguity set}

The controller does not know $q^{\rm true}$. It knows the forecast $f$, the distribution $\mu$ over histories and forecasts, and the certified calibration tests $\Psi$. We therefore define the set of all conditional mean maps consistent with these calibration guarantees.

\begin{definition}[Calibration-induced ambiguity set]
The exact ambiguity set associated with $\Psi$ consists of measurable maps:
\begin{equation}
    \calQ_\Psi
    =
    \left\{
    q:\calH\times\calP\to\calP:
    \E_\mu\left[\psi_i(H,V)^\top(q(H,V)-V)\right]=0,
    \ i=1,\ldots,k
    \right\}.
    \label{eq:ambiguity_exact}
\end{equation}
\end{definition}

The identity map
\begin{equation}
    q_{\rm id}(h,v)=v
\end{equation}
always belongs to $\calQ_\Psi$. Thus $\calQ_\Psi$ is nonempty. As $\Psi$ grows richer, the ambiguity set shrinks. Full calibration corresponds to an infinitely rich test class that forces $q(h,v)=v$ almost surely.

\begin{remark}[Realizability of the ambiguity set]
\label{rem:realizability}
Elements of $\calQ_\Psi$ are calibration-consistent maps into $\calP$; they need not be realizable conditional means of $Y$. If $\calP$ strictly contains the convex hull $\operatorname{conv}(\supp Y)$ of the support of $Y$, then $\calQ_\Psi$ can contain maps taking values outside $\operatorname{conv}(\supp Y)$, so the worst-case model, and hence the robust controller, may be conservative. Taking $\calP=\operatorname{conv}(\supp Y)$ removes this slack; otherwise the calibration-induced robust controller is conservative by construction.
\end{remark}

\begin{definition}[Approximate calibration ambiguity set]
For tolerances $\delta=(\delta_1,\ldots,\delta_k)$, define the set of measurable maps
\begin{equation}
    \calQ_{\Psi,\delta}
    =
    \left\{
    q:\calH\times\calP\to\calP:
    \left|
    \E_\mu\left[\psi_i(H,V)^\top(q(H,V)-V)\right]
    \right|\le \delta_i,
    \ i=1,\ldots,k
    \right\}.
    \label{eq:ambiguity_approx}
\end{equation}
\end{definition}

\subsection{Examples of control-relevant tests}

\paragraph{Global bias tests.}
Taking $\psi_j(h,v)=e_j$, where $e_j$ is the $j$th coordinate vector, enforces zero mean error in coordinate $j$.

\paragraph{Forecast-bin tests.}
For bins $B_1,\ldots,B_m\subseteq\calP$, tests of the form
\begin{equation}
    \psi_{j,\ell}(h,v)=\indic\{v\in B_\ell\}e_j
\end{equation}
enforce bin-wise calibration.

\paragraph{Operating-regime tests.}
For a family of closed-loop regimes $G\subseteq\calH$, tests
\begin{equation}
    \psi_{G,j}(h,v)=\indic\{h\in G\}e_j
\end{equation}
enforce calibration separately in state-space or history-space regions.

\paragraph{Sensitivity-weighted tests.}
For a controller sensitivity $r_h(a)$, the test
\begin{equation}
    \psi(h,v)=r_h(a)
\end{equation}
enforces unbiasedness of the forecast residual in the direction that affects the utility of plan $a$.

\paragraph{Active-set tests.}
If $\sigma(h,v)$ is the active set or qualitative mode of the MPC optimizer, then
\begin{equation}
    \psi_\sigma(h,v)=\indic\{\sigma(h,v)=\sigma\}r_h(\pi_{\rm nom}(h,v))
\end{equation}
requires calibration separately across active-set regimes.

\section{Decision-Calibrated Predictive Control}
\label{sec:decision_calibration}

The calibration-induced ambiguity set~\eqref{eq:ambiguity_exact} describes all conditional future models consistent with the certified tests. We now identify when the nominal controller is already justified by those tests and when the test class leaves an exposed sensitivity direction.

\subsection{Why ordinary decision calibration is insufficient in control}

In a one-shot decision problem with fixed utility $u(a,p)$, decision calibration is based on decision regions
\begin{equation}
    R_a=\{p:u(a,p)\ge u(a',p)\ \text{for all }a'\}.
\end{equation}
Calibration on the indicators $\indic\{V\in R_a\}$ is sufficient for plug-in best-response optimality \citep{kiyani2026partial}.

In predictive control, the utility changes with the history $h$. The same forecast value $v$ can have different downstream meaning depending on the state, constraints, or active set. Thus, unweighted calibration on decision regions or forecast bins may fail to preserve the utility of the chosen control plan.

\begin{example}[Unweighted calibration can miss decision-relevant bias]
\label{ex:weighted_needed}
Let $H\in\{1,2\}$ with equal probability and let the scalar forecast be $V=1/2$. Consider residuals $Y-V=+\varepsilon_{\rm ex}$ when $H=1$ and $Y-V=-\varepsilon_{\rm ex}$ when $H=2$, for some $0<\varepsilon_{\rm ex}<1/2$. Then the forecast is globally unbiased:
\begin{equation}
    \E[Y-V]=0.
\end{equation}
Suppose a nominal plan $a$ has sensitivity $r_H(a)=+1$ when $H=1$ and $r_H(a)=-1$ when $H=2$. Then
\begin{equation}
    \E[r_H(a)(Y-V)]=\varepsilon_{\rm ex}\neq 0.
\end{equation}
Thus the forecast error cancels statistically but not in the utility direction of the controller. A controller that trusts only the unweighted calibration certificate may be systematically misled.

Because $V=1/2$ is constant, every history-blind forecast-bin or static decision-region test reduces to the mean condition $\E[Y-V]=0$. A history-aware regime test also detects this bias: $\E[\indic\{H=1\}(Y-V)]=\varepsilon_{\rm ex}/2$. The example therefore separates history-blind tests from history-aware tests, rather than showing that sensitivity weighting is the only remedy. Corollary~\ref{cor:bin_separation} gives the more general within-cell construction: any partition that leaves sensitivity variation inside a cell can miss a residual in that direction.
\end{example}

This example motivates sensitivity-weighted calibration.

\subsection{Sensitivity-weighted decision calibration}

Recall that the nominal controller selects
\begin{equation}
    \pi_{\rm nom}(h,v)\in\argmax_{a\in\calA(h)}U_h(a,v).
\end{equation}
By the affine utility assumption,
\begin{equation}
    U_h(\pi_{\rm nom}(h,v),p)
    =
    c_h(\pi_{\rm nom}(h,v))
    +r_h(\pi_{\rm nom}(h,v))^\top p.
\end{equation}
The relevant sensitivity test is therefore
\begin{equation}
    \psi_{\rm nom}(h,v)=r_h(\pi_{\rm nom}(h,v)).
    \label{eq:nominal_sensitivity_test}
\end{equation}

\begin{definition}[Sensitivity-weighted decision calibration]
\label{def:decision_calibration}
A forecaster $f$ is \emph{decision-calibrated for the nominal predictive controller} if
\begin{equation}
    \E_\mu\left[
    r_H(\pi_{\rm nom}(H,V))^\top(Y-V)
    \right]=0.
    \label{eq:sensitivity_decision_calibration}
\end{equation}
Equivalently, the nominal sensitivity test $\psi_{\rm nom}$ is calibrated.
\end{definition}

For a single context-independent loss and its induced decision rule, this is the accurate-loss-estimation condition of \citet{zhao2021decisioncalibration}. In the binary-outcome case its weight is the loss difference used by decision outcome indistinguishability \citep{gopalan2023loss}. We use the sensitivity notation to expose the selected controller's dependence on history, feasible plans, and the forecast interface; the scalar identity itself is established decision-calibration structure. It certifies the evaluation of this selected plan and is weaker than calibration of every outcome coordinate within every decision region. In particular, it supplies no general swap-regret guarantee.

This is the scalar selected-plan condition. In practice one often wants a richer, multicalibrated version over operating regimes:
\begin{equation}
    \E_\mu\left[
    \indic\{H\in G\}
    r_H(\pi_{\rm nom}(H,V))^\top(Y-V)
    \right]=0,
    \qquad G\in\calG.
    \label{eq:group_decision_calibration}
\end{equation}
Equation~\eqref{eq:group_decision_calibration} requires decision calibration separately in each regime $G$.

\subsection{A sensitivity-span characterization}

Definition~\ref{def:decision_calibration} gives one sufficient calibration test, but it does not say whether the test class needs to be larger or smaller.  The next result answers that question by giving a geometric \emph{necessary-and-sufficient} condition.  The certified calibration tests must span the nominal MPC sensitivity field $g_{\rm nom}(h,v)=r_h(\pi_{\rm nom}(h,v))$; if they do not, an explicit calibration-preserving residual direction exists that strictly lowers the value of the nominal controller.  The first part of the theorem is the equivalence (an exact invariance characterization, for the fixed selected plan, under bounded calibration-consistent perturbations), and the second part exhibits the loss-inducing perturbation when the span condition fails.

Let
\begin{equation}
    \calZ = \calH\times\calP,
    \qquad
    g_{\rm nom}(h,v)=r_h(\pi_{\rm nom}(h,v)),
\end{equation}
and consider the Hilbert space $L_2(\mu;\R^d)$ with inner product
\begin{equation}
    \langle a,b\rangle_\mu
    =
    \E_\mu[a(H,V)^\top b(H,V)].
\end{equation}
Let
\begin{equation}
    \calS_\Psi
    =
    \overline{\Span}\{\psi_1,\ldots,\psi_k\}
    \subset L_2(\mu;\R^d)
    \label{eq:test_span}
\end{equation}
be the linear span of the certified calibration tests. Because $\Psi=\{\psi_1,\ldots,\psi_k\}$ is finite, $\calS_\Psi$ is finite-dimensional and therefore closed, so the closure in~\eqref{eq:test_span} is redundant and is written only for emphasis.

\begin{assumption}[Interior perturbations and boundedness]
\label{assumption:interior_perturbations}
There is a radius $\rho_{\rm int}>0$ such that
\begin{equation}
    V+z\in\calP
    \qquad\text{whenever}\qquad
    \|z\|_2\le \rho_{\rm int},
\end{equation}
for $\mu$-almost every $(H,V)$. Moreover, the test functions $\psi_1,\ldots,\psi_k$ and the nominal sensitivity field $g_{\rm nom}$ are essentially bounded, that is $\psi_1,\ldots,\psi_k,g_{\rm nom}\in L_\infty(\mu)$ (in particular square-integrable). Boundedness of the test functions and compactness of $\calP$ are the standing regularity conditions of Assumption~\ref{assumption:regularity}; they are assumed once and used consistently here and in Section~\ref{sec:robust}.
\end{assumption}

\begin{remark}[Scope of the interior condition]
The interior condition is local. For probability simplices, use free probability coordinates so that the forecast domain is full-dimensional; the condition then requires a uniform distance from its relative boundary. No simplex has an ambient Euclidean ball in the redundant full-coordinate representation. Boundary forecasts require a separate feasible-cone analysis, which is not established here.
\end{remark}

\begin{theorem}[Sensitivity-span characterization]
\label{thm:sensitivity_span}
Under Assumption~\ref{assumption:interior_perturbations}, the following are equivalent:
\begin{enumerate}[leftmargin=*]
    \item $g_{\rm nom}\in\calS_\Psi$.
    \item For every perturbation $\xi\in L_2(\mu;\R^d)$ satisfying
    \begin{equation}
        \E_\mu[\psi_i(H,V)^\top\xi(H,V)]=0,
        \qquad i=1,\ldots,k,
        \label{eq:delta_calibration_constraints}
    \end{equation}
    and $V+\xi\in\calP$ almost surely, the nominal controller's plan value is unchanged (exactly, by affinity of the utility):
    \begin{equation}
        \E_\mu[g_{\rm nom}(H,V)^\top\xi(H,V)]=0.
        \label{eq:nominal_invariant_all_delta}
    \end{equation}
\end{enumerate}
Moreover, if $g_{\rm nom}\notin\calS_\Psi$, write
\begin{equation}
    g_\perp = g_{\rm nom}-\Proj_{\calS_\Psi}g_{\rm nom}.
\end{equation}
Then $g_\perp\neq 0$ is essentially bounded. For every $0<\alpha\le\rho_{\rm int}/\|g_\perp\|_\infty$, the perturbation
\begin{equation}
    \xi_\alpha(H,V)=-\alpha g_\perp(H,V)
\end{equation}
induces a calibration-consistent conditional model $q_\alpha=V+\xi_\alpha\in\calQ_\Psi$ satisfying
\begin{equation}
    \E_\mu[U_H(\pi_{\rm nom}(H,V),q_\alpha(H,V))]
    =
    \E_\mu[U_H(\pi_{\rm nom}(H,V),V)]
    -\alpha\|g_\perp\|_{L_2(\mu)}^2.
    \label{eq:strict_loss_from_missing_span}
\end{equation}
\end{theorem}

\begin{proof}
The finite test span $\calS_\Psi$ is closed. If $g_{\rm nom}\in\calS_\Psi$, calibration-consistent perturbations are orthogonal to $g_{\rm nom}$, so affinity gives exact fixed-plan invariance. Conversely, if $g_{\rm nom}\notin\calS_\Psi$, let $g_\perp=g_{\rm nom}-\Proj_{\calS_\Psi}g_{\rm nom}$. The projection is a finite linear combination of bounded tests, so $g_\perp$ is bounded and nonzero. For $0<\alpha\le\rho_{\rm int}/\|g_\perp\|_\infty$, the perturbation $\xi=-\alpha g_\perp$ is feasible and satisfies every calibration moment. Yet
\[
\E_\mu[g_{\rm nom}^\top\xi]
=-\alpha\|g_\perp\|_{L_2(\mu)}^2<0.
\]
Affinity gives the stated value loss. Appendix~\ref{app:sensitivity_span_proof} supplies further projection and bounded-density details.
\end{proof}

The invariance in part~2 is exact: for the fixed plan $\pi_{\rm nom}(H,V)$ the utility is affine in the forecast, so the value change equals $\E_\mu[g_{\rm nom}^\top\xi]$ with no higher-order remainder. Section~\ref{sec:replanning_regret} separately treats a first-order change in replanned value. Section~\ref{sec:operating_transfer} treats specified changes of the operating distribution exactly through density ratios and additional product tests; arbitrary deployment changes remain outside these statements.

Theorem~\ref{thm:sensitivity_span} identifies the test directions required by the selected control plan. For an $h$-independent utility and a finite action library, the sensitivity field is constant on each decision region. In MPC, $g_{\rm nom}$ may vary with state, active constraints, costates, and safety margins. The certified tests must therefore span this history-dependent field, not merely the forecast bins.

The next corollary specializes the abstract span condition to the practically common case of forecast-bin calibration: the span condition fails as soon as $g_{\rm nom}$ has nonzero conditional variance inside any populated bin, and the proof exhibits the residual that the numerical study realizes in Section~\ref{sec:experiments}.

\begin{corollary}[Bin calibration can miss heterogeneous control sensitivities]
\label{cor:bin_separation}
Consider the scalar case $Y,V\in[0,1]$, with $V\in[\rho_{\rm int},1-\rho_{\rm int}]$ almost surely for some $\rho_{\rm int}>0$, and a forecast-bin partition $\{B_1,\ldots,B_J\}$. Suppose the certified test class contains only bin indicators, so calibration enforces
\begin{equation}
    \E[Y-V\mid V\in B_j]=0,
    \qquad j=1,\ldots,J.
\end{equation}
Let $s(H,V)=g_{\rm nom}(H,V)$ be an essentially bounded scalar utility or cost sensitivity. If $s$ has nonzero conditional variance inside some bin, then there exists a residual field $\xi(H,V)$ such that
\begin{equation}
    \E[\xi(H,V)\mid V\in B_j]=0
    \qquad\text{for every }j,
\end{equation}
while
\begin{equation}
    \E[s(H,V)\xi(H,V)]>0.
\end{equation}
One such choice is, within each bin,
\begin{equation}
    \xi(H,V)=\lambda_j\left(s(H,V)-\E[s(H,V)\mid V\in B_j]\right),
    \qquad V\in B_j,
    \label{eq:sensitivity_aligned_residual}
\end{equation}
with $\lambda_j>0$ sufficiently small to preserve feasibility.
\end{corollary}

Corollary~\ref{cor:bin_separation} is the template for the numerical study in Section~\ref{sec:experiments}. It shows that ordinary reliability diagrams can be correct while the controller experiences a systematic error in the direction that affects safety or cost.

\subsection{Replanned values and the gap to oracle planning}
\label{sec:replanning_regret}
The span condition is an orthogonality characterization of fixed-plan value. It does not by itself bound the benefit of choosing a different plan with the true conditional mean. This distinction can be expressed exactly through the convex forecast-value function
\[
 \val_h(p)=\max_{a\in\calA(h)}U_h(a,p).
\]

\begin{proposition}[Directional derivative of the replanned value]
\label{prop:replanning_directional}
Under Assumption~\ref{assumption:regularity}, suppose the maximizer at $V$ is unique $\mu$-almost surely. Let $d_0\in L_\infty(\mu;\R^d)$ be fixed, with $V+t d_0\in\calP$ almost surely for all sufficiently small $t\ge0$. Then
\begin{equation}
 \lim_{t\downarrow0}\frac{\E_\mu[\val_H(V+t d_0)-\val_H(V)]}{t}
 =\E_\mu[g_{\rm nom}(H,V)^\top d_0(H,V)].
 \label{eq:replanned_directional}
\end{equation}
Consequently, if $g_{\rm nom}\in\calS_\Psi$ and $\E_\mu[\psi_i^\top d_0]=0$ for every $i$, the expected replanned value has zero first-order change along this direction, with the operating law held fixed.
\end{proposition}
\begin{proof}
The pointwise maximum is differentiable at $V$ in every feasible direction under uniqueness, and its derivative is $r_H(\pi_{\rm nom}(H,V))^\top d_0$, by the maximum-envelope theorem \citep{danskin1967maxmin,bonnans2000perturbation}. For every $p,p'\in\calP$, maximizing the affine utility gives
\[
 |\val_h(p)-\val_h(p')|\le b(h,v)\|p-p'\|_2.
\]
The difference quotient in~\eqref{eq:replanned_directional} is therefore dominated by the integrable random variable $b\|d_0\|_\infty$. Dominated convergence justifies the expectation-level limit. The span conclusion follows by orthogonality.
\end{proof}
This is a directional statement along $t d_0$, not a uniform differentiability claim over all perturbation fields in an unspecified norm. It does not differentiate the operating law or establish a receding-horizon performance bound.

\begin{proposition}[Oracle-plan regret as a generalized Bregman gap]
\label{prop:bregman_regret}
Under Assumption~\ref{assumption:regularity}, fix any measurable nominal maximizer $a_V=\pi_{\rm nom}(H,V)$, without requiring uniqueness. For any measurable $q(H,V)\in\calP$, its pointwise regret relative to planning with $q$ is
\begin{align}
 \calR_H(q,V)
 &:=\val_H(q)-U_H(a_V,q)\notag\\
 &=\val_H(q)-\val_H(V)-g_{\rm nom}(H,V)^\top(q-V)\ge0.
 \label{eq:generalized_bregman_regret}
\end{align}
If the nominal sensitivity moment is calibrated against $q-V$, then
\begin{equation}
 \E_\mu[\calR_H(q,V)]=\E_\mu[\val_H(q)-\val_H(V)].
\end{equation}
Calibration alone imposes no upper bound on this regret.
\end{proposition}
\begin{proof}
Affinity gives $U_H(a_V,q)=\val_H(V)+g_{\rm nom}^\top(q-V)$. Since $\val_H(q)$ maximizes over plans, this affine function supports $\val_H$ at $V$; hence $g_{\rm nom}\in\partial\val_H(V)$ and the difference is nonnegative. Taking expectations gives the identity. For a concrete example, let $a,p\in[-1,1]$ and $U(a,p)=a p-a^2/2$. At $V=0$ the unique nominal plan is $a_V=0$, its sensitivity is zero, and its value is unchanged for every $q$. Nevertheless, at $q=1$ the oracle plan gains $1/2$. Scaling the utility makes that gain arbitrarily large without violating the zero nominal-sensitivity moment.
\end{proof}
Equation~\eqref{eq:generalized_bregman_regret} is the Bregman divergence when $\val_h$ is differentiable at $V$ \citep{bregman1967relaxation}, and the corresponding subgradient-defined gap otherwise. The relation between supporting decision values, proper scores, and such divergences is classical \citep{savage1971elicitation,gneiting2007strictly}. We use this identity to distinguish calibrated anticipated plan value from small oracle-plan regret, rather than as a new regret principle.

\begin{proposition}[A sufficient uniform smoothness condition]
\label{prop:smooth_value_condition}
In addition to Assumption~\ref{assumption:regularity}, suppose plans lie in a finite-dimensional Euclidean space and $\calA(h)$ is convex and independent of the forecast argument. For all $h$ and $p\in\calP$, assume $a\mapsto U_h(a,p)$ is $m$-strongly concave for a common $m>0$, and
\[
 \|r_h(a)-r_h(a')\|_2\le K\|a-a'\|_2
\]
for a common finite $K$. Then the unique optimizer satisfies
\[
 \|a_p-a_q\|_2\le(K/m)\|p-q\|_2,
 \qquad
 \|\nabla\val_h(p)-\nabla\val_h(q)\|_2\le(K^2/m)\|p-q\|_2.
\]
Gradients here are taken in the affine hull of $\calP$, or in ambient coordinates when $\calP$ is full-dimensional.
\end{proposition}
\begin{proof}
Strong concavity and constrained optimality give
$U_h(a_p,p)-U_h(a_q,p)\ge(m/2)\|a_p-a_q\|^2$ and the analogous inequality with $p,q$ exchanged. Their sum is
\[
 m\|a_p-a_q\|^2
 \le(r_h(a_p)-r_h(a_q))^\top(p-q)
 \le K\|a_p-a_q\|\|p-q\|.
\]
This proves optimizer Lipschitzness. Uniqueness and the envelope formula identify the value gradient with the sensitivity, giving the second bound.
\end{proof}
\begin{corollary}[Second-order bound for oracle-plan regret]
\label{cor:smooth_regret}
Under Proposition~\ref{prop:smooth_value_condition},
\begin{equation}
 0\le\E_\mu[\calR_H(q,V)]
 \le\frac{K^2}{2m}\E_\mu\|q-V\|_2^2.
 \label{eq:smooth_regret_bound}
\end{equation}
\end{corollary}
\begin{proof}
Integrating the Lipschitz gradient along the segment from $V$ to $q$ bounds the supporting-plane error in~\eqref{eq:generalized_bregman_regret} by $(K^2/2m)\|q-V\|^2$. Taking expectations proves the result.
\end{proof}
The Lipschitz sensitivity condition is essential to this conclusion. Strong concavity in the plan alone is insufficient: $U(a,p)=-a^2+p\sqrt a$ on $a,p\in[0,1]$ is strongly concave in $a$, but its maximized value is $3(p/4)^{4/3}$, whose derivative is not Lipschitz at zero. Neither~\eqref{eq:smooth_regret_bound} nor the directional derivative result is a trajectory-safety or stability statement.

\subsection{A minimax consequence of sensitivity calibration}

\begin{theorem}[Decision calibration implies nominal MPC minimax optimality]
\label{thm:decision_calibration}
Assume the affine utility model~\eqref{eq:affine_utility} and Assumption~\ref{assumption:regularity}. Suppose the calibration class $\Psi$ contains the nominal sensitivity test $\psi_{\rm nom}$ in its linear span; that is, there exist coefficients $\beta_1,\ldots,\beta_k\in\R$ such that
\begin{equation}
    \psi_{\rm nom}(h,v)=\sum_{i=1}^k\beta_i\psi_i(h,v)
    \qquad \mu\text{-a.s.}
    \label{eq:span_condition}
\end{equation}
Then nominal MPC is minimax optimal over the calibration-induced ambiguity set:
\begin{equation}
    \pi_{\rm nom}
    \in
    \argmax_\pi
    \inf_{q\in\calQ_\Psi}
    \E_\mu\left[U_H(\pi(H,V),q(H,V))\right].
    \label{eq:nominal_minimax}
\end{equation}
\end{theorem}

\begin{proof}
Because $\psi_{\rm nom}$ lies in the span of the calibrated tests, every $q\in\calQ_\Psi$ satisfies
\begin{equation}
    \E_\mu\left[
    \psi_{\rm nom}(H,V)^\top(q(H,V)-V)
    \right]=0.
\end{equation}
Using $\psi_{\rm nom}(H,V)=r_H(\pi_{\rm nom}(H,V))$, we obtain
\begin{align}
&\E_\mu\left[U_H(\pi_{\rm nom}(H,V),q(H,V))\right]\notag\\
&\qquad=
\E_\mu\left[U_H(\pi_{\rm nom}(H,V),V)\right]
\end{align}
for all $q\in\calQ_\Psi$. Therefore
\begin{equation}
    \inf_{q\in\calQ_\Psi}
    \E_\mu\left[U_H(\pi_{\rm nom}(H,V),q(H,V))\right]
    =
    \E_\mu\left[U_H(\pi_{\rm nom}(H,V),V)\right].
    \label{eq:nominal_invariance}
\end{equation}
For any other forecast-based policy $\pi$, the identity map $q_{\rm id}(h,v)=v$ belongs to $\calQ_\Psi$, hence
\begin{equation}
    \inf_{q\in\calQ_\Psi}
    \E_\mu\left[U_H(\pi(H,V),q(H,V))\right]
    \le
    \E_\mu\left[U_H(\pi(H,V),V)\right].
\end{equation}
Because every candidate policy is feasible in $\calA(H)$, pointwise optimality of $\pi_{\rm nom}$ under the forecast $V$ gives
\begin{equation}
    U_H(\pi(H,V),V)
    \le
    U_H(\pi_{\rm nom}(H,V),V)
    \qquad \mu\text{-a.s.}
\end{equation}
Taking expectations and combining with~\eqref{eq:nominal_invariance} proves the claim.
\end{proof}

\subsection{Interpretation}

Theorem~\ref{thm:sensitivity_span} supplies the organizing characterization, and Theorem~\ref{thm:decision_calibration} is its minimax-optimality consequence. It gives a sufficient condition under which robustification is unnecessary for this minimax criterion; failure of the span condition alone does not imply that nominal MPC is non-minimax:
\begin{equation}
\boxed{
\begin{gathered}
\text{sensitivity-weighted decision calibration}\\
\Longrightarrow\ \text{nominal MPC is minimax optimal}.
\end{gathered}
}
\end{equation}

The theorem does not require full calibration of the learned model. It does not require $q^{\rm true}(h,v)=v$ for all histories and forecasts. It also does not assert that every minimax optimizer is pointwise identical to nominal MPC. It asserts only that nominal MPC itself is minimax-optimal over the ambiguity set once the forecast residual is calibrated in the utility direction of the nominal controller.

Under the certificate, the nominal controller's realized expected plan utility equals its anticipated value $\E_\mu[U_H(\pi_{\rm nom}(H,V),V)]$, by the invariance~\eqref{eq:nominal_invariance}. The forecast-based controller is therefore unbiased about its own anticipated plan performance.

\begin{remark}[Relation to one-shot decision calibration]
If $U_h(a,p)=u(a,p)$ is independent of $h$ and the action set is finite, fix a measurable tie-breaking partition $\{R_a\}$ for the nominal decision. Then
\begin{equation}
    g_{\rm nom}(v)=\sum_a\indic\{v\in R_a\}r(a),
\end{equation}
where $u(a,p)=c(a)+r(a)^\top p$. Classical region-wise vector calibration, $\E[\indic\{V\in R_a\}(Y-V)]=0$ for every $a$, implies the scalar selected-plan condition. The converse need not hold: the scalar condition tests the sum of the chosen utility directions, whereas region-wise vector calibration also tests directions relevant to alternative actions. Thus the conditions are not generally equivalent, and their swap-regret consequences must not be identified. In control the field $r_h(a)$ additionally varies with the observed history.
\end{remark}

\begin{remark}[Relation to contextual minimax calibration]
\label{rem:oneshot_minimax}
The plug-in minimax argument of \citet{kiyani2026partial} also applies when the decision problem has an observed context $h$ and utility $U_h(a,p)$. History dependence alone does not create a new minimax principle. The control-specific content here is the identification of the selected-plan field $g_{\rm nom}$, the necessity construction for tests that do not span it, and the resulting bin-calibration separation. When utility is independent of history and the action library is finite, decision-region tests recover the corresponding special case. In a contextual problem the tests may depend on the context; forecast-only bins need not suffice.
\end{remark}

\section{Calibration-Constrained Adversarial Values}
\label{sec:robust}
Partial calibration defines a convex adversarial problem. Under explicit measurable and integrable regularity, its value has an attained finite-dimensional dual. Policy recovery is a saddle-point question and must be distinguished from the adversarial value calculation.

The belief-dual form is the partial-calibration construction of \citet{kiyani2026partial}. The result below supplies explicit conditions for primal and multiplier attainment, handles compact nonconvex plan sets through relaxed policies, and distinguishes a saddle-compatible policy from an arbitrary pointwise best response. These qualifications matter: compactness alone does not justify applying a convex minimax theorem to a nonconvex deterministic policy set, and optimality against one adverse belief alone does not establish the reverse saddle inequality.

Write $z=(h,v)$, let
\begin{equation}
 \val_h(p)=\max_{a\in\calA(h)}U_h(a,p),
 \qquad F(q)=\E_\mu[\val_H(q(H,V))],
 \label{eq:value_function}
\end{equation}
and define $\calD=\{q\in L_\infty(\mu;\R^d):q(z)\in\calP\ \mu\text{-a.s.}\}$. The exact ambiguity set is
\[
 \calQ_\Psi=\{q\in\calD:T(q-V)=0\},\qquad
 T\xi=\bigl(\E_\mu[\psi_i^\top\xi]\bigr)_{i=1}^k.
\]
The value of a policy is
\begin{equation}
 W_\Psi(\pi)=\inf_{q\in\calQ_\Psi}\E_\mu[U_H(\pi(H,V),q(H,V))].
 \label{eq:robust_controller}
\end{equation}
For randomized plans, the utility in this expression is averaged over the plan distribution.

\begin{assumption}[Uniform relative-interior forecast]
\label{assumption:uniform_relative_interior}
Let $L=\operatorname{span}(\calP-\calP)$ be the direction space of the affine hull of $\calP$. There is $\rho_{\rm int}>0$ such that
\[
 V(z)+u\in\calP\quad\text{for every }u\in L,\ \|u\|_2\le\rho_{\rm int},
 \qquad\mu\text{-a.e. }z.
\]
\end{assumption}
This condition is relative to the affine hull and applies to interior probability forecasts without introducing redundant probability directions. It is a constraint qualification, not a statement that the forecast is calibrated.

\begin{theorem}[Attained adversarial duality and relaxed saddle policies]
\label{thm:dual}
Under Assumptions~\ref{assumption:regularity} and~\ref{assumption:uniform_relative_interior}, the adversarial optimum is attained, and
\begin{align}
 \min_{q\in\calQ_\Psi} F(q)
 &=\max_{\lambda\in\R^k}G(\lambda),\label{eq:attained_dual_value}\\
 G(\lambda)
 &=\E_\mu\left[\min_{p\in\calP}
       \{\val_H(p)+p^\top s_\lambda(H,V)\}
       -V^\top s_\lambda(H,V)\right],\notag\\
 s_\lambda(h,v)&=\sum_{i=1}^k\lambda_i\psi_i(h,v).
\end{align}
For primal and dual optima $q^{\rm adv},\lambda^\star$,
\begin{equation}
 q^{\rm adv}(h,v)\in\argmin_{p\in\calP}
 \{\val_h(p)+p^\top s_{\lambda^\star}(h,v)\}
 \quad\mu\text{-a.s.},
 \label{eq:q_star}
\end{equation}
and $T(q^{\rm adv}-V)=0$. Conversely, a measurable pointwise minimizer at an optimal multiplier is primal optimal if it also satisfies these global moments.

When measurable randomized plans are allowed, the maximin policy optimum is attained and equals~\eqref{eq:attained_dual_value}. There is a relaxed saddle pair $(\pi^\star,q^{\rm adv})$, and its policy is supported on
\begin{equation}
 \argmax_{a\in\calA(h)}U_h(a,q^{\rm adv}(h,v))
 \quad\mu\text{-a.s.}
 \label{eq:robust_best_response}
\end{equation}
A general best response to a worst-case map need not be maximin. If the maximizer in~\eqref{eq:robust_best_response} is unique almost surely, its deterministic selection is the maximin policy. Otherwise a saddle-compatible selection, possibly randomized, is required.
\end{theorem}

\begin{proof}
First, $\calD$ is bounded and weak-* closed in $L_\infty$, hence weak-* compact. To see closedness directly, membership in a fixed compact convex $\calP$ is equivalent to the integrated supporting-halfspace inequalities
\[
 \int_E u^\top q\,d\mu\le\mu(E)\max_{p\in\calP}u^\top p
\]
for every measurable $E$ and every vector $u$ with rational coordinates. These inequalities define weak-* closed halfspaces and separate any map leaving $\calP$ on a set of positive measure. The calibration functionals are weak-* continuous because $\psi_i\in L_1(\mu)$. Thus $\calQ_\Psi$ is weak-* compact and nonempty, since $q=V$ is feasible.

For every measurable deterministic policy $a(z)$, the map
\[
 q\longmapsto\E_\mu[c_H(a(z))+r_H(a(z))^\top q(z)]
\]
is weak-* continuous. The envelope in Assumption~\ref{assumption:regularity} makes its coefficients integrable. Measurable pointwise maximization gives $F$ as the supremum of these continuous affine functionals. Therefore $F$ is weak-* lower semicontinuous. It is finite and uniformly bounded on $\calD$ by an integrable envelope, so it attains its minimum on $\calQ_\Psi$.

For dual attainment, consider $T$ on $L_\infty(\mu;L)$ and let $R$ be its range in $\R^k$. This finite-dimensional space accounts for all effective moment constraints. Choose preimages of a basis of $R$ to obtain a bounded linear right inverse $S:R\to L_\infty(\mu;L)$. For all sufficiently small $y\in R$, $\|Sy\|_\infty\le\rho_{\rm int}$ and $V+Sy\in\calD$. The convex perturbation value
\[
 \phi(y)=\inf\{F(q):q\in\calD,\ T(q-V)=y\}
\]
is therefore finite on a neighborhood of zero in $R$ and is bounded below. It has a finite-dimensional supporting subgradient $b_R$ at zero \citep{rockafellar1970convex}. For every $q\in\calD$,
\[
 F(q)\ge\phi(0)+b_R^\top T(q-V).
\]
Extend $b_R$ to $\R^k$ and set $\lambda^\star=-b_R$. The Lagrangian $F(q)+\lambda^{\star\top}T(q-V)$ is at least $\phi(0)$ everywhere and equals it at a primal optimum. This proves zero duality gap and multiplier attainment, including when the original tests are linearly redundant.

For fixed $\lambda$, the Lagrangian integrand is measurable in $z$ and continuous in $p$ on compact $\calP$. Measurable minimization gives a selector attaining its pointwise minimum, so minimizing the integral equals integrating the minima, yielding $G$. At a primal-dual optimum, the nonnegative difference between the Lagrangian integrand and its pointwise minimum has zero integral. It vanishes almost surely, proving~\eqref{eq:q_star}. The converse follows by feasibility and equality to the attained dual value.

For completeness, policy attainment can be proved without assuming it. Let
\[
 C(z)=\operatorname{co}\{(c_h(a),r_h(a)):a\in\calA(h)\}\subset\R^{d+1},
\]
and let $\mathcal K$ be its integrable measurable selections. This is a convex, norm-closed, uniformly integrable subset of $L_1(\mu;\R^{d+1})$: the envelope bounds every selection, and a norm-convergent sequence has an almost-everywhere convergent subsequence whose limit lies in the closed set $C(z)$. The Dunford--Pettis compactness criterion makes $\mathcal K$ weakly compact. Every such payoff profile is realized by a measurable mixture of at most $d+2$ feasible plans. Indeed, finite-dimensional convex-hull representation supplies the actions and weights pointwise, and the measurable compact selection theorem supplies them measurably on the completed probability space \citep{aliprantis2006infinite}.

For $w=(w_0,w_1)\in\mathcal K$, the payoff $\E[w_0+w_1^\top q]$ is affine and separately continuous in the weak topology of $L_1$ and the weak-* topology of $L_\infty$. The compact convex minimax theorem \citep{sion1958minimax} therefore gives
\[
 \max_{w\in\mathcal K}\min_{q\in\calQ_\Psi}\E[w_0+w_1^\top q]
 =\min_{q\in\calQ_\Psi}\max_{w\in\mathcal K}\E[w_0+w_1^\top q]
 =\min_{q\in\calQ_\Psi}F(q).
\]
The maximum is attained because the infimum of continuous affine payoffs is upper semicontinuous on weakly compact $\mathcal K$. Any optimal pair satisfies the saddle inequalities. Its expected nonnegative pointwise best-response gap is zero, so the implementing mixture is supported on~\eqref{eq:robust_best_response}. Uniqueness makes that mixture a point mass. This proves the policy assertions.
\end{proof}

\paragraph{Convex deterministic plans.}
If the feasible plans are convex subsets of a finite-dimensional Euclidean space and $U_h(\cdot,p)$ is concave for every $p$, a deterministic maximin policy exists as well. Take the barycenter of the relaxed saddle policy. It is feasible and measurable, and Jensen's inequality gives utility at least that of the mixture for every $q\in\calQ_\Psi$. Its worst-case value therefore attains the same minimax value. This establishes existence of a suitable deterministic selection without asserting that every best response to $q^{\rm adv}$ is suitable.

\paragraph{Why pointwise recovery needs care.}
For a two-outcome probability vector $p=(p_1,p_2)$, let the feasible actions be $\{-1,1\}$ and $U(a,p)=a(p_1-p_2)$, with no effective calibration constraints. Then
\[
 \max_{a\in\{-1,1\}}\min_{p\in\Delta_2}U(a,p)=-1
 <0=\min_{p\in\Delta_2}\max_{a\in\{-1,1\}}U(a,p).
\]
The equally weighted randomized action attains zero; no deterministic saddle exists in this action set. This example satisfies the linear-in-probability convention and shows why relaxation cannot be omitted in the general compact-plan statement.

Let $a\in[-1,1]$, $p\in[0,1]$, and $U(a,p)=a(2p-1)$, with no effective calibration constraints. The adversarial optimum is $q^{\rm adv}=1/2$. Every action best responds there, but $a=1$ has worst-case value $-1$, while the maximin action $a=0$ has value zero. Thus a worst-case model and a pointwise best-response set do not by themselves specify a robust controller. Likewise, selecting pointwise inner minimizers at a dual optimum does not replace verification of the global calibration moments.

\paragraph{Approximate calibration.}
For $\calQ_{\Psi,\delta}=\{q\in\calD:|T_i(q-V)|\le\delta_i\}$ with $\delta_i\ge0$, the attained value dual is
\begin{equation}
 \min_{q\in\calQ_{\Psi,\delta}}F(q)
 =\max_{\lambda\in\R^k}\left\{G(\lambda)-\sum_i\delta_i|\lambda_i|\right\}.
 \label{eq:penalized_dual}
\end{equation}
One proof introduces slack variables $s_i\in[-\delta_i,\delta_i]$. On the range of the affine map $(q,s)\mapsto T(q-V)-s$, the pair $(V,0)$ has a feasible relative neighborhood: the forecast has its uniform relative-interior ball, and every nonzero slack interval has zero in its relative interior. The preceding supporting-subgradient argument applies. Minimizing the Lagrangian over each slack interval contributes $-\delta_i|\lambda_i|$. Primal feasibility and the same saddle-policy conditions remain necessary for policy recovery. These results concern the stated operating law; estimating the dual from data requires a separate statistical analysis.

\section{Active-Set Calibration for MPC}
\label{sec:active_set}

Sensitivity-weighted decision calibration can be implemented in several ways. The most control-native version is active-set calibration.

\subsection{MPC regimes and critical regions}

Many MPC problems partition the history-forecast space into qualitative regimes. For a strictly convex parametric quadratic program with fixed Hessian and constraint normals, and affine parameter dependence in its linear term and right-hand side, the optimizer is piecewise affine, with pieces determined by active constraint sets; this is the foundational structure exploited by explicit / multiparametric MPC \citep{bemporad2002explicit}. More generally, $\sigma(h,v)$ may denote:
\begin{itemize}[leftmargin=*]
    \item the active constraint set of a convex MPC problem;
    \item the branch of a mixed-integer planner;
    \item the mode of a hybrid system;
    \item whether a safety filter is active;
    \item whether an obstacle-avoidance or collision-avoidance constraint is active;
    \item whether the system is near a state or input constraint.
\end{itemize}
Let $\Sigma$ be a finite set of such regimes and let
\begin{equation}
    \calR_\sigma = \{(h,v):\sigma(h,v)=\sigma\}.
\end{equation}

\subsection{Active-set decision calibration}

\begin{definition}[Active-set decision calibration]
A forecaster is active-set decision-calibrated if, for every $\sigma\in\Sigma$,
\begin{equation}
    \E_\mu\left[
    \indic\{(H,V)\in\calR_\sigma\}
    r_H(\pi_{\rm nom}(H,V))^\top(Y-V)
    \right]=0.
    \label{eq:active_set_calibration}
\end{equation}
\end{definition}

This condition says that forecast errors must be unbiased in the nominal controller's sensitivity direction separately in each active-set regime.

Refining the regime partition can improve control relevance but increases the number of tests. Conditional within-regime errors depend on the number of visits to that regime; the unnormalized indicator-weighted moments used here also depend on its probability mass. A rare regime does not automatically dominate the global tolerances. The partition should be chosen independently of the held-out certification outcomes or accompanied by a separate uniform-validity argument.

\begin{remark}[Active-set calibration is sufficient]
\label{cor:active_set}
If finitely many active-set regions $\{\calR_\sigma\}_{\sigma\in\Sigma}$ partition the closed-loop support of $(H,V)$, then active-set decision calibration implies sensitivity-weighted decision calibration. Consequently, nominal MPC is minimax optimal over the ambiguity set generated by the active-set tests.
\end{remark}

\begin{proof}
Summing~\eqref{eq:active_set_calibration} over $\sigma\in\Sigma$ gives
\begin{equation}
    \E_\mu\left[
    r_H(\pi_{\rm nom}(H,V))^\top(Y-V)
    \right]=0,
\end{equation}
because exactly one active-set indicator is equal to one on the support. The result follows from Theorem~\ref{thm:decision_calibration}.
\end{proof}

\subsection{Why active-set calibration is a natural control analogue of multicalibration}

In fairness-motivated multicalibration, the groups are subpopulations such as demographic or computationally identifiable sets. In predictive control, the relevant groups are not demographic. They are operating regimes of the closed-loop system. Active-set calibration is therefore a control-native form of multicalibration:
\begin{equation}
\boxed{
\begin{gathered}\text{calibrate within regimes where the controller}\\
\text{has qualitatively different sensitivities.}\end{gathered}
}
\end{equation}

This viewpoint is especially important near constraints. A forecaster may appear reliable on average while being under-confident or over-confident exactly when a state constraint is active. Active-set calibration targets precisely this failure mode.

\subsection{Relative coordinates and constructive finite test classes}
\label{sec:constructive_span}
The sensitivity-span characterization is an organizing orthogonality principle. Its usefulness depends on identifying test classes that cover the sensitivity field of a specified controller. Probability forecasts first require removal of redundant coordinates.

\begin{lemma}[Affine-hull reduction]
\label{lem:affine_hull_reduction}
Suppose $Y,V\in\operatorname{aff}(\calP)$ almost surely. Let $\operatorname{aff}(\calP)=p_0+L$, and let the columns of $B\in\R^{d\times r}$ be an orthonormal basis of $L$. Set $\widetilde p=B^\top(p-p_0)$, $\widetilde Y=B^\top(Y-p_0)$, $\widetilde r_h(a)=B^\top r_h(a)$, and $\widetilde\psi_i=B^\top\psi_i$. Then
\[
 r_h(a)^\top(Y-V)=\widetilde r_h(a)^\top(\widetilde Y-\widetilde V),
 \qquad
 \psi_i^\top(Y-V)=\widetilde\psi_i^\top(\widetilde Y-\widetilde V).
\]
Uniform relative-interior feasibility in $\calP$ becomes ordinary interior feasibility in its $r$ free coordinates. The span theorem therefore applies to the reduced sensitivity and reduced tests. Components normal to $L$ do not require calibration.
\end{lemma}
\begin{proof}
Both $Y-V$ and every feasible perturbation belong to $L$, on which $BB^\top$ is the identity. Substitution proves both identities, and the orthonormal coordinate map preserves Euclidean balls in $L$.
\end{proof}
For a probability simplex, $r=d-1$; in the binary scalar parametrization the interior condition is $V\in[\rho_{\rm int},1-\rho_{\rm int}]$. Clipping forecasts to such an interval supplies this feasibility margin, but does not establish any calibration moment.

\paragraph{What regime conditioning can miss.}
History-aware tests need not be sensitivity weights to detect a bias. In particular, a regime indicator can detect a residual that cancels globally. The relevant distinction is between a chosen coarse test class and the variation it leaves unresolved. In the scalar probability case let $\mathcal C$ be a finite partition generated by forecast bins and prescribed regimes, let $s=g_{\rm nom}$ be bounded, and suppose $V\in[\rho_{\rm int},1-\rho_{\rm int}]$ almost surely. The residual field
\begin{equation}
 \xi=-\alpha\{s-\E[s\mid\mathcal C]\},\qquad\alpha>0,
 \label{eq:coarse_regime_separation}
\end{equation}
preserves all $\mathcal C$-cell calibration moments and is feasible for sufficiently small $\alpha$. Nevertheless,
\[
 \E[s\xi]=-\alpha\E[\operatorname{Var}(s\mid\mathcal C)]<0
\]
whenever some populated cell has nonconstant sensitivity. This separates the specified coarse regime tests from a test resolving their within-cell sensitivity variation. It does not assert that sensitivity weighting is the only history-aware way to detect the residual.

\begin{theorem}[Finite quadratic tests for a fixed-Hessian MPC subclass]
\label{thm:quadratic_test_span}
Let $\theta=(x,v)\in\Theta\subset\R^{n_\theta}$ be a finite-dimensional context/forecast parameter, with $\Theta$ compact. Suppose the nominal optimization, after dropping terms independent of the plan, is the strictly convex quadratic program
\begin{align}
 \min_a\quad&\tfrac12a^\top Q a+(F\theta+f)^\top a,\notag\\
 \text{subject to}\quad&Ga\le b+E x,
 \qquad Q\succ0,
 \label{eq:fixed_hessian_parametric_qp}
\end{align}
with fixed matrices, nonempty compact feasible sets, and no forecast dependence in the feasible set. Suppose every component of $r_h(a)$ is a polynomial of total degree at most two in $(a,x)$. Then a finite measurable partition $\{\mathcal R_\sigma\}_{\sigma\in\Sigma}$ of $\Theta$ exists such that the unique optimizer is affine and the nominal sensitivity is quadratic on each cell. Consequently the bounded tests
\begin{equation}
 \psi_{\sigma,\nu,j}(h,v)
 =\indic\{\theta(h,v)\in\mathcal R_\sigma\}\theta(h,v)^\nu e_j,
 \quad |\nu|\le2,\quad j=1,\ldots,d,
 \label{eq:quadratic_region_tests}
\end{equation}
span $g_{\rm nom}$ exactly in $L_2(\mu)$. There are at most
\[
 |\Sigma|\,d\binom{n_\theta+2}{2}
\]
such tests before removing linear dependencies. In relative probability coordinates, replace $d$ by the dimension of the affine hull and use the reduced sensitivity of Lemma~\ref{lem:affine_hull_reduction}.
\end{theorem}
\begin{proof}
The polyhedral normal-cone optimality condition provides multipliers on active constraints. A minimal conic representation of that normal uses linearly independent active rows, say $G_I$. For each such subset $I$, the KKT system is
\[
 \begin{bmatrix}Q&G_I^\top\\G_I&0\end{bmatrix}
 \begin{bmatrix}a\\\lambda_I\end{bmatrix}
 =\begin{bmatrix}-F\theta-f\\b_I+E_I x\end{bmatrix}.
\]
The matrix is nonsingular because $Q\succ0$ and $G_I$ has full row rank. Hence both $a$ and $\lambda_I$ are affine in $\theta$. The requirements $Ga\le b+Ex$ and $\lambda_I\ge0$ define a polyhedral parameter region on which this affine optimizer is valid. There are finitely many active subsets, and their regions cover $\Theta$. When regions overlap the optimizer agrees by uniqueness; a fixed tie-breaking order makes them a disjoint measurable partition, including their boundaries. This is the explicit quadratic-program structure \citep{bemporad2002explicit}.

Substituting the affine optimizer into a degree-two polynomial in $(a,x)$ yields a degree-two polynomial in $\theta$. Expand each sensitivity component in the monomials $\theta^\nu$, $|\nu|\le2$, on each partition cell. These finitely many scalar coefficients give the exact span representation using~\eqref{eq:quadratic_region_tests}. Compactness of $\Theta$ bounds all tests. Linear projection to relative coordinates preserves their polynomial degree.
\end{proof}

\paragraph{A nonempty scenario subclass.}
One sufficient mode-utility form is
\[
 \ell_m(a,x)=-\tfrac12a^\top Q a-a^\top(F_x x+f_m)-d_m(x),
\]
where $Q\succ0$ is common across modes and each $d_m$ is quadratic. Mixing with probabilities $v_m$ leaves the plan Hessian fixed and gives linear plan coefficient $F_xx+\sum_m v_m f_m$, affine in $(x,v)$. The term $\sum_m v_m d_m(x)$ is independent of the plan. With common polyhedral constraints depending affinely on $x$, this model satisfies the theorem. The exact span guarantee concerns this specified structural subclass; the test count alone is not a statistical sample-complexity comparison.

\paragraph{Why arbitrary scenario quadratics are excluded.}
Probability weighting can instead produce $Q(v)=\sum_m v_m Q_m$, or terms bilinear in $x$ and $v$ in the plan coefficient. A fixed active set then need not give an affine optimizer. For example, with $a\in[-2,2]$ and
\[
 J_0(a)=a^2,\qquad J_1(a)=2(a-1)^2,
\]
minimizing $(1-p)J_0+pJ_1$ gives $a^\star(p)=2p/(1+p)$ throughout $p\in[0,1]$, with a single inactive-constraint region. For the scalar forecast interface $U(a,p)=-\{(1-p)J_0(a)+pJ_1(a)\}$, the utility coefficient is $r(a)=J_0(a)-J_1(a)$. Evaluated at the selected plan, it is
\[
 r(a^\star(p))=\frac{2p^2+4p-2}{(1+p)^2},
\]
which is rational rather than quadratic. Thus arbitrary probability-weighted quadratic scenario MPC does not inherit the monomial span guarantee merely from quadratic costs and polyhedral constraints.

\section{Approximate Span and Calibration}
\label{sec:approx}

A single bound accounts for moment tolerances and imperfect representation of the nominal sensitivity. Exact representation is its zero-error case.

\begin{theorem}[Approximate span with representation error]
\label{thm:approx_rep}
Assume the affine utility model~\eqref{eq:affine_utility}, Assumption~\ref{assumption:regularity}, and $g_{\rm nom}=\psi_{\rm nom}\in L_2(\mu;\R^d)$. Let $\beta_1,\ldots,\beta_k\in\R$ be any coefficients and set
\begin{equation}
    \varepsilon_{\rm rep}=\Big\|\psi_{\rm nom}-\sum_{i=1}^k\beta_i\psi_i\Big\|_{L_2(\mu)},
    \qquad
    D=\operatorname{diam}(\calP)=\sup_{p,p'\in\calP}\|p-p'\|_2.
\end{equation}
Define $\eta'=\sum_{i=1}^k|\beta_i|\delta_i+\varepsilon_{\rm rep}D$. Then nominal MPC is $\eta'$-minimax optimal over $\calQ_{\Psi,\delta}$:
\begin{equation}
    \inf_{q\in\calQ_{\Psi,\delta}}
    \E[U_H(\pi_{\rm nom}(H,V),q(H,V))]
    \ge
    \sup_\pi\inf_{q\in\calQ_{\Psi,\delta}}
    \E[U_H(\pi(H,V),q(H,V))]-\eta'.
\end{equation}
\end{theorem}

The square-integrability condition is local to the $L_2$ approximation and exposure results in this section. Assumption~\ref{assumption:regularity} alone supplies only an $L_1$ envelope; without the additional condition, $\varepsilon_{\rm rep}$ can be infinite and the displayed bound is empty. The adversarial duality result retains its weaker $L_1$ assumptions.

\begin{proof}
Fix $q\in\calQ_{\Psi,\delta}$ and split the nominal sensitivity moment as
\begin{equation}
    \E[\psi_{\rm nom}^\top(q-V)]
    =\E\Big[\big(\textstyle\sum_i\beta_i\psi_i\big)^\top(q-V)\Big]
    +\E\big[(\psi_{\rm nom}-\textstyle\sum_i\beta_i\psi_i)^\top(q-V)\big].
\end{equation}
The first term is at most $\sum_i|\beta_i|\,|\E[\psi_i^\top(q-V)]|\le\sum_i|\beta_i|\delta_i$. For the second, Cauchy--Schwarz gives
\begin{equation}
    \big|\E[(\psi_{\rm nom}-\textstyle\sum_i\beta_i\psi_i)^\top(q-V)]\big|
    \le\varepsilon_{\rm rep}\,\|q-V\|_{L_2(\mu)}\le\varepsilon_{\rm rep}D,
\end{equation}
because $q(H,V),V\in\calP$ gives $\|q(H,V)-V\|_2\le D$ almost surely. Hence $|\E[\psi_{\rm nom}^\top(q-V)]|\le\eta'$, so $\E[U_H(\pi_{\rm nom},q)]\ge\E[U_H(\pi_{\rm nom},V)]-\eta'$. The identity map $q_{\rm id}\in\calQ_{\Psi,\delta}$ and pointwise optimality of $\pi_{\rm nom}$ under $V$ give $\inf_q\E[U_H(\pi,q)]\le\E[U_H(\pi_{\rm nom},V)]$ for every $\pi$. Combining proves the claim.
\end{proof}

\subsection{Optimizing the certificate and measuring uncovered sensitivity}
\label{sec:optimized_certificate}
The representation in Theorem~\ref{thm:approx_rep} need not be chosen in advance. Throughout this subsection, assume $g_{\rm nom}\in L_2(\mu;\R^d)$. This local square-integrability condition makes the representation errors and Gram moments finite and the Hilbert-space projections well defined. The standing $L_1$ envelope alone does not imply it; the bounded tests in Assumption~\ref{assumption:regularity} already belong to $L_2(\mu;\R^d)$. For population moment tolerances $\delta_i\ge0$, define
\begin{equation}
 \eta_{\rm opt}(\delta)
 =\min_{\beta\in\R^k}
 \left\{\sum_{i=1}^k\delta_i|\beta_i|
 +D\left\|g_{\rm nom}-\sum_{i=1}^k\beta_i\psi_i\right\|_{L_2(\mu)}\right\},
 \qquad D=\operatorname{diam}(\calP).
 \label{eq:optimized_certificate}
\end{equation}

\begin{proposition}[Best representation-based bound]
\label{prop:optimized_certificate}
Under the assumptions of Theorem~\ref{thm:approx_rep}, with $g_{\rm nom}\in L_2(\mu;\R^d)$, its minimax suboptimality bound holds with $\eta_{\rm opt}(\delta)$. The minimum in~\eqref{eq:optimized_certificate} is attained, including with linearly redundant tests or zero tolerances. It is a convex optimization problem determined by the population Gram moments of $g_{\rm nom}$ and $\psi_i$.
\end{proposition}
\begin{proof}
The approximate-span bound holds for every coefficient vector, so it holds at the infimum. To see attainment when $D>0$, quotient out
\[
 N_0=\{\beta:\textstyle\sum_i\beta_i\psi_i=0,
                \ \delta_i\beta_i=0\ \text{for all }i\}.
\]
The objective is invariant along $N_0$. On a complementary finite-dimensional subspace, $\sum_i\delta_i|\beta_i|+D\|\sum_i\beta_i\psi_i\|_{L_2}$ is a norm. By the triangle inequality, the objective is bounded below by that norm minus $D\|g_{\rm nom}\|_{L_2}$ and is therefore coercive there. Continuity gives a minimizer. If $D=0$, $\beta=0$ attains the minimum. Finally,
\[
 \left\|g_{\rm nom}-\sum_i\beta_i\psi_i\right\|_{L_2}^2
 =a_g-2b_g^\top\beta+\beta^\top\Gamma\beta,
\]
where $a_g=\E\|g_{\rm nom}\|^2$, $(b_g)_i=\E[\psi_i^\top g_{\rm nom}]$, and $\Gamma_{ij}=\E[\psi_i^\top\psi_j]$. The norm and weighted absolute values have a convex conic epigraph representation.
\end{proof}
The coefficient tradeoff is between statistical moment tolerances and representation error. A least-squares projection alone minimizes the second term, and can be suboptimal when its coefficients amplify poorly measured moments. Replacing the population Gram moments with training-sample moments produces a candidate representation, not automatically a validated population certificate. Section~\ref{sec:empirical} gives an independent-validation construction.

\begin{definition}[Uncovered sensitivity energy]
\label{def:exposure_index}
For $g_{\rm nom}\in L_2(\mu;\R^d)$, let $g_\perp=g_{\rm nom}-\Proj_{\calS_\Psi}g_{\rm nom}$. If $\|g_{\rm nom}\|_{L_2}>0$, define
\begin{equation}
 \mathcal E_\Psi
 =\frac{\|g_\perp\|_{L_2(\mu)}^2}{\|g_{\rm nom}\|_{L_2(\mu)}^2}
 \in[0,1].
 \label{eq:exposure_index}
\end{equation}
Set $\mathcal E_\Psi=0$ when $g_{\rm nom}=0$ almost surely.
\end{definition}
This index describes coverage of the sensitivity field by the specified test span. It uses no forecast outcomes and does not measure the true residual, oracle-plan regret, or the frequency of harmful errors.

\begin{proposition}[Correctly normalized nullspace exposure]
\label{prop:nullspace_exposure}
Suppose $g_{\rm nom}\in L_2(\mu;\R^d)$. For $r\ge0$, the unconstrained Hilbert-space perturbation problem satisfies
\begin{equation}
 \sup_{\xi\in\calS_\Psi^\perp,\ \|\xi\|_{L_2}\le r}
       -\E_\mu[g_{\rm nom}^\top\xi]
 =r\|g_\perp\|_{L_2}.
 \label{eq:nullspace_exposure}
\end{equation}
If $g_\perp\ne0$ is bounded and a uniform interior radius $\rho_{\rm int}$ is available in the relevant coordinates, the maximizing direction is also feasible in $\calP$ whenever
\[
 r\le\rho_{\rm int}\frac{\|g_\perp\|_{L_2}}{\|g_\perp\|_\infty}.
\]
Along the differently normalized direction $\xi=-\alpha g_\perp$, the loss is $\alpha\|g_\perp\|_{L_2}^2$.
\end{proposition}
\begin{proof}
Only $g_\perp$ pairs with a vector in $\calS_\Psi^\perp$. Cauchy--Schwarz gives the upper bound, attained by $\xi=-r g_\perp/\|g_\perp\|_{L_2}$ when $g_\perp\ne0$. The displayed radius condition bounds its pointwise norm by $\rho_{\rm int}$. The zero-field case and the $\alpha$-normalization follow directly.
\end{proof}
The loss per unit $L_2$ perturbation is therefore $\|g_\perp\|_{L_2}$, rather than its square. The energy ratio~\eqref{eq:exposure_index} is useful for comparing test spans under one operating law and one sensitivity definition. Estimating it by regressing sensitivities on tests requires independent assessment of the residual energy; the in-sample regression residual can be optimistic. Changing the operating law or the controller changes the quantity being measured.

\section{Finite-Sample Population Certificates}
\label{sec:empirical}
Independent episodes provide a direct connection from observed calibration moments to the population operating law. Dependence within an episode, including overlapping prediction windows, is allowed. The statistical unit is the episode rather than the individual time step.

Condition on independent training data that fix the forecaster, controller used to collect calibration data, test class, and any candidate representations. Let $M$ calibration episodes be independent and identically distributed, each supplying $L_{\rm ep}$ usable decision points with complete outcomes $(H_{e,t},V_{e,t},Y_{e,t})$. Let $\mu$ be the law at a uniformly random usable step of a fresh episode from this same collection policy. We assume $Y,V\in\calP$ almost surely and use the marginal of $\mu$ on $(H,V)$ for plan-value expectations. Within an episode no temporal independence is required.

For the $k\ge1$ fixed tests, define
\begin{equation}
 A_{e,i}=\frac1{L_{\rm ep}}\sum_{t=1}^{L_{\rm ep}}
       \psi_i(H_{e,t},V_{e,t})^\top(Y_{e,t}-V_{e,t}),
 \qquad
 \widehat m_i=\frac1M\sum_{e=1}^M A_{e,i}.
 \label{eq:episodic_moments}
\end{equation}
Suppose $|A_{e,i}|\le B_i$ for known bounds. A sufficient bound is $B_i=D\|\psi_i\|_\infty$, where $D=\operatorname{diam}(\calP)$; an empirical maximum is not a substitute for such an almost-sure bound.

\begin{proposition}[Episode-level population moment bounds]
\label{prop:episodic_moments}
For any $\zeta\in(0,1)$, with probability at least $1-\zeta$, simultaneously for every $i$,
\begin{equation}
 \left|\E_\mu[\psi_i^\top(Y-V)]\right|
 \le\delta_i
 :=|\widehat m_i|+B_i\sqrt{\frac{2\log(2k/\zeta)}{M}}.
 \label{eq:episodic_confidence}
\end{equation}
Consequently the true conditional mean $q^{\rm true}(h,v)=\E[Y\mid H=h,V=v]$ belongs to $\calQ_{\Psi,\delta}$ on this event.
\end{proposition}
\begin{proof}
The variables $A_{e,i}$ are i.i.d. across episodes, lie in $[-B_i,B_i]$, and have mean $\E_\mu[\psi_i^\top(Y-V)]$. Hoeffding's inequality and a union bound over $k$ tests yield~\eqref{eq:episodic_confidence} \citep{hoeffding1963probability}. Since $\calP$ is closed and convex, $q^{\rm true}\in\calP$ almost surely. Conditional expectation identifies its calibration moments with those of $Y-V$, giving ambiguity-set membership.
\end{proof}

Representation error must also be controlled under $\mu$. A representation fitted and assessed on the same observations need not have the stated population error. The following finite-candidate construction provides one complete certificate while allowing a data-dependent final choice.

\begin{theorem}[A validated population minimax certificate]
\label{thm:episodic_certificate}
Before the validation episodes are drawn, fix $J\ge1$ coefficient vectors $\beta^{(1)},\ldots,\beta^{(J)}$, possibly using independent training data. For each candidate assume a known bound
\[
 \left\|g_{\rm nom}(h,v)-\sum_i\beta_i^{(j)}\psi_i(h,v)\right\|_2\le C_j
 \quad\mu\text{-a.s.}
\]
Define
\begin{align}
 \widehat R_j^{\,2}
 &=\frac1{M L_{\rm ep}}\sum_{e,t}
 \left\|g_{\rm nom}(H_{e,t},V_{e,t})-
                  \sum_i\beta_i^{(j)}\psi_i(H_{e,t},V_{e,t})\right\|_2^2,\notag\\
 \overline R_j
 &=\left[\min\left\{C_j^2,\ \widehat R_j^{\,2}
       +C_j^2\sqrt{\frac{\log(2J/\zeta)}{2M}}\right\}\right]^{1/2},
 \label{eq:representation_validation}\\
 \delta_i
 &=|\widehat m_i|+B_i\sqrt{\frac{2\log(4k/\zeta)}{M}},\notag\\
 \widehat\eta_j&=\sum_i\delta_i|\beta_i^{(j)}|+D\overline R_j.
 \label{eq:validated_eta}
\end{align}
Under the affine interface and the assumptions of Theorem~\ref{thm:approx_rep}, with probability at least $1-\zeta$, all of the following hold simultaneously:
\begin{enumerate}[leftmargin=*]
\item $q^{\rm true}\in\calQ_{\Psi,\delta}$ and
$\|g_{\rm nom}-\sum_i\beta_i^{(j)}\psi_i\|_{L_2(\mu)}\le\overline R_j$ for every candidate;
\item nominal MPC has minimax suboptimality at most $\widehat\eta_j$ over $\calQ_{\Psi,\delta}$ for every candidate;
\item its anticipated expected plan value has bias bounded by
\begin{equation}
 \left|\E_\mu\left[U_H(\pi_{\rm nom}(H,V),Y)
                         -U_H(\pi_{\rm nom}(H,V),V)\right]\right|
 \le\min_j\widehat\eta_j.
 \label{eq:validated_plan_bias}
\end{equation}
\end{enumerate}
In particular, selecting the candidate with smallest $\widehat\eta_j$ after validation preserves these guarantees. The bounds hold conditionally on the independent training data and hence also unconditionally.
\end{theorem}
\begin{proof}
Use Hoeffding's inequality with failure budget $\zeta/2$ for all moment bounds. For each candidate, the per-episode mean of its squared representation residual lies in $[0,C_j^2]$ and has expectation equal to the squared population $L_2$ error. The one-sided Hoeffding bound with failure budget $\zeta/(2J)$ gives~\eqref{eq:representation_validation}; truncating the upper bound at the known $C_j^2$ remains valid. A union bound gives the stated simultaneous event.

On that event Theorem~\ref{thm:approx_rep} gives the minimax suboptimality bound for every fixed candidate. For the bias claim, let $r_j=g_{\rm nom}-\sum_i\beta_i^{(j)}\psi_i$. Affinity and conditional expectation give
\begin{align*}
 \left|\E_\mu[g_{\rm nom}^\top(Y-V)]\right|
 &=\left|\E_\mu[g_{\rm nom}^\top(q^{\rm true}-V)]\right|\\
 &\le\sum_i|\beta_i^{(j)}|\delta_i
       +\|r_j\|_{L_2}\|q^{\rm true}-V\|_{L_2}
 \le\widehat\eta_j.
\end{align*}
All candidates satisfy the inequalities on one event, so taking their minimum or selecting one after validation is legitimate.
\end{proof}
The candidate family may include least-squares projections, regularized fits, and solutions of the empirical version of~\eqref{eq:optimized_certificate} at several penalties. A finite search over independently trained candidates is what is validated here. Arbitrarily optimizing coefficients on the validation episodes would require an additional uniform generalization argument. One valid choice is $C_j=\|g_{\rm nom}\|_\infty+\sum_i|\beta_i^{(j)}|\|\psi_i\|_\infty$, using known almost-sure bounds for the terms. This choice can be conservative, affecting usefulness rather than validity.

\paragraph{Validating an exposure diagnostic.}
Suppose a sensitivity regression $\widehat\beta$ is fixed by independent training, $\|g_{\rm nom}\|\le G_0$, and its representation residual is bounded by $C_0$. On independent episodes let $\widehat E_g$ and $\widehat E_r$ be the corresponding sample means of $\|g_{\rm nom}\|^2$ and $\|g_{\rm nom}-\sum_i\widehat\beta_i\psi_i\|^2$. With failure probability at most $\zeta_{\rm exp}$, both
\[
 L_g=\max\left\{0,\widehat E_g-G_0^2\sqrt{\frac{\log(2/\zeta_{\rm exp})}{2M}}\right\}
 \le\|g_{\rm nom}\|_{L_2}^2
\]
and
\[
 \|g_\perp\|_{L_2}^2
 \le\|g_{\rm nom}-\sum_i\widehat\beta_i\psi_i\|_{L_2}^2
 \le U_r:=\min\left\{C_0^2,\widehat E_r+C_0^2\sqrt{\frac{\log(2/\zeta_{\rm exp})}{2M}}\right\}
\]
hold. The same episode-level Hoeffding argument proves these two one-sided bounds. If $L_g>0$, they give $\mathcal E_\Psi\le\min\{1,U_r/L_g\}$; otherwise the generic upper bound remains one. This assesses uncovered sensitivity, rather than certifying any actual harmful residual.

\paragraph{Scope of the population law.}
Here $\mu$ is the fresh-episode occupancy law of the frozen collection pipeline. The theorem closes the population-moment step for that law; it does not replace a transfer argument when deployment changes the controller, forecaster, initial-state law, or environment. Section~\ref{sec:operating_transfer} provides such an argument for a declared density-ratio family with a frozen pipeline and unchanged conditional outcome law. Within-episode dependence does not reduce the number of independent statistical units below $M$, but neither does a long episode create $M L_{\rm ep}$ independent units. Small certified anticipated-value bias does not control the generalized Bregman regret of Proposition~\ref{prop:bregman_regret}, recursive feasibility, or trajectory safety.

\subsection{Joint bin-mean and sensitivity correction}
\label{sec:per_bin_correction}
The correction used here fits the ordinary bin moment and a sensitivity direction together. Choose one fixed standardization
\[
 s_i=(s_i^{\rm raw}-m_0)/\sigma_0,\qquad\sigma_0>0,
\]
and freeze $m_0,\sigma_0$ before held-out evaluation. The same standardized sensitivity is used in fitting, deployment, and reported residuals for this fitted test. The raw selected-plan utility bias is a separate quantity and is reported on its original utility scale. For a nonempty calibration bin $I_b=\{i:V_i\in B_b\}$, write
\[
 \Delta_i=Y_i-V_i,\qquad \overline\Delta_b=\frac1{n_b}\sum_{i\in I_b}\Delta_i,
 \qquad \overline s_b=\frac1{n_b}\sum_{i\in I_b}s_i,
 \qquad \widetilde{s}_i=s_i-\overline s_b.
\]
Consider an unpenalized intercept and a ridge-regularized slope,
\begin{equation}
 (a_b,\gamma_b)\in\argmin_{a,\gamma}
 \left\{\frac12\sum_{i\in I_b}(\Delta_i-a-\gamma \widetilde{s}_i)^2
                         +\frac{\kappa_b}2\gamma^2\right\},
 \qquad\kappa_b\ge0,
 \label{eq:joint_bin_fit}
\end{equation}
with unprojected forecast
\begin{equation}
 p_i^0=V_i+a_b+\gamma_b(s_i-\overline s_b).
 \label{eq:per_bin_correction}
\end{equation}
The fitted bin center $\overline s_b$ and coefficients are frozen when the correction is applied to new contexts.

\begin{lemma}[Bin and sensitivity normal equations]
\label{thm:per_bin_correction}
For each nonempty bin,
\begin{equation}
 a_b=\overline\Delta_b,\qquad
 \gamma_b=\frac{\sum_{i\in I_b}\widetilde{s}_i \Delta_i}
                    {\sum_{i\in I_b}\widetilde{s}_i^2+\kappa_b},
 \label{eq:gamma_b}
\end{equation}
when the denominator is positive; set $\gamma_b=0$ when both numerator and denominator vanish. Then
\begin{align}
 \sum_{i\in I_b}(Y_i-p_i^0)&=0,\label{eq:bin_mean_identity}\\
 \sum_{i\in I_b}\widetilde{s}_i(Y_i-p_i^0)
 =\sum_{i\in I_b}s_i(Y_i-p_i^0)&=\kappa_b\gamma_b.
 \label{eq:per_bin_identity}
\end{align}
For zero ridge, the unprojected fit satisfies both moments exactly, including a zero-variance bin under the stated convention. For positive ridge the bin-mean moment remains exact and the weighted moment has the displayed ridge remainder.
If $s_i^{\rm raw}=m_0+\sigma_0s_i$ is the raw utility coefficient, its corresponding calibration-sample moment is $\sum_i s_i^{\rm raw}(Y_i-p_i^0)=\sigma_0\kappa_b\gamma_b$, because the intercept moment vanishes.
\end{lemma}
\begin{proof}
Because $\sum_{i\in I_b}\widetilde{s}_i=0$, the intercept normal equation gives $a_b=\overline\Delta_b$. The slope normal equation gives $(\sum_i \widetilde{s}_i^2+\kappa_b)\gamma_b=\sum_i \widetilde{s}_i \Delta_i$. These are precisely~\eqref{eq:bin_mean_identity} and the first weighted identity. Since $s_i=\widetilde{s}_i+\overline s_b$, the zero bin-mean residual also gives the identity for $s_i$. If $\sum_i \widetilde{s}_i^2=\kappa_b=0$, every $\widetilde{s}_i=0$ and the chosen zero slope satisfies all identities.
\end{proof}
With zero ridge this is the least-squares projection onto the span of the bin indicator and its within-bin sensitivity direction. Positive ridge shrinks the slope and does not generally cancel its weighted moment. Fitting only a slope without an intercept need not preserve the bin mean, and centering the sensitivity within the bin makes the two normal equations explicit. The deterministic identities concern the fitting sample; they are not population or deployment guarantees.

\paragraph{Projection and fallbacks.}
If deployment requires probabilities in $[\ell,u]\subseteq[0,1]$, let $p_i^{\rm clip}=\clip(p_i^0,\ell,u)$ and $d_i=p_i^0-p_i^{\rm clip}$. On the calibration sample the exact identities become
\begin{align}
 \sum_{i\in I_b}(Y_i-p_i^{\rm clip})&=\sum_{i\in I_b}d_i,\notag\\
 \sum_{i\in I_b}\widetilde{s}_i(Y_i-p_i^{\rm clip})
 &=\kappa_b\gamma_b+\sum_{i\in I_b}\widetilde{s}_i d_i,\notag\\
 \sum_{i\in I_b}s_i(Y_i-p_i^{\rm clip})
 &=\kappa_b\gamma_b+\sum_{i\in I_b}s_i d_i.
 \label{eq:joint_clipping_identity}
\end{align}
Thus clipping can break both fitted moments; for example the last absolute residual is bounded by $\kappa_b|\gamma_b|+\sum_i|s_i d_i|$, not by the ridge term alone. On the raw utility scale the exact identity is $\sum_i s_i^{\rm raw}(Y_i-p_i^{\rm clip})=\sigma_0\kappa_b\gamma_b+\sum_i s_i^{\rm raw}d_i$. Empty bins and bins excluded by a declared minimum-count rule require a specified fallback and are outside the lemma's fitted-bin identities. Sample counts and clipping contributions should be reported when these corrections are evaluated.

\begin{remark}[Policy dependence]
\label{rem:per_bin_correction_scope}
The sensitivity field, bins, and standardization used for fitting must be specified. If changing the forecast changes the selected plan, the deployed plan's utility coefficient can differ from the fitting coefficient even at the same history. Independent held-out moment checks and, where needed, operating-law transfer bounds address distinct issues from the calibration-sample identities. The lemma does not establish a self-consistent fixed point of the correction or a trajectory-safety guarantee.
\end{remark}

\subsection{Certificate transfer under distribution shift}
\label{sec:transfer}

The preceding certificates use a specified operating law. Their transfer to a
different law is a change-of-measure question about the same sensitivity field.
For probability laws $\mu,\nu$, use the convention
\[
 d_{\rm TV}(\mu,\nu)=\sup_A|\mu(A)-\nu(A)|
       =\tfrac12\int|d\mu-d\nu|.
\]

\begin{proposition}[Transfer of a fixed sensitivity moment]
\label{prop:transfer}
Let $\phi(H,V,Y)=g_{\rm nom}(H,V)^\top(Y-V)$ and suppose
$|\E_\mu\phi|\le\eta$. Suppose $\phi$ takes values in $[L,U]$ almost
surely under both $\mu$ and $\nu$. Then
\begin{equation}
 |\E_\nu\phi|\le\eta+(U-L)d_{\rm TV}(\mu,\nu).
 \label{eq:transfer_bound}
\end{equation}
In particular, if $Y,V\in\calP$, $D=\operatorname{diam}(\calP)$, and
$\|g_{\rm nom}\|_2\le G$ on the supports of both laws, one may take
$U-L=2GD$. For laws $\mu_1,\ldots,\mu_T$ and a common range $[L,U]$,
\begin{equation}
 \left|\sum_{t=1}^T\E_{\mu_t}\phi\right|
 \le T\eta+(U-L)\sum_{t=1}^T d_{\rm TV}(\mu,\mu_t).
 \label{eq:transfer_cumulative}
\end{equation}
\end{proposition}

\begin{proof}
Subtract the constant $(L+U)/2$ from $\phi$ in the integral against
$\nu-\mu$. The resulting integrand has absolute value at most
$(U-L)/2$, so the change in expectation is at most
$(U-L)d_{\rm TV}(\mu,\nu)$. The triangle inequality proves the first
claim and its sum proves the second. Finally,
$|g_{\rm nom}^\top(Y-V)|\le GD$ gives the stated uniform range.
\end{proof}

If $\phi$ is $K_\phi$-Lipschitz in a specified metric on $(H,V,Y)$, the
same change in expectation is at most $K_\phi W_1(\mu,\nu)$. These bounds
transfer a fixed plan-value moment. A controller change can alter both its
sensitivity field and its operating law, so applying the result then requires
accounting for both changes.

\subsection{Direct evaluation tests and their statistical cost}
\label{sec:direct_certificate}
For one fixed selected-plan field, using a larger exact representation cannot by itself improve the elementary coordinate-wise certificate. This comparison clarifies why structured tests need an additional purpose.

\begin{proposition}[A direct test versus an exact test decomposition]
\label{prop:direct_test_comparison}
Suppose $g_{\rm nom}=\sum_{i=1}^k\beta_i\psi_i$ exactly, all tests are fixed independently of validation, and $|A_{e,i}|\le B_i$ for every episode. Put $X_e=\sum_i\beta_i A_{e,i}$ and let $B_g\le\sum_i|\beta_i|B_i$ be a valid known bound on $|X_e|$. At a common failure probability $\zeta$, the direct one-test Hoeffding certificate is no larger than the coordinate-wise exact-representation certificate:
\begin{align}
 &|\overline X|+B_g\sqrt{\frac{2\log(2/\zeta)}{M}}\notag\\
 &\qquad\le\sum_i|\beta_i|
 \left\{|\widehat m_i|+B_i\sqrt{\frac{2\log(2k/\zeta)}{M}}\right\}.
\end{align}
\end{proposition}
\begin{proof}
The direct empirical moment equals $\overline X=\sum_i\beta_i\widehat m_i$. Apply the triangle inequality to this sum, use the assumed bound on $B_g$, and note that $\log(2/\zeta)\le\log(2k/\zeta)$. Each construction has its own stated coverage event.
\end{proof}

Known full ranges can be sharper than symmetric absolute bounds. Variance-sensitive confidence radii can further improve numerical precision: equation~\eqref{eq:transfer_eb_radius} gives the empirical Bernstein construction used below. These are classical concentration choices, not a new statistical calibration principle. A direct moment still concerns one fixed field and one source law. Section~\ref{sec:operating_transfer} gives larger classes a different role: covering and designing tests for a family of operating laws, with simultaneous post-validation queries.

\section{Reusable Certificates for Specified Operating Shifts}
\label{sec:operating_transfer}
For one frozen controller and one operating law, the single test $g_{\rm nom}$ is sufficient. A larger test class requires a different purpose. We now specify a family of changes in the frequencies of operating contexts, characterize the additional calibration information that family requires, and optimize the missing test directions. The resulting certificate can be queried over a continuum of operating laws using one source validation sample.

The change-of-measure principle and transfer through density-ratio test classes are established ideas in multicalibration and universal adaptability \citep{kim2022adaptability,wu2024bridging,ye2026boosting}. Our use concerns the products of those ratios with a predictive controller's selected-plan sensitivity. The control-specific results below identify their exact coverage requirement, give a sharp spectral audit and a budgeted test-augmentation rule, and connect that rule to explicit-MPC test classes. The statistical result certifies a specified family; it does not assert that calibration survives an arbitrary deployment change.

\subsection{The target family and the required products}
\label{sec:transfer_products}
Write $z=(h,v)$ and $g=g_{\rm nom}$. The forecaster, controller, feasible-plan correspondence, and utility interface are frozen. Let $K(dy\mid z)$ be the source conditional law of the utility statistic, with conditional mean $q(z)$. For a bounded, measurable $w\ge0$ satisfying $\E_\mu w=1$, define the target context law
\[
 d\mu_w(z)=w(z)d\mu(z).
\]
The target joint law is $\mu_w(dz)K(dy\mid z)$: only the frequencies of contexts change. In particular, neither the conditional outcome law nor the selected plan at a given context changes. Its selected-plan evaluation bias is
\begin{equation}
 B_w(q)=\E_{\mu_w}[g^\top(q-V)]
       =\langle wg,q-V\rangle_\mu.
 \label{eq:transfer_bias_identity}
\end{equation}
The first equality concerns utility evaluated under a new context law; the second is the usual change-of-measure identity. A changed feedback controller need not generate a target law of this form with the same conditional kernel.

\begin{theorem}[Exact coverage of an operating-law family]
\label{thm:transfer_span}
Use free probability coordinates as in Lemma~\ref{lem:affine_hull_reduction}. Under Assumption~\ref{assumption:interior_perturbations}, let $\mathcal W$ be any family of bounded density ratios as above, and let $\calS_\Psi$ be the source test span. The following are equivalent:
\begin{enumerate}[leftmargin=*]
\item $wg\in\calS_\Psi$ for every $w\in\mathcal W$;
\item $B_w(q)=0$ for every $w\in\mathcal W$ and every $q\in\calQ_\Psi$.
\end{enumerate}
If coverage fails at $w$, put $a_w=wg-\Proj_{\calS_\Psi}(wg)$. This field is bounded and nonzero. For every
\[
 0<\alpha\le\rho_{\rm int}/\|a_w\|_\infty,
\]
the conditional mean $q_\alpha=V-\alpha a_w$ belongs to the source ambiguity set, but
\begin{equation}
 B_w(q_\alpha)=-\alpha\|a_w\|_{L_2(\mu)}^2<0.
 \label{eq:transfer_adverse_direction}
\end{equation}
\end{theorem}
\begin{proof}
If $wg$ is in the source span, the source calibration equalities make it orthogonal to $q-V$, which gives $B_w(q)=0$ by~\eqref{eq:transfer_bias_identity}. Conversely, the projection of bounded $wg$ onto the finite span of bounded tests is bounded. Thus $a_w$ is bounded, orthogonal to every source test, and the stated interior margin makes $V-\alpha a_w$ feasible. Its target bias is $-\alpha\langle wg,a_w\rangle_\mu=-\alpha\|a_w\|_{L_2(\mu)}^2$.
\end{proof}
The quantifiers matter. A source law may have $\E_\mu[g^\top(q-V)]=0$ while a target law has nonzero bias, even though the conditional outcome kernel is unchanged. Under Assumption~\ref{assumption:regularity}, exact coverage also preserves nominal plug-in minimax optimality under each $\mu_w$ against the \emph{same source ambiguity set} $\calQ_\Psi$: the nominal policy has invariant expected utility for every $q$ in that set, while the identity mean $q=V$ bounds every competing policy from above, exactly as in Theorem~\ref{thm:decision_calibration}. This statement does not replace source calibration by an unverified promise under the target law.

\subsection{A sharp audit and a budget for additional tests}
\label{sec:transfer_spectral}
Let $\varphi=(\varphi_1,\ldots,\varphi_s)^\top$ be fixed bounded scalar functions with $\E_\mu\varphi=0$. Consider
\begin{equation}
 w_t=1+t^\top\varphi,\qquad
 \mathcal T=\{t\in\R^s:t^\top C t\le R_{\rm sh}^2\},\qquad C\succ0,
 \label{eq:ellipsoidal_shift_family}
\end{equation}
where $R_{\rm sh}>0$ and
\[
 R_{\rm sh}\operatorname*{ess\,sup}_{z}
       \|C^{-1/2}\varphi(z)\|_2\le1.
\]
Then every $w_t$ is nonnegative and normalized. The functions and their centering, the reference law used for normalization, and the shift budget are part of the specified family; estimating them introduces additional uncertainty that is not hidden in the certificate.

Suppose $g\in\calS_\Psi$, so the source tests already cover the unshifted controller. Let $P_\perp$ denote orthogonal projection onto $\calS_\Psi^\perp$ and define
\begin{equation}
 a_\ell=P_\perp(\varphi_\ell g),\qquad
 A t=\sum_{\ell=1}^s t_\ell a_\ell,\qquad
 K_{\ell m}=\langle a_\ell,a_m\rangle_\mu.
 \label{eq:transfer_exposure_matrix}
\end{equation}
All these fields are bounded. The matrix $K$ uses contexts and selected-plan sensitivities, but no outcomes. It measures missing \emph{coverage}, not the actual forecast error. For example, if
\[
 G_{ij}=\langle\psi_i,\psi_j\rangle_\mu,\quad
 H_{i\ell}=\langle\psi_i,\varphi_\ell g\rangle_\mu,\quad
 J_{\ell m}=\langle\varphi_\ell g,\varphi_m g\rangle_\mu,
\]
then
\begin{equation}
 K=J-H^\top G^\dagger H,
 \label{eq:transfer_gram_schur}
\end{equation}
where $G^\dagger$ is the Moore--Penrose inverse. This formula remains valid with redundant source tests, because it computes the orthogonal projection onto their span.

\begin{theorem}[Worst-shift exposure and optimal test augmentation]
\label{thm:spectral_transfer}
Let $\lambda_1\ge\cdots\ge\lambda_s\ge0$ be the eigenvalues of $C^{-1/2}KC^{-1/2}$, and set $\lambda_{s+1}=0$. Under Assumption~\ref{assumption:interior_perturbations} and the preceding conditions:
\begin{enumerate}[leftmargin=*]
\item The largest uncovered target sensitivity is
\begin{equation}
 \max_{t\in\mathcal T}\|P_\perp(w_tg)\|_{L_2(\mu)}
       =R_{\rm sh}\sqrt{\lambda_1}.
 \label{eq:worst_shift_exposure}
\end{equation}
If $\lambda_1>0$ and $v_1$ is a unit top eigenvector, one maximizing shift is $t^\star=R_{\rm sh}C^{-1/2}v_1$.
\item If $\kappa\ge0$ bounds the $L_2(\mu)$ size of a source-calibration-preserving mean perturbation, the relaxed worst-case target bias satisfies
\begin{equation}
 \sup_{\substack{t\in\mathcal T,\ \xi\in\calS_\Psi^\perp\\
                  \|\xi\|_{L_2(\mu)}\le\kappa}}
       |\langle w_tg,\xi\rangle_\mu|
       =\kappa R_{\rm sh}\sqrt{\lambda_1}.
 \label{eq:spectral_residual_exposure}
\end{equation}
When $\lambda_1>0$, the adverse perturbation
\[
 \xi^\star=-\kappa\,At^\star/\|At^\star\|_{L_2(\mu)}
\]
also satisfies the mean-domain constraint $V+\xi^\star\in\calP$ if
\begin{equation}
 \kappa\le\rho_{\rm int}
       \frac{\|At^\star\|_{L_2(\mu)}}{\|At^\star\|_\infty}.
 \label{eq:spectral_feasible_radius}
\end{equation}
At such radii the relaxed value is attained by an admissible source model. No such attainment is asserted at larger radii.
\item Let $b\in\{0,\ldots,s\}$. Among all additions of at most $b$ independent scalar calibration directions, the smallest possible worst-shift uncovered norm is
\begin{equation}
 \inf_{\substack{\mathcal U\subset\calS_\Psi^\perp\\
                       \dim\mathcal U\le b}}
 \max_{t\in\mathcal T}
 \|P_{(\calS_\Psi+\mathcal U)^\perp}(w_tg)\|_{L_2(\mu)}
       =R_{\rm sh}\sqrt{\lambda_{b+1}}.
 \label{eq:budgeted_transfer_tests}
\end{equation}
An optimum is supplied by the first $b$ nonzero left singular directions of the finite-rank operator $AC^{-1/2}$. These directions are bounded and may themselves be used as vector tests. In particular, $\operatorname{rank}K$ is the minimum number of additional independent directions that gives exact coverage of the whole family.
\end{enumerate}
\end{theorem}
\begin{proof}
Since $g\in\calS_\Psi$, the uncovered field of $w_tg$ is $At$. Set $u=C^{1/2}t$. Then $\|At\|_{L_2(\mu)}^2=u^\top C^{-1/2}KC^{-1/2}u$ with $\|u\|_2\le R_{\rm sh}$. The Rayleigh quotient gives~\eqref{eq:worst_shift_exposure}. Cauchy--Schwarz gives the upper bound in~\eqref{eq:spectral_residual_exposure}; $\xi^\star$ attains it. Its essential supremum is at most $\rho_{\rm int}$ under~\eqref{eq:spectral_feasible_radius}, which proves the admissibility assertion by the interior assumption. If $\lambda_1=0$, every uncovered product is zero and both exposure values are zero.

For the budgeted statement, an arbitrary added test space $\mathcal U^\prime$ can first be projected onto $\calS_\Psi^\perp$: $\calS_\Psi+\mathcal U^\prime=\calS_\Psi+P_\perp\mathcal U^\prime$ and $\dim(P_\perp\mathcal U^\prime)\le\dim\mathcal U^\prime$. Thus the restriction in~\eqref{eq:budgeted_transfer_tests} loses no generality. Put $M_0=AC^{-1/2}$. Its singular values are $\sqrt{\lambda_j}$. Adding the span of its first $b$ left singular directions leaves operator norm $\sqrt{\lambda_{b+1}}$. If $\lambda_{b+1}=0$, this upper construction gives zero and nonnegativity completes the argument. Otherwise $b<s$, and any $\mathcal U$ of dimension at most $b$ has a nonzero vector $u$ in the span of the first $b+1$ right singular vectors such that $P_{\mathcal U}M_0u=0$. Normalize $u$ to unit length. Then
\[
 \|(I-P_{\mathcal U})M_0u\|_{L_2(\mu)}=\|M_0u\|_{L_2(\mu)}
       \ge\sqrt{\lambda_{b+1}},
\]
which proves the lower bound. This is the finite-rank singular-value approximation argument \citep{eckart1936approximation}. Each nonzero left singular direction is a finite linear combination of the bounded fields $a_\ell$, so the optimum is available within the allowed bounded test class. Finally, all $s$ products are covered precisely when their projected span has dimension zero; its dimension before augmentation is $\operatorname{rank}K$. Thus fewer than that many directions cannot give exact coverage, and a basis of that projected span does give it.
\end{proof}
The optimization in~\eqref{eq:budgeted_transfer_tests} is a geometric test-design criterion. Adding a direction specifies an additional moment to calibrate or validate; adding a column alone does not make its population moment zero. The criterion also need not minimize worst-case bias over the pointwise-feasible conditional means; Section~\ref{sec:feasible_bias} gives an exact alternative and a counterexample. It does not minimize statistical confidence widths: a normalized singular direction can have a large almost-sure range. Nor does a large exposure imply a harmful error in the actual environment. It identifies the adverse direction that the source tests have not excluded. If the reference Gram moments are estimated, their uncertainty must be controlled before interpreting an empirical eigenvalue as a population certificate.

\subsection{Polynomial closure and its degree cost}
\label{sec:transfer_polynomial}
\begin{corollary}[Control-native product tests]
\label{cor:transfer_polynomial}
Suppose $g$ is polynomial of degree at most $p_g$ on each cell of a fixed finite partition of a compact parameter set $\Theta\subset\R^{n_\theta}$. Suppose each density ratio in a specified family is polynomial of degree at most $p_w$ on those same cells, or on a fixed common refinement. Then region indicators times coordinate monomials of degree at most $p_g+p_w$ span every product $wg$. Let $\widetilde\Sigma$ denote the cells of the partition actually used, including any common refinement. The number of tests before removing linear dependencies is at most
\[
 |\widetilde\Sigma|\,d\binom{n_\theta+p_g+p_w}{p_g+p_w}.
\]
Theorem~\ref{thm:transfer_span} therefore gives exact transfer for all those laws once these source moments are calibrated. For probability forecasts, $d$ denotes the free probability dimension after affine-hull reduction.
\end{corollary}
\begin{proof}
On each cell, multiply the coordinate polynomials for $w$ and $g$. Their products have degree at most $p_g+p_w$. Expanding in monomials gives the representation; compactness bounds the tests. Apply Theorem~\ref{thm:transfer_span}.
\end{proof}
For Theorem~\ref{thm:quadratic_test_span}, a general quadratic sensitivity and affine density ratio require degree three. The original degree-two basis suffices for affine ratios when the sensitivity is additionally piecewise affine, as in the exact-interface example below. That sharper structural property must be verified; it does not follow merely from a fixed nominal Hessian on an arbitrary restricted parameter support.

\begin{example}[Quadratic tests need not transfer through affine ratios]
\label{ex:degree_transfer_failure}
Let $X$ be uniform on $[-1,1]$, $a\in[-1,1]$, $p\in[0,1]$, and
\[
 U_x(a,p)=-a^2/2+p x^2,\qquad V=1/2.
\]
The nominal optimizer is zero, the QP has one critical region, and $g=x^2$. Let $P_3(x)=(5x^3-3x)/2$ and set
\[
 q(x)=1/2+\delta P_3(x),\qquad w(x)=1+\varepsilon x,
 \qquad 0<|\delta|<1/2,\quad 0<|\varepsilon|<1.
\]
The mean is feasible because $|P_3|\le1$. All source polynomial calibration moments of degree at most two vanish by orthogonality of $P_3$. The ratio is positive and normalized. Nonetheless
\[
 B_w(q)=\delta\E[(1+\varepsilon X)X^2P_3(X)]
       =2\varepsilon\delta/35\ne0.
\]
Thus affine operating shifts cannot in general be assigned to the original quadratic test class. The missing product here has degree three.
\end{example}

\subsection{One validation sample and a certified set of shifts}
\label{sec:transfer_validation}
The population audit above asks which information is missing. Validation asks what the observed information certifies. For a fixed family of bounded density ratios $w_t=1+t^\top\varphi$, $t\in\mathcal T$, suppose there are known exact representations
\begin{equation}
 g=\sum_i\beta_{0i}\psi_i,\qquad
 \varphi_\ell g=\sum_i\beta_{\ell i}\psi_i,\qquad
 \beta(t)=\beta_0+\sum_{\ell=1}^s t_\ell\beta_\ell.
 \label{eq:transfer_representation_coefficients}
\end{equation}
Then $w_tg=\sum_i\beta_i(t)\psi_i$. The family can be the small product class $\{g,\varphi_1g,\ldots,\varphi_sg\}$ itself. A structured MPC basis supplies an alternative shared representation and a coverage promise; it is not needed to improve the statistical bound for one fixed family.

Let $A_{e,i}$ and $\widehat m_i$ be the fixed-test episode moments in~\eqref{eq:episodic_moments}. Suppose nonnegative, possibly data-dependent radii $\epsilon_i$ satisfy
\begin{equation}
 \Pr\bigl\{|m_i-\widehat m_i|\le\epsilon_i\ \text{for all }i\bigr\}
       \ge1-\zeta,\qquad
 m_i=\E_\mu[\psi_i^\top(Y-V)].
 \label{eq:transfer_moment_event}
\end{equation}

\begin{theorem}[Uniform certificate and post-validation shift choice]
\label{thm:uniform_transfer_certificate}
Under the independent-episode sampling conditions of Section~\ref{sec:empirical}, freeze the density family, tests and exact representations before drawing validation episodes. On the event in~\eqref{eq:transfer_moment_event}, simultaneously for every $t\in\mathcal T$,
\begin{align}
 |B_{w_t}(q^{\rm true})|
 &\le\widehat C(t),\label{eq:uniform_transfer_certificate}\\
 \widehat C(t)
 &:=|\beta(t)^\top\widehat m|
       +\sum_i\epsilon_i|\beta_i(t)|.\notag
\end{align}
Consequently, for a prescribed tolerance $\tau\ge0$, every shift in the random set
\begin{equation}
 \widehat{\mathcal T}_\tau
    =\{t\in\mathcal T:\widehat C(t)\le\tau\}
 \label{eq:certified_shift_set}
\end{equation}
has target evaluation bias at most $\tau$ on the same event, including a shift chosen from that set after inspecting the validation outcomes. There is no union bound over the number of shift queries.

The function $\widehat C$ is convex and piecewise affine. If $\mathcal T$ is a polytope, selecting a member of $\widehat{\mathcal T}_\tau$ to optimize a linear operating objective is a linear program. For a one-dimensional interval $[-T,T]$,
\begin{equation}
 \sup_{|t|\le T}\widehat C(t)
       =\max\{\widehat C(-T),\widehat C(T)\}.
 \label{eq:transfer_endpoint_certificate}
\end{equation}
\end{theorem}
\begin{proof}
The unchanged conditional kernel and the exact representations give $B_{w_t}(q^{\rm true})=\beta(t)^\top m$. On the coordinate-wise simultaneous event,
\[
 |\beta(t)^\top m|
 \le|\beta(t)^\top\widehat m|
       +\sum_i|\beta_i(t)|\epsilon_i
\]
for every coefficient vector, so it holds for all $t$, including a data-selected one. The right-hand side is the exact supremum of $|\beta(t)^\top u|$ over the rectangular moment confidence set $|u_i-\widehat m_i|\le\epsilon_i$; it retains cancellation in the empirical combined moment. Convexity and piecewise affinity follow from the affine map $t\mapsto\beta(t)$ and absolute values. Introduce variables above both signs of each affine term to obtain a linear-program representation of the sublevel set. Finally, a convex function on an interval is bounded by the maximum of its endpoint values.
\end{proof}
The confidence rectangle may contain moment vectors that no conditional outcome model realizes; the certificate is sufficient and can be conservative.
\begin{remark}[Coverage does not determine statistical efficiency]
\label{rem:product_basis_efficiency}
The raw product class is sufficient for coverage, but it need not give the smallest rectangular confidence certificate among bases of that span. For two equiprobable contexts $x\in\{-1,1\}$, let $g=1+0.9x$, $\varphi=-x$, $V=1/2$, and $w_t=1-tx$ for $|t|\le1$. The two tests $\{g,\varphi g\}$ each have residual-moment range width $1.9$ for binary $Y$. At $t=1$, their coefficients are $(1,1)$, so the two-test Hoeffding rectangle contributes radius $3.8\sqrt{\log(4/\zeta)/(2M)}$. In the alternative basis $\{g+\varphi g,g-\varphi g\}$ the same target is its first test, whose residual-moment range width is $0.2$. The corresponding radius is $0.2\sqrt{\log(4/\zeta)/(2M)}$, with the same empirical combined moment and confidence allocation. Both bases cover the same family. This does not imply that a large monomial library is preferable; it shows why geometric rank, basis choice, and statistical precision are distinct criteria. Proposition~\ref{prop:direct_test_comparison} concerns one fixed target and does not assert uniform dominance of a raw product basis.
\end{remark}
 If a nonpolyhedral convex admissible family is used, the certified set remains convex but need not be polyhedral. For the ellipsoidal family in~\eqref{eq:ellipsoidal_shift_family}, a linear operating objective instead gives a second-order cone program. An empty certified set is a legitimate outcome.

The before-sampling protocol is a sufficient way to ensure independence of the chosen family and tests from the validation sample, conditional on any separate fitting data. A sample drawn earlier can also be used when this independence is justified. Freezing functions after inspecting that sample is not sufficient by itself; outcome-free construction from adaptively selected validation covariates likewise requires a separate argument. An exploratory reanalysis must not inherit an unconditional coverage claim without this independence or an appropriate simultaneous selection guarantee.

\paragraph{Two finite-sample choices of radii.}
If each episode average $A_{e,i}$ has a known almost-sure range of width $R_i$, Hoeffding gives
\[
 \epsilon_i^{\rm H}=R_i\sqrt{\frac{\log(2k/\zeta)}{2M}}.
\]
For $M\ge2$, let
\[
 \widehat s_i^2=\frac1{M-1}\sum_{e=1}^M(A_{e,i}-\widehat m_i)^2,
 \qquad L=\log(4k/\zeta).
\]
The empirical Bernstein inequality of \citet[Theorem 4]{maurer2009empirical}, applied to each tail after scaling to the known range and then union-bounded, gives the alternative
\begin{equation}
 \epsilon_i^{\rm EB}
   =\sqrt{\frac{2\widehat s_i^2 L}{M}}
       +\frac{7R_i L}{3(M-1)}.
 \label{eq:transfer_eb_radius}
\end{equation}
Either choice separately establishes~\eqref{eq:transfer_moment_event}. Choosing the smaller of the two after seeing the data requires splitting the failure budget or proving joint coverage; it is not implicit in either individual formula. The numerical study reports them separately. Dependence within episodes is allowed, but the variance and the sample count here are across the $M$ independent episode averages. Sample maxima cannot replace the known ranges.

\paragraph{Approximate representations.}
Exact polynomial coverage is not necessary for a conservative certificate. Suppose instead the $s+1$ frozen representations have errors $e_0=g-\sum_i\beta_{0i}\psi_i$ and $e_\ell=\varphi_\ell g-\sum_i\beta_{\ell i}\psi_i$. If a simultaneous event also gives $\|e_\ell\|_{L_2(\mu)}\le r_\ell$ for $\ell=0,\ldots,s$, then on that event
\begin{equation}
 |B_{w_t}(q^{\rm true})|
 \le\widehat C(t)
       +D\left(r_0+\sum_{\ell=1}^s|t_\ell|r_\ell\right),
 \label{eq:approximate_uniform_transfer}
\end{equation}
where $D=\operatorname{diam}(\calP)$. Indeed the unrepresented product is $e_0+\sum_\ell t_\ell e_\ell$, and Cauchy--Schwarz bounds its pairing with $q^{\rm true}-V$ by the displayed term. Componentwise independent-validation bounds of the form~\eqref{eq:representation_validation}, with their own failure budget, supply one way to obtain the $r_\ell$. The combined event must include both sources of uncertainty. A representation trained and evaluated on the same contexts without a generalization argument is not such a certificate.

The extension supplies three distinct operations: identify operating shifts the current tests do not cover, add a prescribed number of directions to minimize that geometric vulnerability, and certify a set of shifts from one independent validation sample. These operations concern expected evaluation of the fixed selected plans. They do not certify oracle regret, constraint-violation risk, recursive feasibility, or a feedback-generated target law. A change in the conditional outcome kernel requires an additional model-shift term and is not covered by~\eqref{eq:transfer_bias_identity}.

\subsection{Probability-feasible bias, beyond the spectral relaxation}
\label{sec:feasible_bias}
The spectral criterion answers which tests best reduce an uncovered $L_2$ norm. A different question is how much evaluation bias remains when every conditional mean must obey its pointwise probability constraints. The distinction affects test selection, not only the numerical scale of a bound. We first identify the relevant geometry, then give a finite-context computation with a controlled maximization over continuous operating shifts. The arguments use standard norm duality, linear-programming duality, and support functions \citep{rockafellar1970convex}; their role here is to turn the controller-specific audit into bounds in utility units.

For a bounded vector field $f$ and the finite test span $\calS_\Psi$, define
\begin{align}
 W_\Psi(f)&=\sup_{q\in\calQ_\Psi}
       |\langle f,q-V\rangle_\mu|,\label{eq:feasible_bias_definition}\\
 d_{1,\Psi}(f)&=\inf_{s\in\calS_\Psi}
       \E_\mu\|f-s\|_2.\notag
\end{align}
Here ``feasible'' refers to the mean domain $\calP$ in the ambiguity set. For a binary statistic with $\calP=[0,1]$, every such conditional mean is realizable by a Bernoulli conditional law. For a larger statistic domain, the realizability qualification in Remark~\ref{rem:realizability} still applies.

\begin{proposition}[The pointwise-constrained geometry]
\label{prop:feasible_l1_sandwich}
Under Assumption~\ref{assumption:interior_perturbations}, with
$D=\operatorname{diam}(\calP)$,
\begin{equation}
 \rho_{\rm int}\,d_{1,\Psi}(f)
 \le W_\Psi(f)\le D\,d_{1,\Psi}(f).
 \label{eq:feasible_l1_sandwich}
\end{equation}
In the scalar binary case $\calP=[0,1]$, $V\equiv1/2$, the exact identity is
\begin{equation}
 W_\Psi(f)=\tfrac12 d_{1,\Psi}(f).
 \label{eq:centered_binary_l1}
\end{equation}
\end{proposition}
\begin{proof}
Every feasible $\xi=q-V$ annihilates the test span and satisfies
$\|\xi(z)\|_2\le D$ almost surely. For any $s\in\calS_\Psi$, H\"older's inequality bounds $|\langle f,\xi\rangle_\mu|$ by $D\E_\mu\|f-s\|_2$. This proves the upper bound. The finite-dimensional span is closed in $L_1$. The quotient-norm duality formula, with the pointwise Euclidean norm and its dual, gives
\[
 d_{1,\Psi}(f)=\sup\{\langle f,\xi\rangle_\mu:
     \E_\mu[\psi_i^\top\xi]=0\ \forall i,
     \|\xi\|_\infty\le1\}.
\]
Indeed the annihilator of the span in the dual space $L_\infty$ represents the norm on the quotient of $L_1$ by that span. Multiplying any such $\xi$ by $\rho_{\rm int}$ makes $V+\rho_{\rm int}\xi$ feasible, proving the lower bound. In the centered binary case the feasible residuals are exactly the annihilating fields with $|\xi|\le1/2$, so the same formula gives equality.
\end{proof}
Thus minimizing uncovered $L_2$ exposure need not minimize the pointwise-feasible worst-case bias. Equation~\eqref{eq:spectral_feasible_radius} remains the precise condition under which the spectral adverse direction is itself feasible.

\begin{example}[A spectral optimum can have much larger feasible bias]
\label{ex:spiky_feasible_design}
Take an odd number $n\ge3$ of equiprobable contexts, $V=1/2$, $g=1$, and the source test $1$. Let $a_n=1.2\sqrt{n-1}$, and let $\varphi_1$ equal $a_n$ at one exceptional context and $-a_n/(n-1)$ elsewhere. Let $\varphi_2$ be zero at that context and take balanced values $+1,-1$ on the remaining contexts. Set $C=I$ and $R_{\rm sh}=1/a_n$. These features are centered, and
$R_{\rm sh}\max_z\|\varphi(z)\|_2\le1$, so every ratio in the shift disk is nonnegative. The exposure matrix is
\[
 K=\diag\{1.44,(n-1)/n\}.
\]
With one extra test the spectral rule therefore adds $\varphi_1$. Yet the exact feasible worst-case biases over the entire disk are
\begin{align}
 \sup_{t\in\mathcal T}W_{\{1,\varphi_1\}}(w_t)
   &=\frac{n-1}{2na_n},\label{eq:spiky_bias_first}\\
 \sup_{t\in\mathcal T}W_{\{1,\varphi_2\}}(w_t)
   &=\frac{1}{2(n-1)}.\label{eq:spiky_bias_second}
\end{align}
To verify the first equality, the constant and $\varphi_1$ moments force the residual at the exceptional context to zero. The remaining $\varphi_2$ moment has magnitude at most $(n-1)/(2n)$, attained by setting each remaining residual to half the sign of $\varphi_2$. For the second equality, the constant moment gives $\E[\varphi_1\xi]=a_n\xi_1/(n-1)$, whose largest magnitude is $a_n/[2(n-1)]$. It is attained by $\xi_1=1/2$ and $\xi_j=-1/[2(n-1)]$ elsewhere, which also satisfies the $\varphi_2$ moment. Optimizing over the disk multiplies each remaining one-coordinate moment by $R_{\rm sh}$.

For $n=101$, the spectral choice has feasible bias $0.0412541$, versus $0.005$ for the other test, although its uncovered $L_2$ norm is smaller ($0.0829198$ versus $0.1$). The ratio of the feasible biases is $(n-1)^2/(na_n)$ and grows without bound as $\sqrt n/1.2$. This is a counterexample to feasible-bias optimality, not to the spectral theorem's stated geometric optimum.
\end{example}

\subsection{Exact finite-context audits and a continuum enclosure}
\label{sec:finite_feasible_audit}
Suppose the context law has finite support $z_1,\ldots,z_N$, with known masses $\mu_j>0$ summing to one. Work in binary probability coordinates, writing $v_j=V(z_j)$, $f_j=f(z_j)$, and $\psi_{ij}=\psi_i(z_j)$. Define
\[
 \ell_v(u)=(1-v)(u)_+ +v(-u)_+,
 \qquad (u)_+=\max\{u,0\}.
\]
This is the support function of the residual interval $[-v,1-v]$, and is an asymmetric absolute-value loss. The next result makes the pointwise probability constraints computationally explicit; it does not posit a new duality principle.

\begin{theorem}[Exact binary feasible-bias dual]
\label{thm:feasible_binary_dual}
For the finite context law above,
\begin{equation}
 W_\Psi(f)=
 \max_{\sigma\in\{-1,1\}}\ \min_{\lambda\in\R^k}
 \sum_{j=1}^N\mu_j\,
 \ell_{v_j}\!\left(\sigma f_j-\sum_{i=1}^k\lambda_i\psi_{ij}\right).
 \label{eq:feasible_binary_dual}
\end{equation}
Each signed primal and dual optimum is attained. This remains true with redundant tests and with boundary forecasts $v_j\in\{0,1\}$; no interior assumption is required for this finite linear program.
\end{theorem}
\begin{proof}
For a fixed sign, maximize $\sum_j\mu_j\sigma f_j\xi_j$ subject to
$-v_j\le\xi_j\le1-v_j$ and $\sum_j\mu_j\psi_{ij}\xi_j=0$ for every $i$. The feasible set is nonempty, since $\xi=0$ is feasible, and compact. Subtracting the calibration constraints with multipliers $\lambda$ gives the dual function
\[
 \sum_j\mu_j\sup_{-v_j\le\xi_j\le1-v_j}
 \left(\sigma f_j-\sum_i\lambda_i\psi_{ij}\right)\xi_j,
\]
which is the displayed sum of $\ell_{v_j}$. The problem and its epigraph dual are finite linear programs. Feasibility and boundedness of the primal imply strong duality and dual attainment. Taking both signs proves~\eqref{eq:feasible_binary_dual}.
\end{proof}
This is the fixed-utility support-function construction of Theorem~\ref{thm:dual} on finite contexts. The finite linear-program argument additionally permits boundary forecasts, for which Assumption~\ref{assumption:uniform_relative_interior} need not hold. Thus it specializes the same duality mechanism without requiring that earlier theorem's interior qualification.

For each fixed operating shift, this gives both a feasible adverse conditional mean and a dual upper bound. A small primal--dual gap can be checked directly. A numerical solver's success flag alone is not a proof of feasibility or an error bound.

For the ellipsoidal family~\eqref{eq:ellipsoidal_shift_family}, assume $g\in\calS_\Psi$ and define the vector field
\[
 F_j=R_{\rm sh}g_j C^{-1/2}\varphi(z_j)\in\R^s,
 \qquad b(\xi)=\sum_j\mu_j F_j\xi_j.
\]
Writing $t=R_{\rm sh}C^{-1/2}u$ gives
\begin{equation}
 W_\Psi^{\rm fam}:=\sup_{t\in\mathcal T}W_\Psi(w_tg)
   =\max_{\substack{-v\le\xi\le1-v\\
                    \sum_j\mu_j\psi_{ij}\xi_j=0\ \forall i}}
        \|b(\xi)\|_2.
 \label{eq:feasible_family_norm}
\end{equation}
The equality uses coverage of $g$ to remove the constant term. The outer maximization need not be convex: a finite directional scan is only a lower bound. The following enclosure controls that missing step.

\begin{theorem}[A finite-direction enclosure of the continuous maximum]
\label{thm:feasible_direction_enclosure}
Let $\mathcal B$ be the compact set of vectors $b(\xi)$ in~\eqref{eq:feasible_family_norm}, expressed in an orthonormal basis of a known $r$-dimensional subspace containing it. For $r\ge1$, choose unit vectors $u_1,\ldots,u_L\in\R^r$ and a known $\gamma>0$ such that
\begin{equation}
 \gamma\{u:\|u\|_2\le1\}
       \subseteq\operatorname{conv}\{u_1,\ldots,u_L\}
       \subseteq\{u:\|u\|_2\le1\}.
 \label{eq:direction_inner_polytope}
\end{equation}
For each direction, let $\underline h_\ell$ and $\overline h_\ell$ be valid lower and upper bounds on the signed linear-program value
$h(u_\ell)=\max_{b\in\mathcal B}u_\ell^\top b$.
Then
\begin{equation}
 \max\{0,\max_\ell\underline h_\ell\}
 \le W_\Psi^{\rm fam}
 \le \gamma^{-1}\max_\ell\overline h_\ell.
 \label{eq:feasible_direction_enclosure}
\end{equation}
If $r=0$, the family bias is exactly zero. For $r=1$, the directions $\{-1,1\}$ give $\gamma=1$ and no outer approximation error.
\end{theorem}
\begin{proof}
The support function $h$ of $\mathcal B$ is convex and positively homogeneous. Its maximum on the convex hull of the $u_\ell$ is $\max_\ell h(u_\ell)$. Its maximum on the unit ball is $\max_{b\in\mathcal B}\|b\|_2=W_\Psi^{\rm fam}$. The two inclusions in~\eqref{eq:direction_inner_polytope} therefore give
$\gamma W_\Psi^{\rm fam}\le\max_\ell h(u_\ell)\le W_\Psi^{\rm fam}$. Substitute the directional bounds. The zero lower bound follows from feasibility of $\xi=0$. No symmetry of $\mathcal B$ is required, so nonconstant and noncentral forecasts are included.
\end{proof}
For a centrally symmetric polytope whose facets are
$\nu_a^\top u\le c_a$ with the origin in its interior, one may use
$\gamma=\min_a c_a/\|\nu_a\|_2$. Thus a refined sphere mesh supplies an explicit relative enclosure, not merely more scan directions. The number of directions grows with the residual dimension; this construction is useful for small declared shift families and is not a dimension-free algorithm.

An analytic construction avoids estimating facet distances. In three dimensions, take spherical angles $\theta=j\pi/32$, $j=0,\ldots,32$, and $\phi=k\pi/32$, $k=0,\ldots,63$, identifying repeated poles. There are $31\cdot64+2=1986$ directions. Every unit vector is within angular distance at most $\pi/64+\pi/64=\pi/32$ of one of them: first move to the nearest latitude, then to the nearest longitude, using path length at most the longitude difference. Therefore their convex hull has support at least $\cos(\pi/32)$ in every unit direction, which implies~\eqref{eq:direction_inner_polytope} with that $\gamma$. In two dimensions, a regular 128-gon gives $\gamma=\cos(\pi/128)$. These radii give relative outer factors below $1.00484$ and $1.00031$, respectively, before directional optimization and arithmetic errors are added.

\paragraph{Projection and arithmetic errors.}
Subtracting test-span components from each coordinate of $F$ does not change $b(\xi)$. Its residual coordinate rank is the relevant $r$. To account for an approximate reduction, suppose
\[
 F_j=\sum_i\psi_{ij}d_i+Q\widetilde F_j+e_j,
 \qquad Q^\top Q=I,
\]
and let
$\delta_F=\sum_j\mu_j\max\{v_j,1-v_j\}\|e_j\|_2$.
For every feasible $\xi$, the discrepancy between $b(\xi)$ and
$Q\sum_j\mu_j\widetilde F_j\xi_j$ is at most $\delta_F$.
Consequently the lower enclosure for the reduced problem may be decreased by $\delta_F$ and truncated at zero, and its upper enclosure increased by $\delta_F$. Approximate rank truncation therefore need not be treated as exact zero. Directional LP bounds, the polytope inradius, and this projection remainder all require controlled numerical errors if a numerical enclosure is to be rigorous in machine arithmetic. Our reported computations explicitly separate the mathematical enclosure from the floating-point tolerances used to evaluate it; they are not formal interval-arithmetic proofs.

\subsection{Validation certificates that respect the mean domain}
\label{sec:feasible_validation}
The same finite-context audit can use noisy calibration evidence. Unlike the rectangular bound of Theorem~\ref{thm:uniform_transfer_certificate}, it removes moment combinations that no probability model can realize. Fix the contexts, forecasts, tests, and shift family independently of the full validation sample. Let $\widehat m_i$ and $\epsilon_i$ have the simultaneous coverage event~\eqref{eq:transfer_moment_event}, and define
\begin{equation}
 \widehat{\calQ}
 =\left\{q\in[0,1]^N:
  \left|\sum_j\mu_j\psi_{ij}(q_j-v_j)-\widehat m_i\right|
       \le\epsilon_i\quad\forall i\right\}.
 \label{eq:feasible_moment_box}
\end{equation}
Membership is a linear feasibility problem. If this set is empty, the procedure reports inconsistent evidence and issues no certificate. An empty set must not be used to certify every operating choice by vacuity.

\begin{theorem}[A feasible-moment envelope and operating-choice linear program]
\label{thm:feasible_validation_envelope}
Suppose $\widehat{\calQ}\ne\varnothing$. For any field $f$, its exact envelope is
\begin{align}
 \widehat W(f)
  &:=\max_{q\in\widehat{\calQ}}
        \left|\sum_j\mu_j f_j(q_j-v_j)\right|\notag\\
  &=\max_{\sigma\in\{-1,1\}}\ \min_{\lambda\in\R^k}
   \left\{\sum_j\mu_j
      \ell_{v_j}\!\left(\sigma f_j-\sum_i\lambda_i\psi_{ij}\right)
      +\lambda^\top\widehat m+\sum_i\epsilon_i|\lambda_i|\right\}.
  \label{eq:feasible_validation_dual}
\end{align}
On the simultaneous coverage event, $q^{\rm true}\in\widehat{\calQ}$ and
\begin{equation}
  |B_{w_t}(q^{\rm true})|\le\widehat W(w_tg)
       \quad\text{for every }t\in\mathcal T.
  \label{eq:feasible_validation_uniform}
\end{equation}
If $w_tg=\sum_i\beta_i(t)\psi_i$, then
\begin{equation}
 \widehat W(w_tg)\le
 |\beta(t)^\top\widehat m|+\sum_i\epsilon_i|\beta_i(t)|
 =\widehat C(t).
 \label{eq:feasible_envelope_dominates_rectangle}
\end{equation}
For affine $w_t$ and a polytope $\mathcal T$, maximizing a linear operating objective subject to $\widehat W(w_tg)\le\tau$ is a linear program. A selected operating shift enjoys~\eqref{eq:feasible_validation_uniform} on the same event even when chosen after validation.
\end{theorem}
\begin{proof}
For each sign, the primal is a feasible linear program on a compact set. Introduce the moment variable $m_i=\sum_j\mu_j\psi_{ij}(q_j-v_j)$ in its interval $[\widehat m_i-\epsilon_i,\widehat m_i+\epsilon_i]$. Its support function at $\lambda$ is $\lambda^\top\widehat m+\sum_i\epsilon_i|\lambda_i|$. Maximization over each probability coordinate gives the $\ell_{v_j}$ term as before. Finite linear-programming duality gives~\eqref{eq:feasible_validation_dual}, with attainment. On the coverage event the true model satisfies every defining inequality of $\widehat{\calQ}$, proving the uniform statement. For an exact representation, the objective is $\beta(t)^\top m$; enlarging the attainable moment set to the full confidence rectangle gives~\eqref{eq:feasible_envelope_dominates_rectangle}.

For the operating-choice problem, use a separate multiplier vector $\lambda^\sigma$ for each sign. Introduce nonnegative variables $a_j^\sigma,b_j^\sigma$ above the positive and negative parts of
$\sigma w_t(z_j)g_j-\sum_i\lambda_i^\sigma\psi_{ij}$, and variables $h_i^\sigma\ge\pm\lambda_i^\sigma$. Impose, for both signs,
\[
 \sum_j\mu_j[(1-v_j)a_j^\sigma+v_jb_j^\sigma]
  +(\lambda^\sigma)^\top\widehat m
  +\sum_i\epsilon_i h_i^\sigma\le\tau.
\]
All constraints are linear in $t$ and the auxiliary variables. Dual attainment makes this formulation equivalent to the envelope constraint. Finally, the coverage event contains the true model independently of how $t$ is subsequently selected.
\end{proof}
The moment-slack penalty uses the same support-function calculation as~\eqref{eq:penalized_dual} and the approximate-calibration dual of \citet[Appendix~B]{kiyani2026partial}. Here the moment intervals are translated by $\widehat m$, producing the term $\lambda^\top\widehat m$. The resulting set may exclude $q=V$ or lie on a boundary face. Consequently~\eqref{eq:feasible_validation_dual} is not a direct invocation of the earlier zero-centered, interior-qualified result; finite linear-programming duality establishes the stated generality. On the confidence event the set contains the true mean and is therefore nonempty. The no-certificate rule specifies the procedure when inconsistent evidence is encountered outside that event.

The improvement over a confidence rectangle may be strict, but need not be: if the worst rectangular moment combination is realizable, the bounds agree. This comparison concerns two certificates built from the same moments and simultaneous event, not a universal preference for a particular test basis. The feasible envelope can be used even without an exact span representation; lack of coverage then leaves an irreducible model ambiguity rather than an unacknowledged representation error. For an ellipsoidal $\mathcal T$, the same formulation adds a second-order cone constraint.

\begin{example}[A strict reduction from coupling feasible moments]
\label{ex:feasible_rectangle_strict}
Take three equiprobable contexts $x\in\{-1,0,1\}$, $V=1/2$, and tests $1,x$. Consider the hypothetical moment rectangle
$m_0\in[0.3,0.4]$, $m_1\in[-0.3,0.3]$. It is nonempty as a probability-model set: $q\equiv0.85$ gives $(m_0,m_1)=(0.35,0)$. For $f=x$, the rectangle bounds $|B|=|m_1|$ by $0.3$. Write $\xi_x=q_x-1/2$. Feasibility requires $\xi_x\in[-1/2,1/2]$ and $\xi_{-1}+\xi_0+\xi_1\ge0.9$. Since $\xi_0\le1/2$,
\[
 |\xi_1-\xi_{-1}|\le0.6,
 \qquad |m_1|\le0.2.
\]
The value $0.2$ is attained by $(\xi_{-1},\xi_0,\xi_1)=(-0.1,0.5,0.5)$ and its reflected counterpart, so the feasible envelope is exactly $0.2$. The standalone probability-domain bound is $\E|x|/2=1/3$; the improvement uses coupling with the intercept moment.

The same coupling can enlarge the set of certified operating choices. Take $g=1$ and $w_t=1+tx$, $0\le t\le1$. The ratios are nonnegative and normalized. The rectangular certificate is $0.4+0.3t$, whereas
\begin{equation}
 \widehat W(w_tg)=0.4+0.1t.
 \label{eq:strict_operating_envelope}
\end{equation}
Indeed $m_1\le1/2-m_0$, so
$m_0+tm_1\le(1-t)0.4+t/2$. Equality is attained by
$(\xi_{-1},\xi_0,\xi_1)=(0.2,0.5,0.5)$. Every feasible value is nonnegative, since $m_0\ge0.3$ and $|m_1|\le0.2$, so the absolute value does not change this maximum. At tolerance $\tau=0.45$, the largest certified shift is $t=1/2$, compared with $t=1/6$ for the rectangle. This deterministic illustration posits interval widths; it does not obtain them from a sample using the equally allocated empirical-Bernstein rule in Section~\ref{sec:fresh_shift_validation}. Theorem~\ref{thm:feasible_validation_envelope} applies to any valid simultaneous moment intervals, including conservative enlargements. The example demonstrates strict improvement for the specified rectangle, not that such improvement occurs in the sampled study below.
\end{example}

\paragraph{Several frozen pipelines and one conditional mean.}
When several forecasts are evaluated on the same contexts and binary outcome, their constraints concern a common $q$, although their residuals are centered at different forecasts. A convenient implementation writes all their moment constraints in raw form
$\ell\le Mq\le u$, intersected with $q\in[0,1]^N$. A residual row $\psi_i$ centered at forecast $v^{(a)}$ contributes
$M_{ij}=\mu_j\psi_{ij}$ and limits
$\widehat m_i\pm\epsilon_i+\sum_j M_{ij}v_j^{(a)}$.
Set $D_0=(M^\top,-M^\top)^\top$ and $b_0=(u^\top,-\ell^\top)^\top$. For a target pipeline $a$, put $c_j(t)=\mu_j w_t(z_j)g_j^{(a)}$. On a nonempty feasible set its signed envelope is equivalently
\begin{equation}
 \min_{\alpha\ge0}
 \left\{b_0^\top\alpha
       +\sum_j\bigl(\sigma c_j(t)-(D_0^\top\alpha)_j\bigr)_+
       -\sigma c(t)^\top v^{(a)}\right\},
 \qquad\sigma\in\{-1,1\}.
 \label{eq:feasible_raw_moment_dual}
\end{equation}
This follows by maximizing each $q_j\in[0,1]$ in the inequality-constrained Lagrangian. The same dual-slack construction gives a linear program for the operating choice. Using the common mean can only restrict the ambiguity set compared with separate pipeline-wise constraints, but strict tightening is not guaranteed.

If the pipelines, their test classes, and their operating families are fixed independently of validation and their moment events share one joint failure budget, the conclusions hold for all of them simultaneously. One may then select a pipeline as well as an operating shift using the resulting envelopes. This statement does not allow new outcome-dependent test classes outside that joint event. Like the preceding certificates, it concerns selected-plan evaluation bias, not oracle regret, realized safety, or a feedback-induced change in the conditional outcome kernel.

\section{Illustrative Specialization: Linear MPC with Mode Forecasts}
\label{sec:linear_mpc}

A finite disturbance-mode forecast gives a direct affine interface for a selected horizon plan.

Consider a linear system
\begin{equation}
    x_{t+1}=Ax_t+Bu_t+Ew_t,
\end{equation}
where the disturbance mode $M_t\in\{1,\ldots,d\}$ determines the disturbance statistics. Let
\begin{equation}
    Y_t=e_{M_t}\in\Delta_d
\end{equation}
be the one-hot mode vector, and suppose the forecaster outputs
\begin{equation}
    V_t=f(h_t)\in\Delta_d.
\end{equation}
For a candidate MPC plan $a$, define the mode-dependent horizon utility
\begin{equation}
    \ell_h(a,m)=-J_h(a,m),
\end{equation}
where $J_h(a,m)$ is the predicted finite-horizon cost if mode $m$ occurs. Then
\begin{equation}
    U_h(a,p)=\sum_{m=1}^d p_m\ell_h(a,m)
    = r_h(a)^\top p,
\end{equation}
with
\begin{equation}
    r_h(a)=(\ell_h(a,1),\ldots,\ell_h(a,d)).
\end{equation}
The nominal MPC law chooses
\begin{equation}
    \pi_{\rm nom}(h,v)\in\argmax_{a\in\calA(h)} r_h(a)^\top v.
\end{equation}
Decision calibration requires
\begin{equation}
    \E\left[
    r_H(\pi_{\rm nom}(H,V))^\top(Y-V)
    \right]=0.
\end{equation}
Active-set decision calibration strengthens this to
\begin{equation}
    \E\left[
    \indic\{\sigma(H,V)=\sigma\}
    r_H(\pi_{\rm nom}(H,V))^\top(Y-V)
    \right]=0,
    \qquad \sigma\in\Sigma.
\end{equation}

This is particularly interpretable: the mode probability forecast need not be calibrated everywhere, but it must be calibrated in the utility direction of the selected MPC plan, and preferably separately across active constraint regimes.

\begin{remark}[A certified instance of Theorem~\ref{thm:dual}]
\label{rem:linear_mpc_instance}
For compact convex Euclidean plan sets and continuous convex mode costs $J_h(a,m)$, utility is concave in the plan and affine in mode probabilities. Under the measurable-envelope conditions and uniformly relative-interior forecasts, Theorem~\ref{thm:dual} supplies attained value duality. Convexity also permits deterministic policy recovery by taking the barycenter of a relaxed saddle policy. Strong convexity of the expected plan cost gives a unique best response. These properties are statements about the declared horizon model, whose mode must describe the same uncertainty used to evaluate the selected plan.
\end{remark}

\section{Exact Finite-Horizon Planning Interface}
\label{sec:experiments}

This study isolates the forecast-to-plan interface under an exact finite-mode
model. Each independent episode has one wind mode, drawn once, and the entire
selected plan is committed. There is no overlapping-horizon reset or
replanning. The aim is to exhibit the calibration geometry and measure its
consequences for this specified planning objective. The forecast and its
hidden error are synthetic; this is not evidence about the prevalence of the
construction in learned models or a validation of closed-loop deployment.

\subsection{Physical model, objective, and context law}
The state is $(z_k,\nu_k)$ and the dynamics during the episode are
\begin{align}
z_{k+1}&=z_k+\Delta t\,\nu_k+\tfrac12\Delta t^2(u_k+w_Y),\\
\nu_{k+1}&=\nu_k+\Delta t(u_k+w_Y).
\end{align}
The observed context is $H=(z_0,\nu_0,B)$, where $B$ indexes the five
forecast-group covariate values. The prescribed synthetic base forecaster is
$V=p_B=f(H)$, deterministic given that context; only its corrections are fitted
from outcomes. Crossing forecast groups with the same physical initial states
therefore does not violate the forecasting information structure.
The mode $Y\in\{0,1\}$ is fixed for all $N$ transitions. For a plan
$a=(u_0,\ldots,u_{N-1})\in[-u_{\max},u_{\max}]^N$, its realised scenario cost is
\begin{equation}
J_m(a;h)=\sum_{k=1}^{N}\left[q_z(z_k-r)^2+q_\nu\nu_k^2+r_u u_{k-1}^2\right]
             +q_f(z_N-r)^2.
\label{eq:exact_scenario_cost}
\end{equation}
There is no soft-wall penalty: the state threshold $z_{\max}$ is an auxiliary
safety endpoint, not a hard state constraint or an unreported cost component.
Both scenario costs share a positive-definite Hessian because $r_u>0$.
The predictive utility is exactly
\begin{equation}
U_h(a,p)=-\{(1-p)J_0(a;h)+pJ_1(a;h)\}.
\label{eq:exact_affine_interface}
\end{equation}
Consequently, if the true mode probability is $q(h)$, the negative of this
utility evaluated at $q(h)$ is the actual conditional expected cost of the
committed plan.

\begin{table}[tbp]
\centering\small
\caption{Complete primary configuration. The context law is uniform on the Cartesian product of the stated supports.}
\label{tab:exact_parameters}
\begin{tabular}{p{0.29\linewidth}p{0.63\linewidth}}
\toprule
Quantity & Value \\
\midrule
Horizon and step & $N=6$, $\Delta t=0.2$ \\
Input and state threshold & $|u_k|\le2$, $z_{\max}=1$ \\
Wind modes & $w_0=-0.2$, $w_1=0.35$ \\
Reference and weights & $r=0.65$; $(q_z,q_\nu,r_u,q_f)=(10,1,0.2,5)$ \\
Initial position & $z_0\in\{0.55,0.60,0.65,0.70,0.75,0.80,0.85,0.90\}$ \\
Initial velocity & $\nu_0\in\{-0.2,0,0.2,0.4,0.6,0.8\}$ \\
Base forecast & $v\in\{0.15,0.325,0.5,0.675,0.85\}$ \\
Context count & $240$, sampled independently with replacement \\
Calibration / confirmation & $10{,}000$ / $2{,}000$ independent episodes \\
Shared normalization & $m_0=0.756574732$, $\sigma_0=2.259265385$ \\
Correction ridge and clipping & $\kappa=0.01$, corrected forecasts in $[0.02,0.98]$ \\
Construction coefficient & $\lambda=0.080$, with the feasibility cap below \\
Main seeds & calibration $1729$, confirmation $271828$, bootstrap $314159$ \\
Bootstrap & $5{,}000$ paired independent-episode resamples \\
\bottomrule
\end{tabular}
\end{table}

\paragraph{Exact bin null and exposed utility direction.}
Let $a_{\rm nom}(h)$ minimize the forecast-based cost and write its \emph{cost}
sensitivity as $s_0(h)=J_1(a_{\rm nom};h)-J_0(a_{\rm nom};h)$.
The utility sensitivity used by the theory is $g_{\rm nom}=-s_0$.
Normalize $\bar s_0=(s_0-m_0)/\sigma_0$ using the \emph{known finite context
law}; this uses no calibration or confirmation outcomes. The same normalization
is supplied to every correction. Define
\begin{equation}
q(h)=v+\lambda\{\bar s_0(h)-\mathbb E[\bar s_0(H)\mid V=v]\}.
\label{eq:exact_construction}
\end{equation}
We choose $\lambda$ as the smaller of $0.10$ and the largest nonnegative
coefficient keeping every finite-support probability in $[0.02,0.98]$.
No additional clipping of $q$ is needed. Thus
$\mathbb E[Y-V\mid V=v]=0$ \emph{exactly under the specified population law},
while $\mathbb E[s_0(Y-V)]=0.027555>0$. This is a deliberate adversarial
construction for forecast-group tests. It is fixed before any candidate
correction acts, so the episode mode law is action-independent conditional on
the common context. The corresponding utility bias has the opposite sign.
The error is deliberately in the fitted bin-intercept-plus-sensitivity-slope
class. This is a correctly specified controlled example, not a test of
robustness to arbitrary misspecification of that correction class.

\subsection{Matched fitting and evaluation}
Every correction receives exactly the same $10{,}000$ context/outcome pairs.
In forecast group $b$, sensitivity correction fits an intercept $a_b$ and slope
$\gamma_b$ by minimizing
\begin{equation}
\frac1{n_b}\sum_{i\in I_b}(Y_i-v_i-a_b-\gamma_b\bar s_{0,i})^2+\kappa\gamma_b^2.
\label{eq:exact_correction_fit}
\end{equation}
The intercept is unpenalized; the equivalent sum-loss ridge parameter is
$n_b\kappa$, not $\kappa$. Ridge is retained as the frozen protocol's
regularization choice; it was not tuned after confirmation. Histogram correction is the intercept-only fit on
these same pairs. Regime correction fits separate intercepts within each
forecast group and $G=\mathbf1\{z_0\ge0.75,\nu_0>0\}$. Corrected probabilities
are clipped to the interval in Table~\ref{tab:exact_parameters}. Nominal uses
$v$ and oracle uses $q(h)$. The robust benchmark solves
$\min_a\max\{J_0(a;h),J_1(a;h)\}$, rather than assuming the dangerous scenario
dominates every feasible plan. It is obtained from the scalar saddle dual and
the unique quadratic-plan best response. Its adversarial mixing probability
is a design weight, not a calibrated probabilistic forecast.

All methods are evaluated on the same independent confirmation contexts and
the same single mode draw per context. We compute both the realised cost
$J_Y(a;h)$ and the exact conditional expected cost
$C(a;h)=(1-q)J_0(a;h)+qJ_1(a;h)$. The latter uses the known simulator probability:
it is a controlled-model evaluation quantity, not a directly observed deployed
performance statistic. The primary endpoint, declared before confirmation
outcomes were drawn, is the paired mean of $C(a_{\rm sens};h)-C(a_{\rm nom};h)$.
The protocol was frozen after a geometry feasibility pilot; this was not an
external preregistration. No parameter was changed after inspecting confirmation
results. The geometry pilot and all its attempted configurations are archived.
There is one plan-selection context and one forecast/outcome pair per episode,
followed by six committed controls; the six transitions are not six independent
calibration observations.

Conditional expectations also evaluate one-step and whole-plan violation
indicators on these common initial states. This removes the occupation-measure
confound of comparing different receding-horizon state distributions. Neither
endpoint is the optimized quadratic cost. Bootstrap draws resample complete
independent confirmation episodes, preserving pairing. These intervals
condition on the one frozen calibration fit; they do not include variability
from repeatedly drawing and refitting a calibration dataset. Since the finite
context law is known, we also enumerate its entire support for exact
conditional-on-fit population values; those values need no Monte Carlo
estimation interval. The sampled-context intervals remain approximate inference
for this same conditional target. Separate exploratory refit replications below
address a different source of uncertainty. The primary comparison forms one
declared family. Fourteen secondary cost/safety comparisons use two-sided
paired Student-$t$ approximations with Holm correction; the approximation,
rather than an exact randomization-test guarantee, is intended.
Of these fourteen comparisons, 5 are identically zero
one-step-risk differences with $p=1$; they remain in the declared family.
The 24 grid
comparisons form a separate Holm family.

\paragraph{Solver and oracle checks.}
The box quadratic program is solved by enumerating active sets and checking
primal feasibility and KKT signs. The largest archived KKT residual is
$1.95\times10^{-14}$; the largest robust saddle gap is $6.72\times10^{-11}$.
Independent general-purpose optimizer checks agree. The oracle's expected
cost never exceeds any candidate's on any support context, up to the numerical
tolerance. This is the appropriate planning-interface dominance check; it does
not imply oracle dominance on a safety endpoint or each realised mode draw.

\subsection{Primary result for the original frozen calibration fit}
For the original calibration fit, the primary expected-cost difference is $-0.003361$, with paired-bootstrap
$95\%$ interval $[-0.003629, -0.003104]$, a $0.119\%$ reduction relative to nominal.
The benefit is small in absolute and relative terms, despite its precision.
For this fit, full-plan violation risk \emph{increases} by
$0.265$ percentage points, interval $[0.117, 0.399]$, with Holm-adjusted
secondary $p=0.00164$. All six methods have the same observed common-state
one-step risk. The realised-cost difference is $-0.003273$, interval
$[-0.007853, 0.001333]$, and remains inconclusive and power-limited for that endpoint.
Thus the study demonstrates a small
benefit for its exact expected-cost objective, not a safety improvement or a
resolved realised-cost advantage. The actually observed full-plan event
frequencies are also higher for sensitivity ($5.95\%$) than nominal
($5.65\%$); these descriptive frequencies use the common single
mode draws, whereas the Holm-adjusted risk comparison above averages the known
conditional event probabilities.

Exact enumeration of all 240 contexts for this same frozen fit gives cost
difference $-0.003296$ and risk difference $+0.259$ percentage
points. Both lie inside the corresponding conditional confirmation intervals.
The cost gain closes $93.1\%$ of the nominal-to-oracle expected-cost
gap. Oracle itself changes exact population risk by
$+0.020$ percentage points relative to nominal: minimizing
the specified cost does not order safety. The sensitivity risk change comes
from 7 increasing and 1 decreasing support contexts, and its sign
is not stable across calibration fits, as the replication study below shows.

\begin{table}[tbp]
\centering\small
\caption{Results conditional on the original frozen fit. Panel (a), empirical ECE, and the residuals in panel (b) use the common confirmation episodes. The Population ECE column in panel (b) and all of panel (c) exactly enumerate the known population. ECE uses ten fixed bins of the forecast actually used; robust instead diagnoses its adversarial design weight. Empirical ECE is descriptive; no population ECE confidence interval is claimed.}
\label{tab:exact_methods}
\textit{(a) Cost and whole-plan safety}\par\smallskip
\begin{tabular}{lrrrr}
\toprule
Method & $\overline C$ & Excess over oracle & Risk (\%) & Observed risk (\%) \\
\midrule
nominal & 2.820303 & 0.003600 & 5.615 & 5.650 \\
histogram & 2.820488 & 0.003785 & 5.680 & 5.600 \\
regime & 2.819798 & 0.003094 & 5.623 & 5.700 \\
sensitivity & 2.816942 & 0.000239 & 5.880 & 5.950 \\
oracle & 2.816704 & 0.000000 & 5.622 & 5.550 \\
robust & 3.132998 & 0.316295 & 4.221 & 4.100 \\
\bottomrule
\end{tabular}
\par\medskip\textit{(b) Calibration and distinct weight fields}\par\smallskip
\begin{tabular}{lrrrr}
\toprule
Method & Empirical ECE & Population ECE & $R_{\rm selected}^{\rm raw}$ & $R_{\rm nominal}^{\rm std}$ \\
\midrule
nominal & 0.0195 & 0.0000 & 0.03343 & 0.01085 \\
histogram & 0.0189 & 0.0058 & 0.03497 & 0.01092 \\
regime & 0.0228 & 0.0102 & 0.02371 & 0.00490 \\
sensitivity & 0.0252 & 0.0055 & 0.00577 & -0.00233 \\
oracle & 0.0251 & 0.0000 & 0.00490 & -0.00192 \\
robust & 0.0863 & 0.0973 & -0.00529 & -0.26370 \\
\bottomrule
\end{tabular}
\par\medskip\textit{(c) Exact population cost and safety for the frozen fit}\par\smallskip
\begin{tabular}{lrrr}
\toprule
Method & Population $C$ & Excess over oracle & Population risk (\%) \\
\midrule
nominal & 2.824499 & 0.003541 & 5.836 \\
histogram & 2.824680 & 0.003722 & 5.917 \\
regime & 2.824149 & 0.003191 & 5.791 \\
sensitivity & 2.821203 & 0.000245 & 6.095 \\
oracle & 2.820958 & 0.000000 & 5.856 \\
robust & 3.138653 & 0.317695 & 4.430 \\
\bottomrule
\end{tabular}
\end{table}

\begin{figure}[tbp]
\centering
\includegraphics[width=\linewidth]{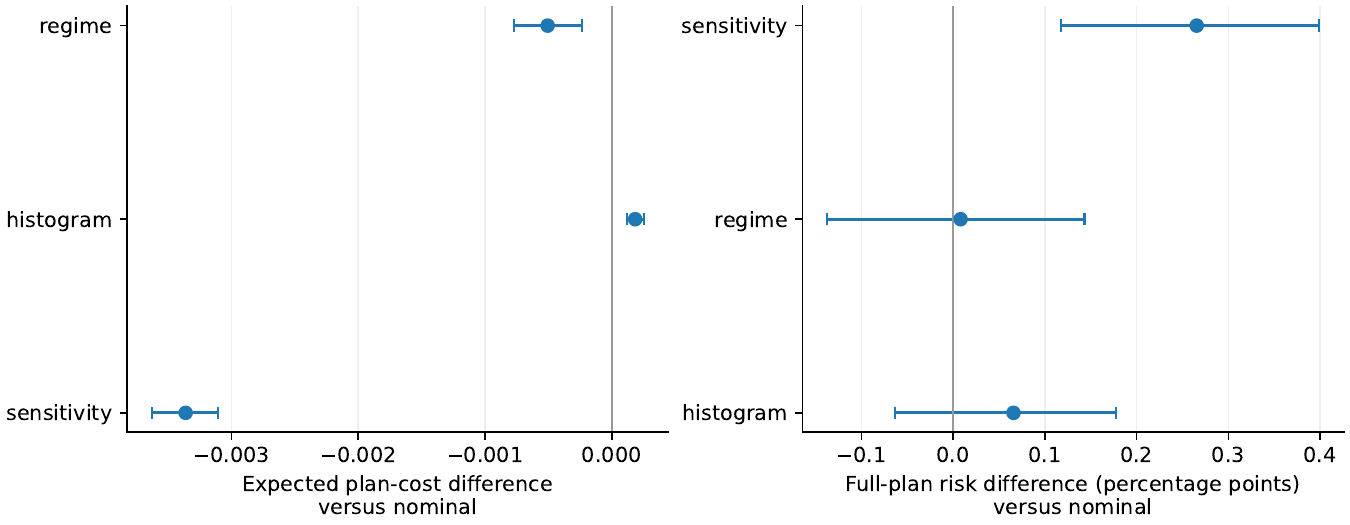}
\caption{Paired differences relative to nominal on common contexts for the original frozen calibration fit. Error bars are pointwise episode-bootstrap $95\%$ intervals conditional on that fit; risk-comparison significance uses the separate conditional Holm-adjusted family. The displayed safety direction does not persist across refits.}
\label{fig:exact_paired}
\end{figure}

\subsection{Calibration and representation diagnostics}
Two weighted quantities must be distinguished. The fit-field diagnostic is
$R_{\rm nominal}^{\rm std}=\mathbb E[\bar s_0(H)(Y-p_{\rm used})]$, using exactly
the shared nominal field in~\eqref{eq:exact_correction_fit}. Before clipping,
the calibration-sample normal equations give
$\sum_{I_b}(Y-\widetilde p)=0$ and
$\sum_{I_b}\bar s_0(Y-\widetilde p)=n_b\kappa\gamma_b$.
Consequently the raw nominal-field sum is
$\sigma_0n_b\kappa\gamma_b$; the intercept removes its constant component.
These are fitting-sample identities, not held-out or changed-plan guarantees.
For actual selected-plan value we instead report
$R_{\rm selected}^{\rm raw}=\mathbb E[(J_1(a)-J_0(a))(Y-p_{\rm used})]$.
The latter is the negative utility-prediction bias of that selected plan, and
its weight changes when the plan changes. It is not the same field as the fit.

The known population nominal raw moment is $0.027555$, and the sensitivity
method's selected-plan population moment is $0.000414$.
The finite-outcome confirmation interval for the nominal raw moment is wider:
$0.03343$ with interval $[-0.00962, 0.07782]$. The shared standardized
nominal-field diagnostic is $0.01085$, interval $[-0.00723, 0.02890]$.
Both intervals include zero. For sensitivity, the corresponding selected raw
and standardized nominal diagnostics are $0.00577$, interval
$[-0.03534, 0.04791]$, and $-0.00233$, interval $[-0.02063, 0.01593]$.
The exact nonzero population construction is therefore distinguished from an
inconclusive finite-outcome test of the nominal moment. No additional evaluation
data were generated to force that diagnostic to become significant.

\begin{figure}[tbp]
\centering
\includegraphics[width=0.9\linewidth]{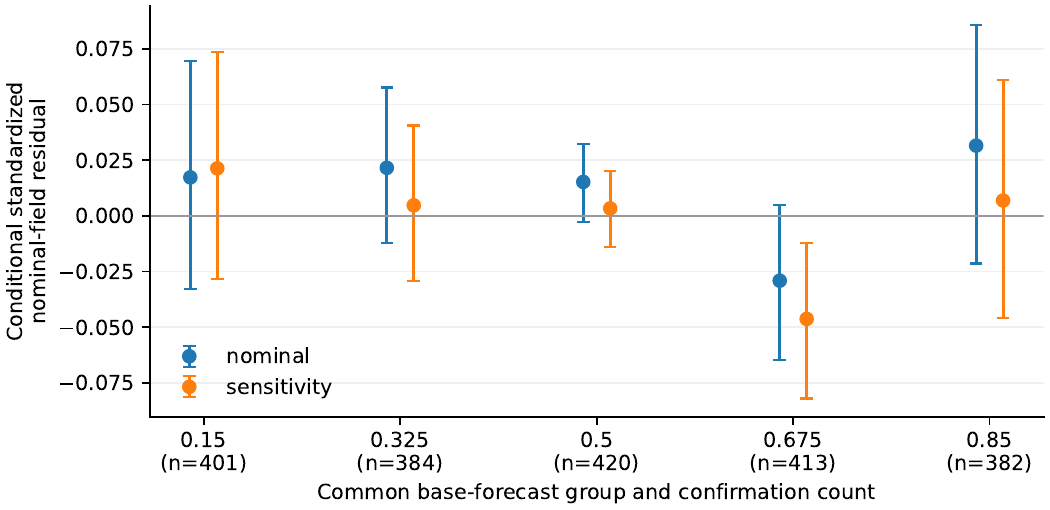}
\caption{Common-base-group conditional means of $\bar s_0(H)(Y-p_{\rm used})$ before and after correction. Counts refer to the same independent confirmation episodes for both methods. Error bars resample whole episodes and recompute the within-group ratio; they do not treat six transitions as separate observations. The field remains the nominal standardized fitting field, not the corrected plan's raw utility coefficient.}
\label{fig:exact_weighted_groups}
\end{figure}

Figure~\ref{fig:exact_weighted_groups} exposes the within-group heterogeneity
behind the aggregate standardized nominal-field diagnostic. Both methods use
the same initial contexts, group membership, and fitting field. The displayed
conditional moments need not improve in every group, and their intervals are
pointwise exploratory intervals rather than simultaneous five-group guarantees.

The nominal forecast-group confirmation counts are $401, 384, 420, 413, 382$.
Per-group binomial tests are valid here because episodes are independent and
the base forecast is constant within a group. Holm correction over those five
tests rejects no group at level $0.05$; non-rejection is not a proof of
calibration. Exact population bin calibration follows instead from the finite
construction. Reliability frequency and signed-gap bands resample independent
episodes in ten fixed equal-width used-forecast bins. Empirical ECE is a
descriptive point diagnostic; exact population ECE is available by enumeration.
We do not report the ordinary percentile bootstrap as a population ECE interval:
the absolute-value statistic is nonsmooth at a zero bin gap, and nominal and
oracle have exact population ECE zero despite positive empirical ECE.
Specifically, let $I_j=\{i:p_i^{\rm used}\in[(j-1)/10,j/10)\}$, with the
right endpoint included in the final bin. Writing $n_j=|I_j|$, the statistic is
\begin{equation}
\operatorname{ECE}(p^{\rm used})=
\sum_{j:n_j>0}\frac{n_j}{M}\left|\overline Y_j-\overline p^{\rm used}_j\right|.
\label{eq:exact_used_ece}
\end{equation}
Empty bins are omitted, and the bin membership is determined by the forecast
actually used, rather than by the original forecast-group label.

\begin{figure}[tbp]
\centering
\includegraphics[width=0.9\linewidth]{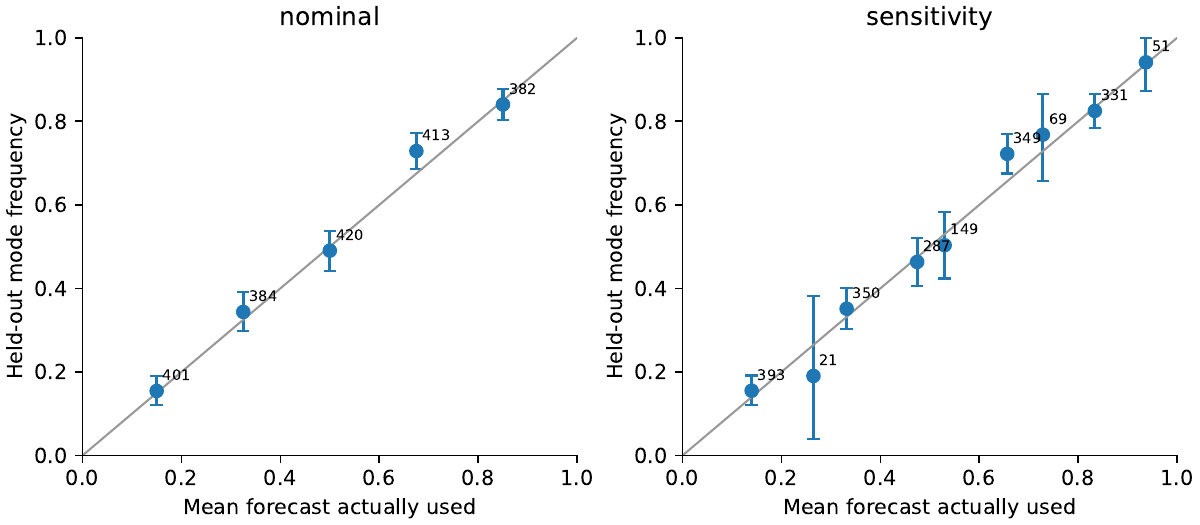}
\caption{Probabilities actually used by nominal and sensitivity methods on common confirmation contexts for the frozen fit. Labels give bin counts; bars are pointwise independent-episode bootstrap intervals for within-bin outcome frequencies, conditional on that fit. These are not ECE confidence intervals. Empirical ECE need not improve when the decision-relevant cost direction improves.}
\label{fig:exact_reliability}
\end{figure}

For a test span $\mathcal S_\Psi$, define the unspanned exposure energy
\[
E_\Psi=\frac{\|g_{\rm nom}-\operatorname{Proj}_{\mathcal S_\Psi}g_{\rm nom}\|_{L_2(\mu)}^2}
{\|g_{\rm nom}\|_{L_2(\mu)}^2}.
\]
Table~\ref{tab:exact_exposure} computes this projection
under the \emph{known complete finite context law}, not by fitting and testing
on the same outcome sample. The quadratic context monomials use
$\theta=(z_0,\nu_0,v)$; the active-set construction multiplies every monomial
of degree at most two by each nominal box-active-set indicator. The exact
representation diagnostics illustrate what the specified tests can span;
test count alone is not a data-efficiency or generalization claim. Observed
moments of these fixed tests on independent confirmation outcomes are archived.
Near-zero representation error establishes coverage, not calibration of those
tests: the nominal population weighted moment is nonzero here. It therefore
does not establish nominal minimax optimality from the exact-calibration
sufficiency theorem.

\begin{table}[tbp]
\centering\small
\caption{Known-population sensitivity representation, with $g_{\rm nom}=-s_0$. Redundant columns explain rank smaller than nominal test count.}
\label{tab:exact_exposure}
\resizebox{\linewidth}{!}{\begin{tabular}{lrrrr}
\toprule
Test class & Columns & Rank & $L_2$ representation error & $E_\Psi$ \\
\midrule
Forecast-group indicators & 5 & 5 & $8.84\times10^{-1}$ & $1.38\times10^{-1}$ \\
Group-by-regime indicators & 10 & 10 & $7.36\times10^{-1}$ & $9.54\times10^{-2}$ \\
Context monomials, degree $\le2$ & 10 & 10 & $1.88\times10^{-1}$ & $6.26\times10^{-3}$ \\
Active-set indicators & 4 & 4 & $2.19\times10^{0}$ & $8.43\times10^{-1}$ \\
Active sets $\times$ monomials, degree $\le2$ & 40 & 35 & $2.82\times10^{-14}$ & $1.40\times10^{-28}$ \\
Groups plus nominal sensitivity & 6 & 6 & $2.23\times10^{-15}$ & $8.77\times10^{-31}$ \\
\bottomrule
\end{tabular}}
\end{table}

\paragraph{A finite-sample certificate and its conservatism.}
For the five forecast-group indicators and the shared standardized nominal sensitivity,
the raw utility sensitivity has the exact representation
$g_{\rm nom}=-m_0\sum_b\mathbf1_b-\sigma_0\bar s$.
Using the independent-episode bound with failure probability $0.05$, the
known absolute bounds are
\[
B=(0.850000, 0.675000, 0.500000, 0.675000, 0.850000, 1.968192)
\]
and the empirical moments are
\[
\widehat m=(0.00092500, 0.00360000, -0.00200000, 0.01111250, -0.00185000, 0.01084807).
\]
With $M=2000$, $k=6$, and $\zeta=0.05$, the
certificate
\[
\eta=\sum_i|\beta_i|\left\{|\widehat m_i|+B_i\sqrt{2\log(2k/\zeta)/M}\right\}
=0.567282.
\]
The known population anticipated-cost bias is only $0.027555$.
Thus the finite-sample result supplies a valid but conservative plan-value
certificate in this example. Its conservatism reflects worst-case bounded
ranges and the absolute representation weights, rather than the observed
variance alone. It is neither an estimate of the correction's
cost benefit nor a trajectory-safety certificate. The exact representation
is available from the known model, so no representation-validation failure
budget is used in this calculation.

\subsection{Controls and parameter sweep}
The zero-shift control has $q=v$, so nominal already minimizes true expected
cost. Finite-sample fitting can only add regret and the original fit produces a small
positive correction-cost difference. The orthogonal control centers a fixed
sinusoidal context function within each forecast group and projects out the
centered nominal sensitivity. Its exact nominal weighted population moment
is zero; the original sensitivity fit also worsens expected cost there. These
controls delimit the fitted correction's usefulness rather than assert that
all finite-sample pipelines become identical. The easy-safety control uses
$z_0\in\{0.3,0.4,0.5,0.6\}$, $\nu_0\in\{-0.2,0,0.2\}$,
$r=0.45$, $(w_0,w_1)=(-0.1,0.15)$ and $u_{\max}=2.5$.
For the original fit, all methods then have zero whole-plan violation probability while a small
expected-cost improvement remains. Other settings are unchanged. Control
calibration/confirmation/bootstrap seeds use offsets $100$, $200$, and $300$.

\begin{table}[tbp]
\centering\small
\caption{Original frozen control-fit evaluations, $2{,}000$ episodes each. Differences are sensitivity minus nominal; intervals are pointwise paired-bootstrap intervals conditional on each control's fit.}
\label{tab:exact_controls}
\setlength{\tabcolsep}{4pt}
\begin{tabular}{lrrcc}
\toprule
Control & \shortstack{Cost\\difference} & $95\%$ interval & \shortstack{Nominal\\risk (\%)} & \shortstack{Sensitivity\\risk (\%)} \\
\midrule
zero & 0.000492 & $[0.000459, 0.000526]$ & 5.576 & 5.405 \\
orthogonal & 0.000823 & $[0.000270, 0.001380]$ & 5.857 & 5.857 \\
easy & -0.000242 & $[-0.000261, -0.000223]$ & 0.000 & 0.000 \\
\bottomrule
\end{tabular}
\end{table}

The 24-cell sweep crosses $u_{\max}\in\{1.6,2.0,2.4\}$,
$w_1\in\{0.25,0.35,0.45,0.55\}$ and $r\in\{0.60,0.70\}$, keeping
$w_0=-0.2$. Each cell refits all corrections on the same $10{,}000$ calibration
pairs within that cell and uses $1{,}000$ independent common confirmation
episodes. Cell $j=0,\ldots,23$ adds $1000+j$ to each main seed and uses $2{,}000$
bootstrap replicates. The construction and common scaler are recomputed from
each specified known context law, without consulting confirmation outcomes.
Cost differences favor sensitivity in $24/24$ cells; $24/24$
remain significant in the conditional two-sided Holm family. Each result
conditions on one frozen fit per cell. These local model variations do not
establish prevalence or universal superiority, and their cost results do not
repair the adverse safety finding for the original primary fit.

\begin{table}[tbp]
\centering\small
\caption{Complete expected-cost parameter sweep. All differences are sensitivity minus nominal. The final column is the Holm-adjusted paired-$t$ approximation across 24 cells; intervals are pointwise paired-bootstrap intervals conditional on one original frozen calibration fit per cell.}
\label{tab:exact_grid}
\begin{tabular}{rrrrrrr}
\toprule
Cell & $u_{\max}$ & $w_1$ & $r$ & Difference & $95\%$ interval & Adjusted $p$ \\
\midrule
1 & 1.6 & 0.25 & 0.60 & -0.001560 & $[-0.001712, -0.001418]$ & $<10^{-4}$ \\
2 & 1.6 & 0.25 & 0.70 & -0.001553 & $[-0.001741, -0.001372]$ & $<10^{-4}$ \\
3 & 1.6 & 0.35 & 0.60 & -0.002089 & $[-0.002306, -0.001896]$ & $<10^{-4}$ \\
4 & 1.6 & 0.35 & 0.70 & -0.002015 & $[-0.002222, -0.001815]$ & $<10^{-4}$ \\
5 & 1.6 & 0.45 & 0.60 & -0.002102 & $[-0.002341, -0.001869]$ & $<10^{-4}$ \\
6 & 1.6 & 0.45 & 0.70 & -0.001868 & $[-0.002137, -0.001612]$ & $<10^{-4}$ \\
7 & 1.6 & 0.55 & 0.60 & -0.001253 & $[-0.001668, -0.000831]$ & $<10^{-4}$ \\
8 & 1.6 & 0.55 & 0.70 & -0.002479 & $[-0.002821, -0.002130]$ & $<10^{-4}$ \\
9 & 2.0 & 0.25 & 0.60 & -0.002529 & $[-0.002823, -0.002264]$ & $<10^{-4}$ \\
10 & 2.0 & 0.25 & 0.70 & -0.003130 & $[-0.003470, -0.002791]$ & $<10^{-4}$ \\
11 & 2.0 & 0.35 & 0.60 & -0.003168 & $[-0.003569, -0.002802]$ & $<10^{-4}$ \\
12 & 2.0 & 0.35 & 0.70 & -0.003256 & $[-0.003648, -0.002873]$ & $<10^{-4}$ \\
13 & 2.0 & 0.45 & 0.60 & -0.003889 & $[-0.004438, -0.003395]$ & $<10^{-4}$ \\
14 & 2.0 & 0.45 & 0.70 & -0.003464 & $[-0.003975, -0.002982]$ & $<10^{-4}$ \\
15 & 2.0 & 0.55 & 0.60 & -0.003271 & $[-0.003860, -0.002731]$ & $<10^{-4}$ \\
16 & 2.0 & 0.55 & 0.70 & -0.004581 & $[-0.005240, -0.003931]$ & $<10^{-4}$ \\
17 & 2.4 & 0.25 & 0.60 & -0.003522 & $[-0.003897, -0.003167]$ & $<10^{-4}$ \\
18 & 2.4 & 0.25 & 0.70 & -0.002141 & $[-0.002371, -0.001924]$ & $<10^{-4}$ \\
19 & 2.4 & 0.35 & 0.60 & -0.004221 & $[-0.004669, -0.003801]$ & $<10^{-4}$ \\
20 & 2.4 & 0.35 & 0.70 & -0.002229 & $[-0.002444, -0.002027]$ & $<10^{-4}$ \\
21 & 2.4 & 0.45 & 0.60 & -0.003280 & $[-0.003721, -0.002877]$ & $<10^{-4}$ \\
22 & 2.4 & 0.45 & 0.70 & -0.001409 & $[-0.001712, -0.001120]$ & $<10^{-4}$ \\
23 & 2.4 & 0.55 & 0.60 & -0.004802 & $[-0.005316, -0.004295]$ & $<10^{-4}$ \\
24 & 2.4 & 0.55 & 0.70 & -0.000580 & $[-0.000970, -0.000190]$ & $0.00321$ \\
\bottomrule
\end{tabular}
\end{table}

\FloatBarrier
\subsection{Exploratory calibration-fit replication}
We independently draw 200 additional calibration datasets of $10{,}000$
episodes for each of the primary, three control, and 24 grid cases. This
exploratory analysis was specified after inspecting the original study; its
stream rule was fixed before these additional draws. It is separate from the
original frozen confirmation protocol and does not replace its data or change
any configuration. The fixed NumPy SeedSequence entropy $2026111401$ uses child keys $(c,r)$ for
zero-based case and replication indices, with the full case order and stream
rule saved before any new outcomes. Within each fit, all corrections again
receive identical pairs and the same original model-known scaler. We evaluate
each fitted method by exact enumeration over that case's entire context law,
so these distributions isolate calibration-fit variation rather than
confirmation-context or mode Monte Carlo error. No new robust solves are
needed. The central ranges below describe the sampled distribution over fits,
not confidence intervals or simultaneous guarantees.

The primary sensitivity cost difference is negative in
$200/200$ replications, with mean $-0.002851$ and sample
standard deviation $0.000328$. The mean closes
$80.5\%$ of the same fixed nominal-to-oracle population cost gap.
The original frozen fit is favorable relative to these exploratory refits:
only 7 of 200 have a more negative exact population
cost difference, an empirical $3.5\%$ lower tail.
The original fit was fixed before confirmation outcomes were drawn.
Across the primary refits, the risk difference has mean
$0.016$ percentage points, standard deviation $0.247$,
and central $95\%$ range $[-0.477, 0.432]$ percentage points.
Risk increases in $115$ fits, decreases in $84$,
and ties in $1$ (absolute tolerance $10^{-12}$).
Thus the sampled fits consistently improve this expected-cost objective,
while safety direction in the primary configuration varies with the calibration sample. These results do
not prove a zero average safety effect or universal cost improvement.

\begin{table}[tbp]
\centering\small
\caption{Exact-law evaluation of 200 independent calibration fits per case. Primary rows compare each correction with nominal; control rows use sensitivity. Risk units are percentage points. The final column counts positive/negative/tied risk differences. Standard deviations describe refitting variability, not uncertainty conditional on one fit.}
\label{tab:exact_refits}
\setlength{\tabcolsep}{4pt}
\begin{tabular}{lrrrrr}
\toprule
Case/method & \shortstack{Mean\\cost difference} & \shortstack{Cost\\SD} & \shortstack{Mean risk\\difference} & \shortstack{Risk\\SD} & \shortstack{Risk\\$+/-/0$} \\
\midrule
histogram & 0.000369 & 0.000253 & 0.049 & 0.176 & 95/73/32 \\
regime & -0.000232 & 0.000355 & -0.082 & 0.171 & 65/135/0 \\
sensitivity & -0.002851 & 0.000328 & 0.016 & 0.247 & 115/84/1 \\
zero control & 0.000669 & 0.000312 & 0.020 & 0.161 & 80/86/34 \\
orthogonal control & 0.000670 & 0.000340 & -0.003 & 0.131 & 78/86/36 \\
easy control & -0.000276 & 0.000070 & 0.000 & 0.000 & 0/0/200 \\
\bottomrule
\end{tabular}
\end{table}

\begin{figure}[tbp]
\centering
\includegraphics[width=\linewidth]{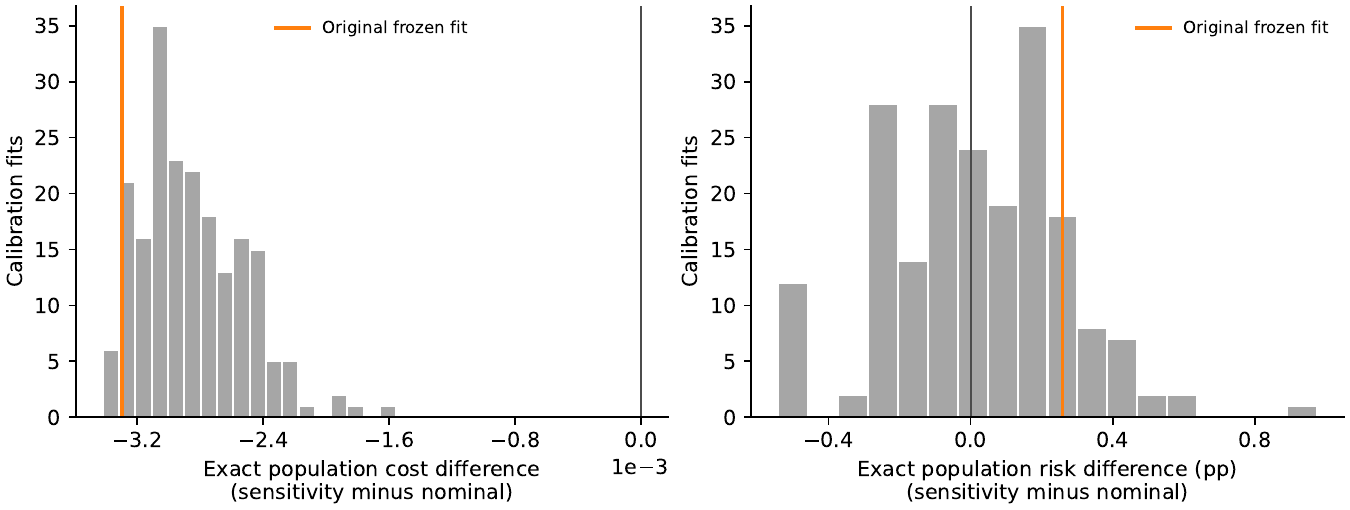}
\caption{Exploratory exact population differences over 200 independent primary calibration refits. The orange line marks the original frozen fit. Cost gains persist in these replications, whereas the threshold-based full-plan risk difference changes sign. These histograms contain no confirmation-sampling noise.}
\label{fig:exact_refits}
\end{figure}
\FloatBarrier

Sensitivity has higher expected cost in every replicated zero-shift and
orthogonal fit, while the easy control retains zero safety risk and improves
cost in all 200 fits. In the grid, all 200 fits improve cost in
$22/24$ cells; cells 22 and 24 improve in 199 and 196 fits,
respectively (one-based table numbering). The per-fit records and all
method/cell distribution summaries are archived, including unfavorable fits.

Table~\ref{tab:exact_grid_refits} makes the configuration dependence of risk
explicit, using the same one-based cell numbering and parameter combinations
as Table~\ref{tab:exact_grid}. Risk decreases in
200/200 fits in cell 18,
196/200 in cell 10, and
195/200 in cell 2, but increases in
183/200 in cell 21. Cells 1, 9, 17, and 19 have no risk
change in any fit. These are descriptive exact-law refit summaries, with no
additional significance tests or uniform direction claim.

\FloatBarrier
\begin{table}[!htbp]
\centering\small
\caption{Exploratory grid replication: 200 independent calibration fits per cell, evaluated over the exact context law. All differences are sensitivity minus nominal. Cost SD is the sample standard deviation across fits; risk units are percentage points. The last column counts risk increases/decreases/ties using tolerance $10^{-12}$. Cell parameters are in Table~\ref{tab:exact_grid}; no new inference family is introduced.}
\label{tab:exact_grid_refits}
\setlength{\tabcolsep}{5pt}
\begin{tabular}{rrrrr}
\toprule
Cell & \shortstack{Mean cost\\difference} & Cost SD & \shortstack{Mean risk\\difference (pp)} & \shortstack{Risk\\$+/-/0$} \\
\midrule
1 & -0.001520 & 0.000185 & 0.000 & 0/0/200 \\
2 & -0.001476 & 0.000200 & -0.181 & 1/195/4 \\
3 & -0.001989 & 0.000317 & -0.050 & 5/139/56 \\
4 & -0.001705 & 0.000296 & -0.119 & 57/107/36 \\
5 & -0.002556 & 0.000388 & 0.138 & 128/70/2 \\
6 & -0.002051 & 0.000434 & -0.341 & 2/168/30 \\
7 & -0.003020 & 0.000503 & 0.163 & 120/64/16 \\
8 & -0.002492 & 0.000561 & 0.397 & 113/86/1 \\
9 & -0.002190 & 0.000221 & 0.000 & 0/0/200 \\
10 & -0.002634 & 0.000244 & -0.388 & 1/196/3 \\
11 & -0.002874 & 0.000308 & -0.067 & 0/136/64 \\
12 & -0.003034 & 0.000331 & -0.475 & 9/172/19 \\
13 & -0.003777 & 0.000420 & 0.210 & 160/40/0 \\
14 & -0.003500 & 0.000471 & -0.176 & 0/151/49 \\
15 & -0.004469 & 0.000589 & -0.052 & 78/102/20 \\
16 & -0.004113 & 0.000576 & 0.190 & 86/106/8 \\
17 & -0.003161 & 0.000263 & 0.000 & 0/0/200 \\
18 & -0.002036 & 0.000241 & -0.632 & 0/200/0 \\
19 & -0.003362 & 0.000351 & 0.000 & 0/0/200 \\
20 & -0.002004 & 0.000376 & -0.416 & 16/172/12 \\
21 & -0.003621 & 0.000449 & 0.354 & 183/15/2 \\
22 & -0.001995 & 0.000513 & -0.010 & 3/8/189 \\
23 & -0.003743 & 0.000617 & -0.410 & 30/154/16 \\
24 & -0.001945 & 0.000649 & -0.447 & 75/120/5 \\
\bottomrule
\end{tabular}
\end{table}
\FloatBarrier

The artifact saves the complete finite context/scenario plan cache, every
independent calibration and confirmation episode, parameters, fit coefficients,
and inference summaries. Original frozen-fit costs, safety probabilities, diagnostics, intervals,
and tables can be reconstructed from those records without rerunning fitting
or selecting a new configuration. The separate replication records can be
regenerated from their archived stream rule. The controlled result concerns this exact
finite-horizon interface; robustness to unknown mode laws, distribution shift,
online replanning, and improved trajectory safety remain separate questions.
\FloatBarrier

\subsection{Direct-test and variance-sensitive certificates on the original sample}
\label{sec:direct_certificate_experiment}
The six-test certificate above is not the most economical way to evaluate this one frozen controller. We recompute direct-test alternatives on the \emph{same original 2,000 confirmation episodes}, without refitting any forecast or drawing new outcomes. The utility moment is $X=-s_0(Y-p)$, with empirical mean $-0.0334268$, unbiased episode variance $0.942730$, and known full range $[-5.089756,1.715366]$. The range enumerates all 240 model contexts and both possible outcomes; it is not estimated from validation extrema. Its width is $6.805122$, whereas the symmetric absolute bound is $5.089756$.

\begin{table}[t]
\centering\small
\caption{Certificates for the original frozen controller and unchanged confirmation sample. Each row separately has 95\% coverage; the rows do not authorize selecting the smallest bound without a joint confidence construction. Values bound the absolute \emph{utility evaluation bias}, whose known population magnitude is $0.027555$.}
\label{tab:direct_certificate_comparison}
\begin{tabular}{lrr}
\toprule
Construction & Absolute-bias upper bound & Signed utility-bias interval\\
\midrule
Six tests, original Hoeffding & $0.567282$ & ---\\
Direct test, symmetric Hoeffding & $0.342559$ & $[-0.342559,0.275705]$\\
Direct test, full-range Hoeffding & $0.240085$ & $[-0.240085,0.173232]$\\
Direct test, empirical Bernstein & $0.132508$ & $[-0.132508,0.065654]$\\
\bottomrule
\end{tabular}
\end{table}

Table~\ref{tab:direct_certificate_comparison} implements Proposition~\ref{prop:direct_test_comparison} and equation~\eqref{eq:transfer_eb_radius} with one test. The tighter direct range and observed variance reduce conservatism, but all three direct intervals still contain zero. Thus this original sample does not detect the planted value bias with these finite-sample guarantees. We retain that limitation. The structured class gains a different function in the following study: it covers an operating-law family that a calibrated single source moment can leave exposed. The new family validation uses its own sample size and confidence allocation, stated explicitly below.

\section{Operating-Law Transfer and Calibration-Test Budgets}
\label{sec:operating_transfer_experiment}

The original study evaluates a fixed operating law. We now exercise the additional
capability of Section~\ref{sec:operating_transfer}: using source-law calibration
evidence to evaluate a specified family of operating laws without collecting
target-law outcomes. The first part is a designed geometric separation on the
same physical model, distinct from the original calibration/cost experiment.
The second part applies the transfer calculation to the independently trained
forecasts of Section~\ref{sec:learned_forecast_experiment}. Neither calculation
changes the controller or the conditional outcome law when the operating law
is reweighted.

\subsection{A source certificate and a transferable certificate}
The source distribution $\mu$ is uniform on the original 240 contexts, the
forecast is the original five-valued $p$, and the nominal plan remains fixed at
each context. Write $g=-s_0$ for its utility sensitivity. For this particular
model, subtracting the two mode costs cancels their common quadratic terms and
leaves an affine function of the plan and the initial state. Hence $g$ is
affine on every critical region. This stronger property is specific to the
model; a general quadratic sensitivity in Theorem~\ref{thm:quadratic_test_span}
would require cubic tests for affine density ratios, as explained in
Corollary~\ref{cor:transfer_polynomial}.

Let
\[
 \varphi_v=\frac{v_0-\E_\mu v_0}{\max|v_0-\E_\mu v_0|},\qquad
 w_t=1+t\varphi_v,\qquad |t|\le0.8.
\]
The weights integrate to one and lie in $[0.2,1.8]$. They define an entire
continuum of target context laws, while the Bernoulli mode probability $q$ is
held fixed at each context. For any fixed residual field $\xi=q-p$ the target
evaluation bias is
\begin{equation}
 B(t)=\E_\mu[w_t g\xi]=m_0+t m_1,
 \qquad m_0=\E_\mu[g\xi],\quad m_1=\E_\mu[\varphi_vg\xi].
 \label{eq:transfer_experiment_bias}
\end{equation}

We compare the source test span
\[
 \calS_0=\Span\{\indic_{b=1},\ldots,\indic_{b=5},g\}
\]
with the augmented structured span $\calS_2$ containing the five bin indicators
and all critical-region monomials of degree at most two. The former has rank 6;
the latter has 45 columns and rank 37 on the finite population. The 40 regional
quadratic columns alone have rank 35. They span $g$ and $\varphi_vg$ exactly,
whereas $\calS_0$ need not span the product. Thus calibration of $g$ alone fixes
$B(0)$, while calibration of the product class fixes $B(t)$ for every declared
target law. The minimal product class for this scalar shift family is just
$\{g,\varphi_vg\}$. The larger regional library supplies a structural coverage
promise; it has no asserted statistical advantage over these two direct tests.

We construct two nonzero residual laws using the known model and context law,
without sampled outcomes:
\begin{enumerate}[label=(\alph*)]
\item \emph{Source-only calibration.} Project $\varphi_vg$ onto
$\calS_0^\perp$ and rescale the nonzero projection to a feasible residual.
All five bin moments and the source sensitivity moment vanish, but
$m_1=0.0617303782$.
\item \emph{Structured calibration.} Project
$\sin(19.7z_0+7.3v_0+3.1p)$ onto $\calS_2^\perp$ and rescale it to a feasible
residual. All structured moments vanish, so both coefficients in
\eqref{eq:transfer_experiment_bias} are zero despite $q\ne p$.
\end{enumerate}
For each law, we normalize the projection by its maximum absolute value and
use $80\%$ of the largest scale preserving $q\in[0.02,0.98]$. The resulting
root-mean-square probability errors are $0.0411491$ and $0.0444220$;
the probability ranges are $[0.046000,0.923569]$ and
$[0.046000,0.935604]$. These laws are deliberately engineered examples of
different calibration promises. They are not evidence that either residual
pattern occurs naturally in trained forecasts.

\subsection{One validation sample for a continuum of target laws}
We draw 100,000 independent source episodes per law. Each episode contains a
uniform context and one mode outcome. The sample size, two residual laws,
shift family, and stream rule were fixed before the final validation draws.
The final entropy is 2026111403, with separate NumPy SeedSequence child streams
for the two laws. A distinct stream with entropy 2026111402 was used for an
earlier implementation check; it is not part of the reported validation sample.
There is no external preregistration claim, adaptive stopping rule, seed
replacement, or final-outcome-dependent setting choice.

Apply the empirical-Bernstein radius~\eqref{eq:transfer_eb_radius} to the two
fixed variables $g(Y-p)$ and $\varphi_vg(Y-p)$ in each law. Their known ranges
are obtained by enumerating the 240 contexts and the two possible mode
outcomes. A two-sided union bound across the four moments gives a single
probability-at-least-$0.95$ event, using $\log(16/0.05)$ in each radius.
On that event, the interval
\[
 \widehat m_0+t\widehat m_1
 \ \pm\ \{r_0+|t|r_1\}
\]
covers $B(t)$ for both laws and every $|t|\le0.8$. This is an application of
Theorem~\ref{thm:uniform_transfer_certificate}; the continuum guarantee does
not come from a union bound over plotted grid points.

\begin{table}[t]
\centering\small
\caption{Designed calibration-transfer separation. Exact values enumerate the
known finite context law. The interval and uniform upper bound share a joint
95\% event across both laws and every $|t|\le0.8$. Finite samples give bounds;
the exact zero moments in the structured law follow from its construction.}
\label{tab:operating_transfer_validation}
\begin{tabular}{lrr}
\toprule
Quantity & Source-only law & Structured law\\
\midrule
Exact source bias $B(0)$ & $0$ & $0$\\
Exact slope $m_1$ & $0.0617304$ & $0$\\
Exact target bias $B(0.8)$ & $0.0493843$ & $0$\\
Estimated target bias & $0.0562548$ & $-0.0038942$\\
Simultaneous interval at $t=0.8$
 & $[0.0388284,0.0736811]$ & $[-0.0217930,0.0140046]$\\
Upper bound on $\sup_{|t|\le0.8}|B(t)|$
 & $0.0736811$ & $0.0217930$\\
\bottomrule
\end{tabular}
\end{table}

The exact source moments of the first law are zero to numerical precision, yet
its bias at $t=0.8$ is $0.0493843$. The simultaneous interval excludes zero
there (Table~\ref{tab:operating_transfer_validation}). This shows why a valid
source-law sensitivity certificate alone cannot be transferred. The structured
law has zero bias throughout the continuum; its finite-sample result is the
nonzero upper bound $0.0217930$, not a statistical proof of exact calibration.

For an illustrative evaluation-bias tolerance $\tau=0.025$, the same confidence
set certifies every $t\in[-0.8,0.8]$ for the structured law. For the source-only
law it certifies the smaller interval $[-0.248124,0.137108]$. Maximizing the
positive frequency shift within these certified sets selects $t=0.8$ and
$t=0.137108$, respectively, after validation without further outcome queries.
The latter has known-model bias $0.008464$, below its certified bound $0.025$.
These choices intersect four affine inequalities from
$|\widehat m_0+t\widehat m_1|+r_0+|t|r_1\le\tau$; they are an explicit
one-dimensional instance of the linear-program certificate. The tolerance is
an illustration of the confidence set, not a new effect-testing endpoint.

\begin{figure}[t]
\centering
\includegraphics[width=0.90\linewidth]{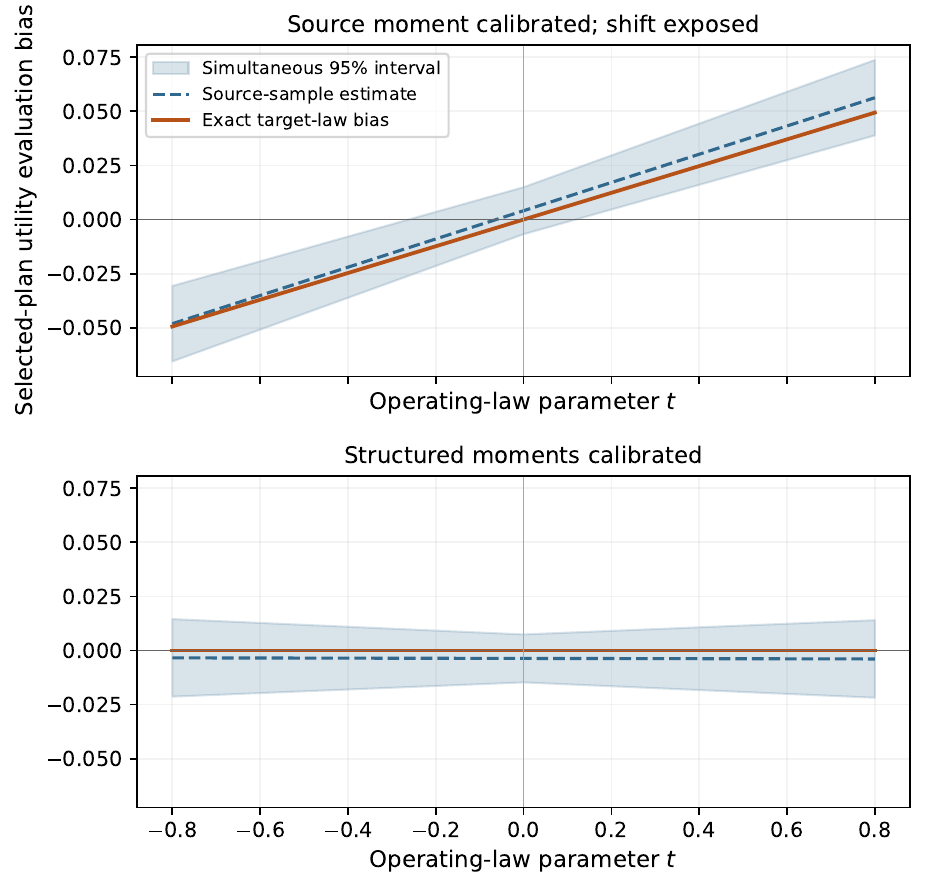}
\caption{One source validation sample evaluates an entire declared operating-law
family. Top: bin and direct-sensitivity calibration under the source law coexist
with nonzero target-law evaluation bias. Bottom: the nonzero residual orthogonal
to the structured class preserves evaluation throughout the family. Shading is
the simultaneous empirical-Bernstein interval, valid jointly across both panels
and all shift parameters. Exact curves integrate the known conditional mode
probabilities; no target outcomes are sampled.}
\label{fig:operating_transfer_bias}
\end{figure}

Figure~\ref{fig:operating_transfer_bias} shows the signed intervals, retaining
cancellation between the estimated moments rather than summing their absolute
values before combining them.

\subsection{An exact calibration-direction budget}
\begin{figure}[t]
\centering
\includegraphics[width=\linewidth]{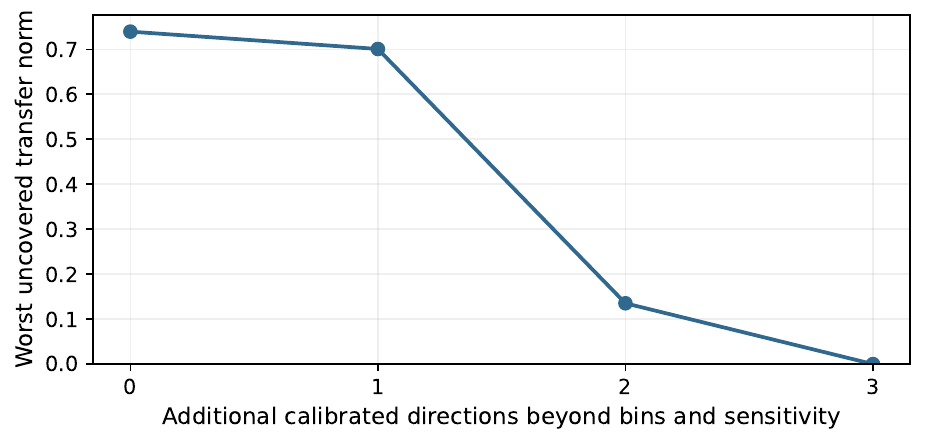}
\caption{Exact worst-shift exposure versus the number of added calibrated
directions. Each direction is fixed using the source context law and controller,
without outcome observations. The curve is the singular-value budget formula
of Theorem~\ref{thm:spectral_transfer}, independently checked by projection of
the three product fields after every augmentation. It is not a claim that these
directions minimize statistical confidence widths.}
\label{fig:operating_transfer_budget}
\end{figure}

We next use the three centered, maximum-absolute-value-normalized covariates
$\varphi=(\varphi_z,\varphi_v,\varphi_p)$ and
$w_t=1+t^\top\varphi$, with
$\|t\|_2\le R=0.8/\sqrt3$. This ellipsoid also guarantees
$w_t\ge0.2$. Starting from $\calS_0$, the exposure matrix of
Theorem~\ref{thm:spectral_transfer} has eigenvalues, in decreasing order,
\[
 2.56264130,\qquad 2.30185700,\qquad 0.08565670.
\]
Its rank is three: three additional calibrated directions are necessary and
sufficient to cover this shift family relative to the six-dimensional source
span. The leading left singular directions give the budget-optimal uncovered
norms in Table~\ref{tab:operating_transfer_budget} and
Figure~\ref{fig:operating_transfer_budget}. Direct projection after adding each
direction agrees with the spectral formula; after all three, the numerical
residual norm is $1.6\times10^{-15}$.

\begin{table}[t]
\centering\small
\caption{Calibration-test design for the three-feature shift ellipsoid. The
last column is the exact spectral worst uncovered $L_2(\mu)$ norm, not an
estimated cost loss or a finite-sample confidence radius.}
\label{tab:operating_transfer_budget}
\begin{tabular}{rrr}
\toprule
Added directions & Total test rank & Worst uncovered transfer norm\\
\midrule
0 & 6 & $0.7393895$\\
1 & 7 & $0.7007588$\\
2 & 8 & $0.1351792$\\
3 & 9 & $0$\\
\bottomrule
\end{tabular}
\end{table}

The regional affine library has 16 columns and represents $g$ to numerical
precision, but the $L_2$ error for $\varphi_vg$ is $0.9247583$. The regional
quadratic library covers all three products to numerical precision. This
degree increase gives the larger library a transfer function that its
fixed-law representation check did not supply. It does not make the 40-column
library the smallest certificate: the three optimal added directions yield
total rank nine. Budget optimality here concerns uncovered $L_2$ geometry;
finite-sample variance, ranges, and calibration effort can favor a different
choice of tests.

\subsection{Exploratory transfer profiles for independently trained forecasts}
The same calculation can be made without a sensitivity-planted residual.
Using the frozen populations and plans from
Section~\ref{sec:learned_forecast_experiment}, define $g$ separately for each
pipeline's selected plan and retain the same velocity density-ratio family.
Table~\ref{tab:learned_operating_transfer} reports exact evaluation biases.
The base forecast's bias changes from $-0.0303145$ to $-0.2163734$ as the law
shifts. Both corrections have smaller biases in this example, but their exact
ordering changes with the operating law: the sensitivity correction has the
smaller bias at $t=-0.8$, while the generic context correction has the smaller
bias at $t=0.8$. A fixed-law comparison therefore does not establish a uniform
ordering across this family.

\begin{table}[t]
\centering\small
\caption{Operating-law profiles of the separate learned-forecast study's frozen
pipelines. The first three columns are exact population utility-evaluation biases,
not sampled estimates. The final column evaluates the uniform empirical-Bernstein formula on the existing 2,000 confirmation episodes. This later analysis is exploratory: independence of its selection from previously inspected validation data was not recorded, so the numerical column is not asserted to have unconditional 95\% coverage. The bounds are conservative and do not establish the ordering of the two corrected pipelines. Section~\ref{sec:fresh_shift_validation} supplies a new validation experiment with a prospectively frozen analysis.}
\label{tab:learned_operating_transfer}
\begin{tabular}{lrrrr}
\toprule
Pipeline & $B(-0.8)$ & $B(0)$ & $B(0.8)$ & Uniform upper bound\\
\midrule
Base & $-0.0303145$ & $-0.1233439$ & $-0.2163734$ & $0.4510366$\\
Sensitivity & $0.0052374$ & $0.0090592$ & $0.0128810$ & $0.1773702$\\
Generic context & $0.0140356$ & $0.0103762$ & $0.0067168$ & $0.2034018$\\
\bottomrule
\end{tabular}
\end{table}

This exploratory analysis reuses frozen forecasts, plans, and the existing
independent confirmation data: there is no retraining, new outcome draw, or
retuning for a favorable target law. The calculation uses the known ranges and the same six-moment empirical-Bernstein formula. Its simultaneous guarantee would require the analysis choices to be independent of the validation sample; the available record does not establish that condition for this reuse. At $t=0.8$, the base
pipeline's interval is $[-0.4510366,-0.0425468]$ and numerically excludes zero; this exploratory calculation is not a confirmatory detection claim. The corrected
pipelines' intervals include zero and remain too wide to certify their exact
ordering. These are statements about evaluation under frozen conditional laws,
not guarantees for new feedback-induced state distributions.

\FloatBarrier

\section{A Probability-Feasible Audit of the Test Budget}
\label{sec:feasible_bias_experiment}
The geometric budget curve in Table~\ref{tab:operating_transfer_budget} does not by itself
quantify the worst evaluation error attainable by a conditional probability.
We therefore audit the same frozen 240-context controller and the same
three-feature operating family directly over $q_j\in[0,1]$. The reference
context law is uniform, the radius is $R_{\rm sh}=0.8/\sqrt3$, and all density
ratios remain between $0.2$ and $1.8$. The initial class consists of the five
forecast-group indicators and $g$, with rank six. For each augmented span
$\mathcal S$, the audited quantity is
\begin{equation}
 W_{\mathcal S}
 =\max_{\substack{\|t\|_2\le0.8/\sqrt3,\;0\le q_j\le1\\
                 q-V\perp\mathcal S}}
    \left|\frac1{240}\sum_{j=1}^{240}
      (1+t^\top\varphi_j)g_j(q_j-V_j)\right|.
 \label{eq:experiment_feasible_bias}
\end{equation}
This is an outcome-free audit of the entire calibration-consistent ambiguity
set, not an estimate of the error of the particular conditional law used in
Section~\ref{sec:operating_transfer_experiment}. The comparison assumes the
additional population moments are calibrated; appending a test to a library
does not establish that premise.

\paragraph{A continuous-family enclosure, rather than a direction scan.}
For a fixed shift direction we solve the binary conditional-mean linear
program of Theorem~\ref{thm:feasible_binary_dual}. A returned equality multiplier supplies a box-support upper bound
whether or not it is an optimal multiplier. For a lower bound, we contract the
primal residual toward zero by a factor $1-10^{-6}$ and bound an exact equality
repair; its worst possible objective loss is subtracted. The saved diagnostic
checks both the interior slack and the correction norm. Thus the solver's
objective alone is not used as a certificate.

We then enclose the \emph{continuous} outer maximization using
Theorem~\ref{thm:feasible_direction_enclosure}. The residual product
span has dimension three, two, one and zero after zero, one, two and three
independent additions. In dimension three we use 32 latitude intervals and
64 longitude intervals, including both poles, for 1,986 directions. Every unit
vector has angular distance at most
$\delta=\pi/(2\cdot32)+\pi/64=\pi/32$ from a grid direction: travel first along
a meridian and then along a latitude, whose arc length is no greater than the
longitude difference. Consequently the directions' convex hull contains the
ball of radius $\cos(\pi/32)$. The largest directional upper bound divided by
this radius bounds the continuous maximum. Dimension two uses a regular
128-gon with radius $\cos(\pi/128)$; dimension one uses the two endpoints.
A separate residual budget covers the numerical projection and rank
reduction. These are deterministic geometric enclosures, not confidence
intervals and not an assertion that the grid maximum is the continuum maximum.

The implementation uses long-double evaluations of box support, a checked
inverse-residual bound for the equality repair, and conservative $10^{-10}$
arithmetic guards. It is numerical verification rather than a formal
interval-arithmetic proof. Across the 2,506 directional LPs in the ten designs,
the largest inner upper--lower gap is $1.86\times10^{-7}$, mainly from the
deliberate contraction; the largest contracted equality residual is below
$3.5\times10^{-14}$. The three-dimensional outer enclosure contributes the
visible width in the first row of Table~\ref{tab:feasible_budget}.

\begin{table}[htbp]
\centering
\small
\begin{tabular}{clrr}
\toprule
Extra moments & Added directions & Uncovered $L_2$ norm & Feasible-bias enclosure\\
\midrule
0 & None & 0.7393895 & $[0.185280,\;0.186178]$\\
1 & First spectral & 0.7007588 & $[0.176456,\;0.176510]$\\
2 & First two spectral & 0.1351792 & $[0.035493,\;0.035494]$\\
3 & All three spectral & $<10^{-12}$ & $[0,\;2\times10^{-10}]$\\
\midrule
1 & Position product & 0.7330117 & $[0.181774,\;0.181830]$\\
1 & Velocity product & 0.7072554 & $[0.179477,\;0.179532]$\\
1 & Forecast product & 0.7014294 & $[0.177523,\;0.177578]$\\
2 & Position, velocity products & 0.1436578 & $[0.037794,\;0.037795]$\\
2 & Position, forecast products & 0.3618577 & $[0.093142,\;0.093143]$\\
2 & Velocity, forecast products & 0.6327772 & $[0.173696,\;0.173697]$\\
\bottomrule
\end{tabular}
\caption{Worst absolute evaluation bias over feasible binary conditional
means and the full declared shift ellipsoid. Endpoints are rounded outward.
The raw alternatives add the indicated fields $\varphi_\ell g$ to the same
rank-six source span. With three additions the exact product span is covered;
the small numerical upper bound records the arithmetic/projection budget.
Spectral optimality concerns the $L_2$ column. No global optimality over all
feasible-bias test designs is inferred from this comparison.}
\label{tab:feasible_budget}
\end{table}

\paragraph{What the additional moments buy.}
Two spectral directions reduce the worst feasible absolute bias by between
$80.84\%$ and $80.94\%$ relative to the initial class, using the unrounded
enclosures. Their upper bound is below the lower bound for every pair of raw
product fields tested here. One spectral direction gives only a modest
reduction, consistent with the two large exposure eigenvalues. This comparison
puts the budget in utility units and shows a quantitative advantage of the
spectral ordering over these specified alternatives in this controller. The
two criteria rank all ten audited designs in the same order here: the
feasible audit adds utility-scale meaning and an admissibility check, rather
than changing this example's ranking. It
does not establish that spectral directions minimize feasible bias over all
possible test choices, or that a forecaster trained to satisfy them performs
better. The exact span, law and coefficients remain known inputs.

\begin{figure}[t]
\centering
\includegraphics[width=0.78\linewidth]{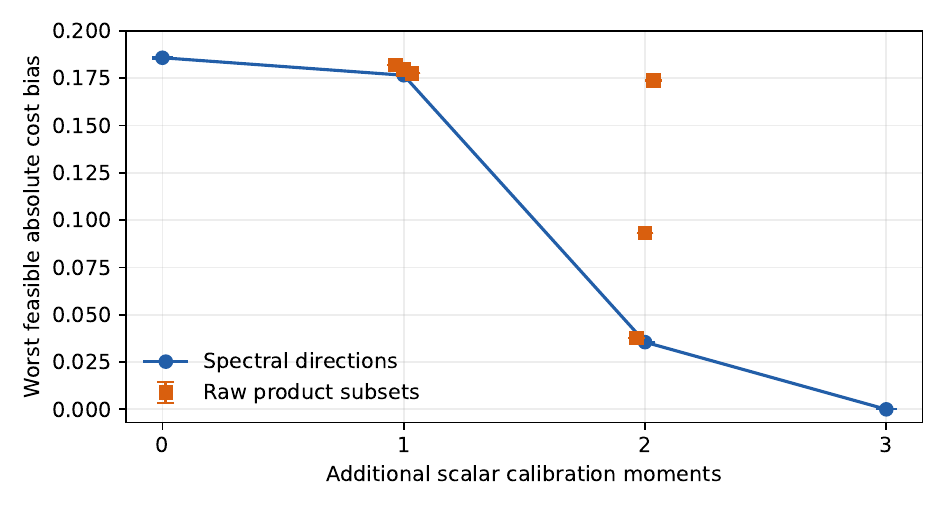}
\caption{Probability-feasible audit of the calibration-test budget. Bars give
the deterministic lower--upper enclosures; most are smaller than the markers.
The orange alternatives are all one- and two-field subsets of the three raw
products. The plotted quantity is an ambiguity-set worst case, not observed
forecast error or a validation confidence width.}
\label{fig:feasible_budget}
\end{figure}

\paragraph{A check that geometric optimality is not being overgeneralized.}
The separate spike-versus-balanced construction of
Example~\ref{ex:spiky_feasible_design} uses $V=1/2$, a constant
sensitivity, and a constant source test. For $n=101$ equiprobable contexts,
adding the spike direction has the smaller uncovered $L_2$ norm, $0.08292$
versus $0.10000$, but its exact probability-feasible worst bias is $0.041254$
versus $0.005000$ when the balanced direction is added. At $n=401$ the feasible
values are $0.020781$ versus $0.001250$. Analytic witnesses and upper bounds
establish these values, and the LP enclosures contain them. Thus the favorable
ordering in Table~\ref{tab:feasible_budget} is specific to the stated controller
and comparisons; the feasible-adversary audit is needed to substantiate it.

\paragraph{Reproduction and scope.}
\texttt{experiments/exact\_interface/feasible\_bias.py} generates the audit and
figure without drawing outcomes or changing earlier results. Its
\texttt{--check} mode reconstructs every design and checks the saved enclosures.
The protocol records the input hashes and numerical geometry, and the saved
summary records all direction-wise lower and upper bounds, feasibility-repair
checks, projection budgets and analytic counterexample values. There is no
validation-sample selection or detection-power claim in this deterministic
study.

\section{An Independently Trained Forecaster with Physical-Covariate Error}
\label{sec:learned_forecast_experiment}

The preceding geometric study constructs an error from the controller's
sensitivity. We now use a probability law specified directly from physical
covariates and an exogenous signal, and train the base forecaster from independent
outcomes. Both ten-dimensional recalibrators improve expected cost, but the
generic contextual recalibrator performs better than the sensitivity-based
one in all 200 calibration refits. Thus the study tests relevance beyond the
planted direction while retaining a strong comparator; it does not establish
superiority of sensitivity-based correction.

\paragraph{Prespecified probability law and trained base model.}
We retain the physical dynamics, six-step committed plans, costs, input bounds
and uniform 240-context product law of Section~\ref{sec:experiments}.
The five original forecast levels are now exogenous signal levels
$c_b\in\{0.15,0.325,0.5,0.675,0.85\}$; they are not supplied as probability
predictions. Define
\[
 t_{\rm sig}=\frac{c_b-0.5}{0.35},\qquad
 \widetilde z=\frac{z_0-0.725}{0.175},\qquad
 \widetilde v=\frac{v_0-0.3}{0.5}.
\]
The mode is drawn once per episode with danger-mode probability
\begin{equation}
 q(h)=\operatorname{logit}^{-1}
 \!\left(-0.15+1.2t_{\rm sig}+0.7\widetilde z
                    +0.9\widetilde v+0.5\widetilde z\widetilde v\right).
 \label{eq:learned_truth}
\end{equation}
No controller sensitivity enters~\eqref{eq:learned_truth}. The formula,
sample sizes, correction spaces, penalty and random-stream rule were saved
before any outcome or fit in this study; they were not changed after evaluation.
Using 20,000 independent training episodes, we fit the unpenalized logistic
maximum-likelihood model
\[
 p(h)=\operatorname{logit}^{-1}
       (\widehat\beta_0+\widehat\beta_1t_{\rm sig}).
\]
The base model deliberately omits the physical covariates. It is a trained,
misspecified forecaster on a synthetic law, rather than an application-data
forecaster. Its selected plan and sensitivity $s_0=J_1(a_p)-J_0(a_p)$ are
recomputed after fitting; no sensitivity from the original constructed
forecast experiment is reused.

\paragraph{Matched correction spaces and data.}
An independent set of 10,000 context/outcome pairs is shared by all
recalibrators. In addition to the base forecast and a global intercept
correction, we compare two rank-ten correction spaces. The sensitivity space
is generated by a constant, four signal-group indicators, and the five
signal-group indicators multiplied by $s_0$. The generic space contains all
monomials of total degree at most two in
$(t_{\rm sig},\widetilde z,\widetilde v)$ and uses no controller sensitivity.
Both spaces receive the same observed covariates; their representations differ. The generic span contains the logit in~\eqref{eq:learned_truth}, so its features are well aligned with this synthetic law. The fitted correction below is additive probability-residual regression, however, not logistic regression in that span; exact logit representation does not make that probability correction correctly specified.

To match regularization in function space, we center and SVD-whiten each
space under the same known uniform context law. Each resulting basis
$\phi=(1,\phi_1,\ldots,\phi_9)$ is orthonormal in $L_2(\mu)$, with
$\E_\mu\phi_j=0$ for $j\ge1$. Normalization uses contexts and the frozen
controller, without calibration or confirmation outcomes. For either space
we fit
\begin{equation}
 \widehat\gamma
 =\argmin_\gamma\left\{
 \frac1{10000}\sum_{i=1}^{10000}
       \bigl(Y_i-p(H_i)-\phi(H_i)^\top\gamma\bigr)^2
       +0.01\sum_{j=1}^9\gamma_j^2\right\}.
 \label{eq:learned_matched_fit}
\end{equation}
Thus the nonconstant penalty is the same population $L_2(\mu)$ penalty for
both rank-ten spaces; the constant is unpenalized. Corrected probabilities
are clipped to $[0.02,0.98]$ before replanning. The global-intercept comparator
uses the same calibration residuals and no penalty. No penalty or feature
selection uses the confirmation data. Selected-plan sensitivities are
recomputed after each correction; the nominal fitting equations alone do not
certify the corrected plan.

\paragraph{Frozen-fit population results.}
Table~\ref{tab:learned_frozen} integrates both modes over the complete known
context law, so its entries have no Monte Carlo confidence intervals. The
available base-to-oracle cost gap is $0.101076$, or $3.347\%$ of base cost.
The sensitivity recalibrator reduces cost by $2.590\%$ and closes $77.4\%$
of this gap. The equally sized generic contextual recalibrator reduces cost
by $3.239\%$ and closes $96.8\%$. The generic method is the better cost
comparator, despite using no controller-specific feature.

\begin{table}[t]
\centering\small
\caption{Independently trained forecaster study: exact population values for
the frozen training and calibration fits. Cost bias is actual minus
anticipated selected-plan cost,
$B_C=\E_\mu[(J_1(a)-J_0(a))(q-p^{\rm used})]$;
the corresponding utility bias is $-B_C$. The oracle minimizes expected
cost, not the separate threshold-risk endpoint.}
\label{tab:learned_frozen}
\begin{tabular}{lrrrr}
\toprule
Method & Cost & Gap closed (\%) & Full-plan risk (\%) & $B_C$\\
\midrule
Trained base & 3.019914 & 0.0 & 6.210 & 0.123344\\
Global intercept & 3.019670 & 0.2 & 6.485 & 0.126232\\
Sensitivity, rank 10 & 2.941704 & 77.4 & 5.356 & $-0.009059$\\
Generic context, rank 10 & 2.922113 & 96.8 & 6.051 & $-0.010376$\\
Oracle & 2.918838 & 100.0 & 5.666 & 0\\
\bottomrule
\end{tabular}
\end{table}

The trained base has exact population ten-bin ECE $0.00537$ while its
selected-plan cost bias is $0.12334$, approximately $4.08\%$ of base cost.
Both recalibrators reduce the absolute value of this bias, although their
ten-bin ECE values increase to $0.01354$ and $0.02009$, respectively. These
are exact model diagnostics under this law, not evidence that ECE generally
worsens when decisions improve. They show that the separation between
forecast-bin calibration and decision-value evaluation can also arise from
an omitted physical covariate rather than a sensitivity-planted error.

\paragraph{Independent confirmation.}
The frozen fits are evaluated on 2,000 new common episodes.
The conditional expected-cost differences from the base are $-0.07543$
for sensitivity and $-0.09545$ for generic context; their paired-bootstrap
95\% intervals are $[-0.08075,-0.06996]$ and $[-0.10056,-0.09014]$.
The realized-cost differences are $-0.07026$ and $-0.08797$, with intervals
$[-0.08946,-0.04873]$ and $[-0.11038,-0.06323]$.
These pointwise intervals use 2,000 common multinomial-bootstrap draws and
condition on the frozen base and correction fits. They are secondary
descriptions of confirmation sampling, not simultaneous guarantees or
intervals for training/calibration-fit variability. The known-law results
in Table~\ref{tab:learned_frozen} remain separate from them.

The directly observed selected-plan cost-bias moment for the base has mean
$0.14371$ and pointwise bootstrap interval $[0.09907,0.18527]$ on these same
confirmation episodes. For sensitivity and generic context, the means are
$0.01490$ and $0.01194$, with intervals $[-0.01903,0.04765]$ and
$[-0.02281,0.04613]$. Thus the base evaluation bias is detected by this pointwise bootstrap diagnostic at the stated sample size; inclusion of zero in the corrected
intervals does not establish exact calibration. These intervals have the
same frozen-fit and pointwise scope as the cost intervals above.

\paragraph{Calibration-refit variability.}
We keep the trained base fixed and generate 200 new independent sets of
10,000 calibration pairs. Every refit shares its observations across
methods and is evaluated exactly under the same law.
Table~\ref{tab:learned_refits} reports the resulting descriptive distributions.
The sensitivity and generic methods reduce cost in all 200 refits. The
generic method also beats sensitivity in all 200, with sensitivity-minus-generic
cost difference having mean $0.019550$, SD $0.000367$, and central 95\%
range $[0.018694,0.020202]$. The mean oracle-gap closures are $77.7\%$
and $97.0\%$. These sampled ranges are not confidence intervals and do
not assert uniform dominance across other laws, forecasters or controller
configurations.

\begin{table}[t]
\centering\small
\caption{Two hundred independent calibration refits conditional on one
trained base forecaster. Cost differences are relative to that base;
Counts give cost decreases and
increases, with ties defined by absolute tolerance $10^{-12}$. Central
95\% ranges describe the sampled refit distribution.}
\label{tab:learned_refits}
\begin{tabular}{lrrrr}
\toprule
Method & Mean cost diff. & Cost SD & Central 95\% range & Cost $-/+$\\
\midrule
Global intercept & $-0.000127$ & 0.000414 & $[-0.000496,0.001123]$ & 142/58\\
Sensitivity & $-0.078507$ & 0.000423 & $[-0.079166,-0.077695]$ & 200/0\\
Generic context & $-0.098057$ & 0.000428 & $[-0.098794,-0.097224]$ & 200/0\\
\bottomrule
\end{tabular}
\end{table}

The mean risk changes across refits are $-0.758$ percentage points for
sensitivity and $-0.186$ for generic context. The sensitivity risk decreases
in 200/200 fits; generic risk decreases in 192 and increases in 8. In this
law, the better expected-cost method does not have the lower threshold risk.
This is a model-dependent observation, not a safety certificate.

\paragraph{Scope and reproducibility.}
The study resolves two limitations of the planted example: the base
forecast is learned from separate outcomes, and the same-dimensional
generic comparator is retained even though it wins. It establishes that
a substantial decision-value error can arise from a standard omitted-covariate
model and can be detected and reduced without constructing the error from
the controller. It does not show that the proposed recalibrator is generally
preferable to contextual regression, that the known-law normalization is
available in applications, or that the result transfers to feedback control.
The 200 refits condition on one trained base; training-fit variability is
not assessed. The committed-plan design and synthetic finite support remain
limitations.

The protocol, all training/calibration/confirmation observations, frozen
population plans, 200 refit records, and checks are included in the
reproducibility archive. Independent physical rollouts of all 1,200 saved
method/context plans agree with scenario costs to $5.69\times10^{-14}$,
with no risk or bound discrepancy. The base likelihood gradient is below
$6.68\times10^{-15}$; both correction spaces have rank ten and population
Gram errors below $1.33\times10^{-14}$. The original constructed study's
source and saved observations are unchanged.

\section{Prospective Validation of Operating Choices for Learned Forecasts}
\label{sec:fresh_shift_validation}
The exploratory reuse of the learned study's confirmation sample in
Section~\ref{sec:operating_transfer_experiment} does not establish that the
subsequent choice of shift analysis was independent of those observations.
We therefore perform a separate prospective validation of the same frozen
base, sensitivity-corrected, and generic-context pipelines. This study asks
whether the operating-family certificate can authorize a specified shift
for learned forecast errors that were not constructed from a sensitivity
field. It also compares the rectangular certificate with the
probability-feasible envelope of
Theorem~\ref{thm:feasible_validation_envelope}.

\paragraph{Protocol fixed before the new outcomes.}
Before drawing this sample, we save the exact hashes of the trained forecasts,
selected plans, conditional probabilities, and analysis script, together with
the sample size, six tested moments, shift family, confidence allocation,
tolerances, selection objective, and random stream. The protocol is a local
prospective record, not an external preregistration. It specifies
$M=100{,}000$ independent source episodes, with contexts uniform over the same
240 points and one Bernoulli draw from the unchanged conditional kernel per
episode. The three pipelines share these observations. There is no retraining,
replanning, field selection, seed replacement, or outcome-dependent change
of sample size. All new observations are retained separately from the earlier
studies.

\paragraph{Design rationale and prior information.}
The study is a controlled synthetic illustration: the frozen conditional law
and the earlier exact-bias profiles were available when the protocol was
chosen. The budget $M=100{,}000$ and primary tolerance $0.04$ are illustrative
design choices, not an externally supplied operating requirement. The saved
record fixes them before the new draw, but does not document a formal
power-based rule or an optimization that selected them. We therefore do not
interpret success at this tolerance as an unexpected or blinded empirical
discovery. The retrospective population calculations below explain why this
budget and tolerance make the primary result highly predictable, while the
fresh observations still supply a valid sample-based certificate. No
retrospective diagnostic changes the original sample, protocol or result.

For each frozen pipeline $m$, use utility sensitivity $g_m=J_{0,m}-J_{1,m}$,
its own forecast $V_m$, and the fixed velocity feature
$\varphi=(\mathrm{velocity}-0.3)/0.5$. The operating family is
$w_t=1+t\varphi$ for $|t|\le0.8$. We test $g_m$ and $\varphi g_m$ for each
pipeline. The empirical-Bernstein radii use $\log(4\cdot6/0.05)$, unbiased
sample variances, and ranges obtained by enumerating all contexts and both
possible outcomes. Thus one event of probability at least $0.95$ covers all
three pipelines, every shift in the interval, and every subsequently queried
tolerance. The primary tolerance is fixed at $\tau=0.04$ before sampling;
$0.01$ and $0.02$ are prespecified secondary tolerances. For each pipeline
and tolerance, the operating objective is to maximize $t$ in the certified
set, with an empty set reported as such. This is a specified illustrative
operating preference, not an inferred improvement in control cost.

\begin{table}[htbp]
\centering\small
\caption{Fresh independent validation of frozen learned pipelines.
The three bias columns are exact known-law utility biases, not estimates.
The uniform bound is from the new six-moment joint 95\% event and applies
simultaneously to all $|t|\le0.8$ and all three rows. The final column gives
the random certified set at the prespecified primary tolerance $0.04$.}
\label{tab:fresh_shift_validation}
\begin{tabular}{lrrrrl}
\toprule
Pipeline & $B(-0.8)$ & $B(0)$ & $B(0.8)$ & Uniform bound & Certified set\\
\midrule
Base & $-0.030314$ & $-0.123344$ & $-0.216373$ & $0.227915$ & $\varnothing$\\
Sensitivity & $0.005237$ & $0.009059$ & $0.012881$ & $0.031881$ & $[-0.8,0.8]$\\
Generic context & $0.014036$ & $0.010376$ & $0.006717$ & $0.033817$ & $[-0.8,0.8]$\\
\bottomrule
\end{tabular}
\end{table}

\paragraph{The certificate in the realized sample.}
Table~\ref{tab:fresh_shift_validation} shows that the fresh sample certifies
the whole family for both corrected pipelines at tolerance $0.04$. The
selected operating choice is therefore $t=0.8$ for each. The base pipeline
has an empty certified set, even though its exact bias at $t=-0.8$ is below
the tolerance: its finite-sample bound there is $0.043470$. Failure to certify
is not proof of failure of the population condition. Conversely, no pipeline
has a nonempty certified set at either stricter tolerance $0.01$ or $0.02$.
These unfavorable outcomes are retained; the tolerance was not adjusted
after inspecting the new observations.

At the selected shift $t=0.8$, the sensitivity and generic certificates are
$0.031881$ and $0.025042$, respectively. Their exact biases are $0.012881$
and $0.006717$. The base pipeline's joint signed interval at this shift is
$[-0.227915,-0.190576]$, detecting a substantial evaluation error. The
certified corrections have small bias at the chosen operating law on the
same confidence event, even though the choice was made after validation.
Their exact expected costs at this common shift are $3.838257$ and $3.814669$,
against oracle cost $3.811880$; their exact full-path violation risks are
$0.091289$ and $0.098030$. These known-law cost and risk values describe the
example and are not consequences of the bias certificate. In particular,
the generic correction has lower cost while the sensitivity correction has
lower threshold risk.

\paragraph{Checking the probability-feasible envelope.}
The same six moment intervals can be imposed on one common conditional mean
vector $q\in[0,1]^{240}$: row $i$ uses its pipeline's forecast offset, while
the underlying physical $q$ is shared. We first check that this set is
nonempty, then solve both signed support problems and the operating-choice
LP for each pipeline. On the joint confidence event, the true mean lies in
this common set, so all selected pipeline--shift pairs remain covered.
We evaluate numerical upper bounds using nonnegative dual multipliers and
the box-support correction for any remaining dual residual, with a
$10^{-10}$ utility allowance; selection reserves $10^{-9}$ from the
tolerance. The largest observed primal constraint residual is
$2.5\times10^{-16}$.

Here the feasible envelope equals the rectangular bounds to the displayed
precision and gives exactly the same operating choices. The probability
constraints and cross-pipeline consistency do not tighten these particular
confidence intervals. This null result distinguishes a valid improvement
in the optimization formulation from an asserted empirical improvement:
weak dominance does not imply strict improvement. The mean domain can
matter greatly for the worst-case design problem in the preceding study,
while adding it to this sample's confidence set produces no visible gain.

A retrospective geometric check explains this particular null result.
After including the forecast offsets, all $2^6=64$ vertices of the saved
six-moment rectangle admit a numerical witness $q\in[0,1]^{240}$, with
maximum equality residual $3.4\times10^{-16}$ and no observed box violation.
The image of the probability cube under the raw moment matrix is convex.
If it contains every rectangle vertex, it contains the whole rectangle, so
intersecting it with the mean domain cannot improve any linear objective
formed from these six moments. The vertex calculations are floating-point
feasibility checks, not interval-arithmetic proofs. They explain the
observed box, rather than establish no tightening for every validation
sample or every informative sample size. The retained $10^{-10}$ allowance
still applies to numerical support bounds.

\paragraph{Retrospective sampling distribution.}
To separate a realized certificate from the typical behavior of this
synthetic design, we replay a separately declared Monte Carlo diagnostic
with 4,000 independent replications at each specified sample size. This
analysis was introduced after the original results were seen; it is
retrospective, not another prospective confirmation experiment. A
multinomial draw over the 480 context--outcome atoms has the same
distribution of empirical moment means and variances as independent
episodes from the known law. Each
$M=100{,}000$ replication supplies one common sample for all three pipelines,
with the original six-moment confidence allocation. Continuous certificate
minima are evaluated at the endpoints and the knots of
$|\widehat m_0+t\widehat m_1|+\epsilon_0+|t|\epsilon_1$; maxima occur at
an endpoint. The diagnostic therefore does not replace the shift interval
by a plotting grid.

Inserting the exact population moments and variances into the radius
formula gives minimum/maximum certificates $0.018108/0.026927$ for the
sensitivity correction and $0.019522/0.028748$ for the generic correction;
the base minimum is $0.049126$. These are population \emph{plug-in} values,
not expectations of the random certificates: absolute values, square roots
and optimization do not commute with expectation. They show that $0.04$
lies well between the two corrected pipelines and the base in this design.

\begin{table}[htbp]
\centering\small
\caption{Retrospective sampling behavior under the known learned-forecast
law. Each entry is the ordered pair of counts (nonempty certified set, whole
interval certified), out of 4,000 common replications. The original realized validation sample is not replaced.}
\label{tab:sampling_diagnostic_sets}
\begin{tabular}{lrrr}
\toprule
Pipeline & $\tau=0.01$ & $\tau=0.02$ & $\tau=0.04$\\
\midrule
Base & $({0},{0})$ & $({0},{0})$ & $({0},{0})$\\
Sensitivity & $({0},{0})$ & $(3244,13)$ & $(4000,4000)$\\
Generic context & $({0},{0})$ & $(2376,1)$ & $(4000,4000)$\\
\bottomrule
\end{tabular}
\end{table}

Thus both corrected pipelines pass the whole interval at $0.04$ in every
replication, and the base never has a certified shift there. These are high
observed frequencies, not deterministic statements about all possible
streams: for a reported event with zero exceptions in 4,000 trials, the
one-sided 95\% upper exception probability is about $0.00075$. At tolerance $0.02$, nonempty-set
frequencies are $81.10\%$ and $59.40\%$, with Monte Carlo standard errors
$0.62$ and $0.78$ percentage points. The much smaller whole-interval counts
in Table~\ref{tab:sampling_diagnostic_sets} show that intermediate certified
sets can arise in repeated samples. The original sample's empty sets at
$0.02$ are retained; its first moments are approximately $2.09$, $2.27$ and
$1.85$ population standard errors above their true values for the base,
sensitivity and generic pipelines. These correlated deviations explain why
that draw was unfavorable for the two corrections at the stricter
tolerance. Joint coverage of all six moments occurs in $3997/4000$
replications ($99.925\%$, Monte Carlo standard error $0.043$ percentage
points), consistent with a conservative nominal 95\% guarantee. This
empirical coverage frequency is not substituted for that guarantee.

\paragraph{Additional interpretation of the original confirmation samples.}
For the learned base pipeline's original 2,000 episodes, a separate
one-moment empirical-Bernstein calculation gives cost-bias interval
$[0.048135,0.239277]$. It excludes zero without relying on the pointwise
bootstrap approximation. This separately valid 95\% interval is not part
of the new six-moment event, and the two constructions are not combined by
selecting their smaller radius.

For the original constructed study in Section~\ref{sec:experiments},
holding its observed variance $0.942730$ fixed, the empirical-Bernstein
radius first falls below the exact bias magnitude $0.027555$ at
$M=15{,}522$. The analogous full-range Hoeffding and normal-approximation
crossings are $112{,}500$ and $4{,}770$. These are radius-versus-signal
planning calculations, not detection-power guarantees, externally validated
sample-size recommendations, or stopping rules. A future sample has a
random center and variance, so radius crossing alone does not guarantee
that its interval will exclude zero.

A separate retrospective power calculation makes the distinction
quantitative. After the learned-pipeline diagnostic, the same declared
simulation stream supplies 4,000 independent multinomial replications of
the original constructed law at each sample size in
Table~\ref{tab:sampling_diagnostic_power}. The success event is that the
original one-moment 95\% empirical-Bernstein interval excludes zero.
The roughly 50\% detection frequency at the radius crossing is therefore
consistent with, rather than a contradiction of, the planning calculation.
These model-specific frequencies are Monte Carlo estimates; they do not
establish uniform power guarantees or a rule for sequential stopping.

\begin{table}[htbp]
\centering\small
\caption{Retrospective detection frequency for the original constructed
study. Each row uses 4,000 independent replications. Standard errors quantify
Monte Carlo uncertainty in percentage points.}
\label{tab:sampling_diagnostic_power}
\begin{tabular}{rrrr}
\toprule
Episodes & Detections & Frequency & Monte Carlo SE\\
\midrule
$15{,}522$ & $1998$ & $49.95\%$ & $0.79$\\
$22{,}700$ & $3214$ & $80.35\%$ & $0.63$\\
$30{,}000$ & $3793$ & $94.83\%$ & $0.35$\\
\bottomrule
\end{tabular}
\end{table}

\paragraph{Scope and reproducibility.}
The fresh sample resolves the independence issue for this new analysis:
its entire analysis was fixed before its own validation data were drawn.
It demonstrates a valid, computable uniform operating-family certificate
for the frozen learned pipelines in the specified synthetic law, including
a validation-selected shift. In this realized sample the successful sets
are entire intervals, so selecting $t=0.8$ does not itself demonstrate a
nontrivial improvement in operating performance. The
$100{,}000$-episode budget is substantial, the contexts and normalization law
are still synthetic and known, and the conditional outcome kernel and
committed plans remain fixed. Practical informativeness at realistic sample budgets, with estimated
context frequencies or density ratios, has not been established. This is
not evidence of robustness to arbitrary feedback-induced shifts or
uncertain density ratios. The protocol, all fresh
episodes, summaries and a saved-data verifier are retained in the companion
artifact; the verifier checks input hashes and recomputes both certificate
constructions without redrawing outcomes. The separate retrospective
sampling diagnostic has its own script, hashed inputs, protocol and saved
summary. Its verifier replays the declared diagnostic stream without
writing files or replacing the original observations. The simulation
entropy is $2026111501$, with PCG64 and no spawned child stream: the draws
are, in order, 4,000 learned-law replications at $100{,}000$ episodes and
4,000 original-law replications at each of $15{,}522$, $22{,}700$ and
$30{,}000$ episodes. The full sampling diagnostics retain success counts
and exact binomial intervals, including for frequencies zero and one.

\FloatBarrier

\section{Discussion and Limitations}
\label{sec:discussion}

\paragraph{What a forecast-interface certificate resolves.}
The framework separates the moment information needed to evaluate a chosen plan from the information needed to select an oracle plan. Sensitivity calibration removes a specified expected evaluation bias. Oracle regret also depends on how the optimizer and optimal value respond to forecast error. The replanning and regret results state the regularity needed to relate these quantities. The operating-law scope of these statements is fixed in Section~\ref{sec:setup}.

\paragraph{What the larger test class buys.}
A direct test of $g_{\rm nom}$ is sufficient for one fixed-law moment and can have a tighter elementary confidence bound than an exact multi-test decomposition. Larger classes acquire a role when calibration evidence is reused over specified target contexts. The products $w g_{\rm nom}$ identify the required information; the exposure matrix and singular directions quantify missing coverage and allocate a test budget. The regional monomial library is a constructive coverage class, not necessarily the smallest representation or the most statistically efficient choice. Adding a direction does not establish its calibration; outcomes must still validate or train the corresponding moments.

\paragraph{Interpretation of the numerical studies.}
The fixed-law exact-interface study and the transfer separation deliberately construct residual geometry. They demonstrate logical and quantitative consequences, not the prevalence of those residuals in applications. The operating-shift validation uses 100,000 episodes per law, and the original 2,000-episode sample remains insufficient to detect its planted bias with the reported finite-sample bounds. Variance-sensitive confidence improves precision without removing the cost of certification.

The independently learned forecast instead omits physical covariates from a prespecified probability model. Its cost-evaluation error is materially larger than in the original construction and is detected by the stated pointwise bootstrap diagnostic. Both rank-ten corrections improve cost, but generic context regression is consistently better than the sensitivity representation. The sensitivity representation therefore improves on the misspecified base in this law, but the experiment does not isolate an incremental benefit attributable specifically to controller knowledge. The generic features are well aligned with the synthetic truth, and the results establish no comparative advantage for sensitivity-based recalibration. Exact-law normalization, the finite synthetic support, and conditioning the 200 calibration refits on one trained base remain limitations. The operating-law bias profiles of these frozen pipelines also change their ranking, reinforcing the need to specify the target law. The initial reuse of confirmation data is labeled exploratory; the separate prospectively fixed validation in Section~\ref{sec:fresh_shift_validation} supplies simultaneous guarantees for all declared methods and shifts. That experiment certifies both corrected pipelines at its primary tolerance, but its mean-domain intersection does not tighten the rectangular certificate. A valid optimization formulation is not by itself evidence of an empirical precision gain. The primary sample size and tolerance were specified with the synthetic law available, not supplied by an external operating requirement. The retrospective diagnostics quantify the highly predictable primary result and substantial sampling variation at a stricter tolerance. They do not establish inevitable outcomes, and population plug-in certificates are not expectations of the random bounds. Informativeness at realistic data budgets with estimated context frequencies or density ratios has not been established.

\paragraph{Limits of transfer and test design.}
The transfer theorem freezes the conditional outcome kernel and the selected plan at each context. Estimated density ratios, uncertain context Gram matrices, or changes in the conditional outcome law require additional uncertainty or model-shift terms. The spectral audit measures uncovered sensitivity geometry and the corresponding $L_2$-relaxed bias exposure. Only its stated small-radius condition guarantees attainment by a probability-feasible residual; the actual environment need not point in that adverse direction. Its test budget minimizes uncovered $L_2$ exposure, which need not minimize pointwise-feasible worst-case bias or statistical confidence width. The finite-context audit in Section~\ref{sec:feasible_bias} evaluates the probability-feasible criterion directly and encloses its continuous shift maximum using directional primal--dual bounds and a geometric covering factor. Even an exact selected-plan evaluation certificate does not bound oracle regret or threshold risk.

\paragraph{Robust computation.}
The robust-value duality is a mathematical characterization. Turning it into a deployed controller requires satisfying the primal calibration moments and selecting a saddle-compatible policy, as specified by the duality result. A collection of pointwise inner minimizers alone need not meet those global constraints. Efficient and reliable numerical solution of that constrained problem is a separate algorithmic question.

\paragraph{Extensions.}
The population interface assumes an exogenous future statistic. Action-dependent prediction requires tests for the corresponding action-conditional interface. Repeated controller changes require a justified transfer or renewed certification under the resulting law, as in performative prediction \citep{perdomo2020performative}. Combining these statistical conditions with recursive feasibility or stability requires control-specific assumptions. For strategic agents, the incentive constraints studied by \citet{bergemann2026mechanism} supply an additional requirement beyond statistical forecast reliability. A possible extension would calibrate incentive margins, the payoff differences between compliance and joint reporting/action deviations, using tests conditional on the relevant agent type and information set.

\section{Conclusion}

The selected-plan sensitivity expresses an established decision-calibration requirement in the coordinates of predictive control. For one fixed operating law it is one residual moment. Reusing that evidence under specified changes of context frequency requires coverage of density-weighted sensitivity fields. The new operating-family characterization, spectral audit and budgeted augmentation rule identify that additional information, while polynomial product classes make it constructive for a specified MPC subclass. A signed simultaneous confidence set turns the covered fields into a certificate for a continuum of operating choices from one source validation sample. Intersecting that set with probability-feasible conditional means yields a finite-context linear-program audit and operating-choice formulation. The directional enclosure controls approximation of a continuous shift family, so sampled directions are not mislabeled as an exact global maximum.

The transfer experiment demonstrates why a calibrated source moment can fail after reweighting and how extra directions remove the exposed component. The separately learned-forecast study shows that a substantial decision-evaluation error can arise from omitted physical covariates, without constructing the error from controller sensitivity. Its matched generic contextual recalibrator outperforms the sensitivity recalibrator, so the paper's contribution is the audit, test-design and certification capability rather than a universal recalibration advantage. The probability-feasible study establishes a quantitative reduction in admissible bias from added tests, while fresh fixed-protocol validation certifies operating choices for the learned pipelines at a prespecified illustrative tolerance. Repeated-sampling diagnostics explain the predictability and sampling variation of this known-law demonstration; they do not show practical sample efficiency. Neither result establishes optimal feasible-bias test design or a new general calibration principle. The original small-gain, adverse-risk, and null certificate-tightening findings are retained. Expected plan evaluation, oracle regret, threshold risk, and feedback behavior remain separate scientific requirements.

\section*{Acknowledgment of AI assistance}
Codex assisted with manuscript development, code preparation, and mathematical and numerical checks. An AI-assisted referee review using Claude Opus~5.5 suggested the $L_1$ comparison in Proposition~\ref{prop:feasible_l1_sandwich} and the spike-versus-balanced construction in Example~\ref{ex:spiky_feasible_design}. The manuscript supplies their analytic derivations and witnesses, together with the stated verification. These suggestions are acknowledged as part of the development process; they were not supplied by a journal referee. The author retains responsibility for the arguments, computations, interpretation, and attribution.

\appendix

\section{Proof of the Sensitivity-Span Characterization}
\label{app:sensitivity_span_proof}

We prove Theorem~\ref{thm:sensitivity_span}. Let
\begin{equation}
    \calS_\Psi=\overline{\Span}\{\psi_1,\ldots,\psi_k\}\subset L_2(\mu;\R^d).
\end{equation}
The calibration constraints on a perturbation $\xi$ are
\begin{equation}
    \langle \psi_i,\xi\rangle_\mu=0,
    \qquad i=1,\ldots,k.
\end{equation}
Thus the feasible first-order perturbation directions lie in $\calS_\Psi^\perp$.

If $g_{\rm nom}\in\calS_\Psi$, then for every $\xi\in\calS_\Psi^\perp$,
\begin{equation}
    \E[g_{\rm nom}(H,V)^\top\xi(H,V)]
    =
    \langle g_{\rm nom},\xi\rangle_\mu
    =0.
\end{equation}
This proves one direction.

Before the converse we record a density fact that repairs the passage from bounded feasible perturbations to the full orthocomplement.

\begin{lemma}[Bounded density in the orthocomplement]
\label{lem:bounded_density}
Under Assumption~\ref{assumption:interior_perturbations}, the bounded elements of $\calS_\Psi^\perp$ are dense in $\calS_\Psi^\perp$ in the $L_2(\mu)$ norm.
\end{lemma}
\begin{proof}
Let $f\in\calS_\Psi^\perp$ and for $M>0$ set $f_M:=f\,\indic\{\|f\|_2\le M\}$; then $f_M\to f$ in $L_2(\mu)$ by dominated convergence. The subspace $\calS_\Psi$ is finite-dimensional and spanned by the essentially bounded functions $\psi_1,\ldots,\psi_k$ (Assumption~\ref{assumption:interior_perturbations}). All norms on a finite-dimensional space are equivalent, so on $\calS_\Psi$ the $L_2(\mu)$ and $L_\infty(\mu)$ norms are comparable, and therefore $\Proj_{\calS_\Psi}f_M\in L_\infty(\mu)$ with $L_\infty$ norm controlled by $\|f_M\|_{L_2(\mu)}\le\|f\|_{L_2(\mu)}$. Hence
\begin{equation}
    g_M:=f_M-\Proj_{\calS_\Psi}f_M\in\calS_\Psi^\perp\cap L_\infty(\mu).
\end{equation}
Since $f\in\calS_\Psi^\perp$ we have $\Proj_{\calS_\Psi}f=0$, and continuity of the orthogonal projection gives $\Proj_{\calS_\Psi}f_M\to0$ in $L_2(\mu)$, so $g_M\to f$ in $L_2(\mu)$. Bounded elements are therefore dense in $\calS_\Psi^\perp$.
\end{proof}

Conversely, suppose $\langle g_{\rm nom},\xi\rangle_\mu=0$ for every bounded feasible perturbation $\xi\in\calS_\Psi^\perp$. Every nonzero $g\in\calS_\Psi^\perp\cap L_\infty(\mu)$ yields such a perturbation after scaling: taking $\xi=\beta g$ with $\beta=\rho_{\rm int}/\|g\|_\infty$ gives $\|\xi\|_2\le\rho_{\rm int}$ almost surely, so $V+\xi\in\calP$ by Assumption~\ref{assumption:interior_perturbations}, and $\langle g_{\rm nom},g\rangle_\mu=\beta^{-1}\langle g_{\rm nom},\xi\rangle_\mu=0$. Thus $g_{\rm nom}$ is orthogonal to every bounded element of $\calS_\Psi^\perp$. By Lemma~\ref{lem:bounded_density} these are dense in $\calS_\Psi^\perp$, and $g_{\rm nom}\in L_2(\mu)$, so $\langle g_{\rm nom},\cdot\rangle_\mu$ vanishes on all of $\calS_\Psi^\perp$. Since the finite-dimensional span $\calS_\Psi$ is closed, the Hilbert-space projection theorem gives
\begin{equation}
    (\calS_\Psi^\perp)^\perp=\calS_\Psi,
\end{equation}
and hence $g_{\rm nom}\in\calS_\Psi$.

Now suppose $g_{\rm nom}\notin\calS_\Psi$. Let
\begin{equation}
    g_\perp=g_{\rm nom}-\Proj_{\calS_\Psi}g_{\rm nom}.
\end{equation}
Then $g_\perp\in\calS_\Psi^\perp$ and $g_\perp\neq0$. Moreover $g_\perp$ is essentially bounded: by Assumption~\ref{assumption:interior_perturbations} $g_{\rm nom}\in L_\infty(\mu)$, and $\Proj_{\calS_\Psi}g_{\rm nom}\in L_\infty(\mu)$ because $\calS_\Psi$ is finite-dimensional and spanned by bounded functions, so
\begin{equation}
    \|g_\perp\|_\infty\le\|g_{\rm nom}\|_\infty+\|\Proj_{\calS_\Psi}g_{\rm nom}\|_\infty<\infty,
\end{equation}
where $\|\cdot\|_\infty$ denotes the essential supremum of the Euclidean norm. Choose any $0<\alpha\le\rho_{\rm int}/\|g_\perp\|_\infty$. Then $\|\alpha g_\perp(H,V)\|_2\le\rho_{\rm int}$ almost surely, so Assumption~\ref{assumption:interior_perturbations} guarantees that
\begin{equation}
    q_\alpha(H,V)=V-\alpha g_\perp(H,V)
\end{equation}
lies in $\calP$ almost surely. Since $g_\perp\in\calS_\Psi^\perp$, $q_\alpha$ satisfies all exact calibration moments and hence belongs to $\calQ_\Psi$. Finally, by affine utility,
\begin{align}
&\E[U_H(\pi_{\rm nom}(H,V),q_\alpha(H,V))]
-
\E[U_H(\pi_{\rm nom}(H,V),V)]
\notag\\
&\qquad=
\E[g_{\rm nom}(H,V)^\top(q_\alpha(H,V)-V)]
\notag\\
&\qquad=
-\alpha\E[g_{\rm nom}(H,V)^\top g_\perp(H,V)].
\end{align}
Because $g_\perp$ is orthogonal to $\Proj_{\calS_\Psi}g_{\rm nom}$,
\begin{equation}
    \E[g_{\rm nom}^\top g_\perp]
    =
    \E[\|g_\perp\|_2^2],
\end{equation}
which proves~\eqref{eq:strict_loss_from_missing_span}.

\section{Proof of the Bin-Calibration Separation}
\label{app:bin_separation_proof}

We prove Corollary~\ref{cor:bin_separation}. Let $B_1,\ldots,B_J$ be the forecast-bin partition and define
\begin{equation}
    \xi(H,V)=\lambda_j\left(s(H,V)-\E[s(H,V)\mid V\in B_j]\right),
    \qquad V\in B_j.
\end{equation}
Then for every bin $B_j$,
\begin{equation}
    \E[\xi(H,V)\mid V\in B_j]
    =
    \lambda_j\left(
    \E[s\mid V\in B_j]-\E[s\mid V\in B_j]
    \right)
    =0.
\end{equation}
Thus adding $\xi$ to the conditional mean preserves ordinary bin calibration. However,
\begin{align}
    \E[s(H,V)\xi(H,V)]
    &=
    \sum_{j=1}^J
    \Prob(V\in B_j)\lambda_j
    \E\left[s(H,V)\left(s(H,V)-\E[s\mid V\in B_j]\right)\mid V\in B_j\right]
    \\
    &=
    \sum_{j=1}^J
    \Prob(V\in B_j)\lambda_j
    \Var(s(H,V)\mid V\in B_j).
\end{align}
This quantity is positive whenever at least one populated bin has $\lambda_j>0$ and nonzero conditional variance of $s$. The uniform interior margin $\rho_{\rm int}$ permits positive $\lambda_j$ small enough that $V+\xi\in[0,1]$ almost surely.

\section{Sequential Calibration with Delayed Outcomes}
\label{app:sequential_certificate}

Closed-loop data are generally dependent. The following simple result gives a high-probability certificate under bounded martingale-difference noise.

For a lifted outcome spanning $N$ physical steps, use complete non-overlapping windows. Assume $T\ge N$ and let $J=\lfloor T/N\rfloor$, $t_j=1+(j-1)N$ for $j=1,\ldots,J$, and $T_{\rm eff}=J$. Define $\calG_{j-1}$ as the information available at the start of block $j$, including $H_{t_j}$, $V_{t_j}$ and any randomization used to form them. Let $\calG_j$ include that block's revealed outcome and the information available before the next block. Assume these sigma-algebras form a filtration with $Y_{t_j}$ measurable at the end of block $j$. This indexes information by outcome revelation rather than identifying a future-window outcome with an instantaneous observation.
\begin{equation}
Z_{j,i}=\psi_i(H_{t_j},V_{t_j})^\top(Y_{t_j}-V_{t_j}),\qquad
\widehat m_i^{\rm thin}=\frac1J\sum_{j=1}^J Z_{j,i}.
\label{eq:thinned_moment}
\end{equation}
The tests are fixed before observing these held-out outcomes, and $\psi_i(H_{t_j},V_{t_j})$ is $\calG_{j-1}$-measurable. If tests are learned from the same outcomes, separate uniform or sample-splitting arguments are required.

\begin{assumption}[Bounded sequential residuals]
\label{assumption:bounded}
For each test $i$, $|Z_{j,i}|\le B_i$ almost surely, and the centered block increments $D_{j,i}=Z_{j,i}-\E[Z_{j,i}\mid\calG_{j-1}]$ form a martingale-difference sequence with respect to $\{\calG_j\}_{j=0}^J$.
\end{assumption}

\begin{remark}[What the empirical bound certifies]
Assumption~\ref{assumption:bounded} is an empirical-certification assumption. The bound below controls the gap between observed closed-loop calibration moments and the sequential average of the corresponding conditional moments on the calibration run. It does not by itself certify calibration after a controller change, under heavy-tailed residuals without truncation, or under deployment distributions that differ substantially from the calibration distribution $\mu$.
\end{remark}

\begin{lemma}[Sequential calibration confidence bound]
\label{thm:closed_loop_concentration}
Under Assumption~\ref{assumption:bounded}, with probability at least $1-\zeta$,
\begin{equation}
    \left|
    \frac1J\sum_{j=1}^J\E[Z_{j,i}\mid\calG_{j-1}]
    -
    \widehat m_i^{\rm thin}
    \right|
    \le
    2B_i\sqrt{\frac{2\log(2k/\zeta)}{T_{\rm eff}}}
    \label{eq:azuma_bound}
\end{equation}
simultaneously for all $i=1,\ldots,k$.
\end{lemma}

\begin{proof}
The increments obey $|D_{j,i}|\le 2B_i$. Apply Hoeffding--Azuma \citep{hoeffding1963probability,azuma1967weighted} to the $J$ block increments for each fixed test and take a union bound over $i=1,\ldots,k$.
\end{proof}

The constant $2B_i$ reflects the conditional range of the increments; it can be replaced by a tighter conditional-range bound on $Z_{j,i}$ given $\calG_{j-1}$ whenever that range is smaller than $2B_i$.

\begin{remark}[Unthinned data]
If one uses the full-sample estimator $\widehat m_i=T^{-1}\sum_{t=1}^T Z_{t,i}$, where $Z_{t,i}=\psi_i(H_t,V_t)^\top(Y_t-V_t)$, outcomes from overlapping prediction windows have delayed, shared information. The block-filtration proof above does not directly justify a full-sample radius. A separate delayed-feedback or dependence argument is required; a mixing-based analysis is one possible route, but is not established here.
\end{remark}

The episodic certificate in Section~\ref{sec:empirical} identifies a population operating law directly. The present single-run result instead controls the average of conditional moments on that run; an occupation-law argument is needed to connect that average to a deployment population.

\section{Random-Stream Provenance}
\label{app:random_streams}
Tables~\ref{tab:streams_constructed}--\ref{tab:streams_extended} collect the
random streams underlying the reported numerical studies. The saved protocols
and analysis scripts in the companion artifact define the draw order and
record the numerical environment (NumPy 1.26.4). The episode generators first
draw uniform context indices, then uniforms for the binary outcomes. They use
NumPy's PCG64 generator, initialized either by the integer seed or by the
listed SeedSequence entropy and child key. Every index below is zero-based
unless stated otherwise. Different child keys identify separate pseudorandom
streams; they do not make methods evaluated on the same observations
independent. Shared observations and shared bootstrap weights preserve the
paired comparisons throughout.

\begin{table}[htbp]
\centering\small
\caption{Streams for the constructed exact-interface study and its exploratory
calibration-fit replications. Integer seeds initialize separate PCG64
generators without an explicit child key.}
\label{tab:streams_constructed}
\renewcommand{\arraystretch}{1.15}
\begin{tabular}{p{0.24\linewidth}p{0.24\linewidth}p{0.43\linewidth}}
\toprule
Purpose & Seed or entropy & Use and indexing\\
\midrule
Original calibration & $1729+d$ & $10{,}000$ context/outcome pairs per case.\\
Original confirmation & $271828+d$ & $2{,}000$ episodes for the primary and each control; $1{,}000$ per grid cell.\\
Conditional bootstrap & $314159+d$ & $5{,}000$ paired episode-bootstrap replicates for the primary and each control; $2{,}000$ per grid cell. One set of multinomial weights is reused across methods and endpoints within a case.\\
Exploratory calibration refits & $2026111401$, key $(c,r)$ & $200$ fits per case, $r=0,\ldots,199$, each using $10{,}000$ new pairs. Evaluation is exact enumeration, with no confirmation or bootstrap draws.\\
\bottomrule
\end{tabular}
\end{table}

Here $d=0$ for the primary case; $d=100,200,300$ for the zero,
orthogonal and easy controls, respectively; and $d=1000+j$ for grid cell
$j=0,\ldots,23$. Grid cells follow the lexicographic product of
$u_{\max}\in(1.6,2.0,2.4)$, $w_1\in(0.25,0.35,0.45,0.55)$ and
$r\in(0.60,0.70)$, with the last coordinate varying fastest. The refit index
$c$ follows the order primary, zero, orthogonal, easy, then grid cells 0--23.
These rules are implemented in \texttt{run.py}, \texttt{core.py} and
\texttt{fit\_variability.py} under \texttt{experiments/exact\_interface/}.
The model-defined geometry pilot and exact population comparisons require no
random draws. The direct-certificate and historical empirical-Bernstein
calculations reuse the saved confirmation samples.

\begin{table}[htbp]
\centering\small
\caption{Streams for operating transfer, learned forecasts and sampling
diagnostics. Child keys refer to NumPy SeedSequence. All three pipelines in
the fresh validation use the same observations.}
\label{tab:streams_extended}
\renewcommand{\arraystretch}{1.15}
\begin{tabular}{p{0.25\linewidth}p{0.23\linewidth}p{0.43\linewidth}}
\toprule
Purpose & Entropy and key & Use\\
\midrule
Transfer implementation check & $2026111402$ & Exploratory prototype inspected before the final transfer protocol; excluded from reported validation data. The retained protocol does not specify the child key or full draw record.\\
Final constructed transfer validation & $2026111403$, keys $(0)$, $(1)$ & $100{,}000$ episodes for each of the source-only and structured laws, in that order; no bootstrap.\\
Base-forecaster training & $2026111423$, key $(0,0)$ & $20{,}000$ pairs for the single logistic base fit.\\
Learned correction fits & $2026111423$, keys $(1,r)$ & $r=0$ for the original $10{,}000$-pair correction fit; $r=1,\ldots,200$ for the additional calibration refits. The base fit stays fixed.\\
Learned confirmation & $2026111423$, key $(2,0)$ & $2{,}000$ episodes shared by all frozen learned pipelines.\\
Learned conditional bootstrap & $2026111423$, key $(3,0)$ & $2{,}000$ multinomial episode-bootstrap replicates, with weights shared across methods and reported endpoints.\\
Fresh operating-choice validation & $2026111407$, key $(0)$ & One $100{,}000$-episode sample for the prospectively fixed six-moment analysis; no bootstrap or refitting.\\
Retrospective sampling diagnostics & $2026111501$, no child key & One PCG64 stream; $4{,}000$ multinomial replications at each specified design, in the order described below.\\
\bottomrule
\end{tabular}
\end{table}

The extended studies are implemented in \texttt{operating\_transfer.py},
\texttt{learned\_forecast.py}, \texttt{fresh\_shift\_validation.py} and
\texttt{sampling\_diagnostics.py}. The final transfer protocol preserves the
prototype entropy and its exclusion from the final sample, but does not
specify enough information to replay that implementation check exactly;
the missing child keys are therefore not reported. The final streams are fully
specified. Raw episode records are retained for
the original calibration and validation samples, the learned base-training
sample and the fresh validation sample; the additional calibration refits and
bootstrap resampling are specified by their keys and analysis code. Reusing
the learned confirmation sample for the exploratory shift
analysis consumes no additional random stream.

The retrospective diagnostics use counts over the 480 context/outcome atoms,
ordered $(z_0,0),(z_0,1),\ldots,(z_{239},1)$. They first generate $4{,}000$
replications of the learned-pipeline six-moment design with $M=100{,}000$.
The same generator then continues with $4{,}000$ replications of the original
constructed one-moment design at $M=15{,}522$, $22{,}700$ and $30{,}000$, in
that order. These samples describe sampling distributions under the known
laws; they do not replace the saved prospective validation sample. The
probability-feasible bias audit, angular cover and 64-vertex moment-box check
are deterministic and consume no random stream.

\FloatBarrier

\FloatBarrier

\end{document}